WAVEGUIDE, PHOTODETECTOR, AND IMAGING APPLICATIONS OF
MICROSPHERICAL PHOTONICS

by

Kenneth Wayne Allen Jr.

A dissertation submitted to the faculty of
The University of North Carolina at Charlotte
in partial fulfillment of the requirements
for the degree of Doctor of Philosophy in
Optical Science and Engineering

Charlotte

2014

Approved by:

_______________________________
Dr. Vasily N. Astratov

_______________________________
Dr. Michael A. Fiddy

_______________________________
Dr. Yong Zhang

_______________________________
Dr. Greg Gbur

_______________________________
Dr. Nathaniel M. Fried

ABSTRACT

KENNETH WAYNE ALLEN JR. Waveguide, photodetector, and imaging applications of microspherical photonics. (Under the direction of DR. VASILY N. ASTRATOV)

Dielectric microspheres with diameters ($D$) on the order of several wavelengths of light ($\lambda$) have attracted increasing attention from the photonics community due to their ability to produce extraordinarily tightly focused beams termed "photonic nanojets," to be used as microlenses for achieving optical super-resolution or to develop sensors based on whispering gallery mode resonances. In this dissertation, we study the optical properties of more complicated structures formed by multiple spheres which can be assembled as linear chains, clusters or arrays, integrated with waveguides or embedded inside other materials to achieve new optical properties or device functionalities.

For linear chains of polystyrene microspheres ($n$=1.59), we observed a transition from the regime of geometrical optics (at $D$>20$\lambda$) to the regime of wave optics (at $D$<20$\lambda$). We showed that this transition is accompanied by a dramatic change of focusing and optical transport properties of microsphere-chain waveguides. The results are found to be in qualitative agreement with numerical modeling.

We developed, designed, and tested a single-mode microprobe device based on spheres integrated with a waveguide for ultraprecise laser surgery. Our design is optimized using a hollow-core microstructured fiber as a delivery system with a single-mode Er:YAG laser operating at $\lambda$=2.94 micron. Using a high-index ($n$~1.7-1.9) microsphere as the focusing element we demonstrate experimentally a beam waist of ~4$\lambda$, which is sufficiently small for achieving ultraprecise surgery.

For embedded microspherical arrays, we developed a technology to incorporate high-index ($n$~1.9-2.1) spheres inside thin-films made from polydimethylsiloxane (PDMS). We showed that by using liquid lubrication, such thin-films can be translated along the surface to investigate structures and align different spheres with various objects. Rigorous resolution treatment was implemented and we demonstrated a resolution of ~$\lambda$/7 which can be obtained by such thin-films.

We experimentally demonstrated that microspheres integrated with mid-IR photodetectors produce up to 100 times photocurrent enhancement over a broad range of wavelengths from 2 to 5 microns. This effect is explained by an increased power density produced by the photonic jet coupled to the active device layers through the photodetector mesas. The photocurrent gain provided by photonic jets is found to be in good agreement with the numerical modeling.

# DEDICATION

*I dedicate this dissertation to my…*

*…wife for her unconditional support, encouragement, and unwavering faith in me.*

*…parents for their guidance and persistent love.*

*…sisters for being my first teachers.*

## ACKNOWLEDGMENTS

Foremost I wish my deepest gratitude to my advisor, Dr. Vasily N. Astratov. Over the course of the past 5 years he has patiently and continuously supported my research efforts with immense knowledge and insightful discussions. During this time period, he advised me through my bachelor's thesis, master's thesis, and doctoral dissertation with the utmost enthusiasm. He has diligently prepared me for all of my scientific presentations, ensuring that the information is disseminated as clearly and effectively as possible. It has been my privilege to have learned from his experience, often expressed in long-winded anecdotes. Without his guidance and impetus the work presented in this dissertation could not have been accomplished.

I would also like to thank my committee members, Dr. Michael A. Fiddy, Dr. Greg Gbur, Dr. Nathaniel M. Fried, and Dr. Yong Zhang for serving on my committee, their time and comments are much appreciated. I would like to thank Dr. Glenn Boreman for providing me with advice and support regarding my future career opportunities. I am grateful to Dr. Pedram Leilabady, Dr. Susan Trammell and Dr. Angela Davies for their invaluable advice and guidance through my undergraduate and graduate years.

I am grateful to Dr. Michael A. Fiddy for his insights on super-resolution in our colloquial conversations, and to Dr. Nathaniel M. Fried for stimulating conversations on surgical applications of lasers.

I am very thankful to Dr. Nicholaos I. Limberopoulos, Dr. Augustine M. Urbas, and Dr. Dean P. Brown for enabling my internship at the Air Force Research Laboratory at Wright-Patterson Air Force Base. I am also thankful for the stimulating interactions I experienced during my internship with Dr. Dennis E. Walker Jr., Dr. Gamini Ariyawansa, Dr. Jarrett H. Vella, Joshua M. Duran, Imran Vakil, Michael Noyola, and John Goldsmith.

My research benefited from multiple collaborations of the Mesophotonics Laboratory. These include our collaborations with ophthalmologists, Dr. Howard Ying (Johns Hopkins University) and Dr. Andrew Antoszyk (Charlotte Ear Nose and Throat Associates), our collaboration with an optoelectronics expert, Dr. Anatolie Lupu (Institute for Fundamental Electronics, Paris), and our collaboration with the producers of the microstructured optical fibers, Dr. Alexey F. Kosolapov and his team (Fiber Optics Research Center, Moscow).

During my graduate studies I had the privilege of being a part of an interactive research environment in the Mesophotonics Laboratory. I am grateful for the relationships forged from this experience with Dr. Oleksiy Svitelsky, Dr. Arash Darafsheh, S. Adam Burand, Yangcheng Li, Farzaneh Abolmaali, and Navid Farahi. I would also like to thank fellow graduate students Jason S. D. Roberts, Jason R. Case, Mark Green, Frances Bodrucki, and Max Burnett, for their support and productive discussions.

I am grateful to Dr. Lou Deguzman, Dr. Robert Hudgins, and Scott Williams for their patience while teaching me how to operate the equipment in the UNCC Optics center. Also, I am thankful to Mark Clayton, Wendy Ramirez, and Elizabeth Butler for assisting me throughout numerous endeavors.

I was supported by the GASP award throughout my graduate studies. I am thankful for the Department of Physics and Optical Sciences for supporting me with teaching assistantships. My graduate research assistantships in Dr. Vasily N. Astratov's

Mesophotonics Laboratory were supported through his grants from the National Science Foundation, National Institutes of Health, and Army Research Office.

Lastly, none of this would have been possible without the love, support, and patience of my family to whom this dissertation is dedicated.

TABLE OF CONTENTS

## LIST OF FIGURES

sphere, through the objective lens then the emitted light is collected for imaging through the same objective.

## LIST OF ABBREVIATIONS

| | |
|---|---|
| 2-D | two-dimensional |
| 3-D | three-dimensional |
| ~ | approximately |
| ≈ | approximately equal to |
| $\alpha$ | attenuation coefficient |
| a | radius |
| Au | gold |
| au | arbitrary units |
| As | arsenic |
| AFM | atomic force microscopy |
| $B_i$ | incident magnetic field |
| $BaTiO_3$ | barium titanate |
| C | cadmium |
| CaFl | calcium fluoride |
| CCD | charge-coupled device |
| $\Delta$ | change |
| dB | decibel |
| °C | degrees Celsius |
| $D$ | diameter |
| $\eta$ | efficiency parameter |
| $E_i$ | incident electric field |
| $\varepsilon$ | permittivity |

Er                     erbium

*ex vivo*              Latin: (out of the living)

FDTD                   finite-difference time-domain

FOV                    field-of-view

FEM                    finite element method

FPA                    focal plane array

FTIR                   Fourier transform infrared

FWHM                   full width-half maximum

FM                     fundamental mode

*g*                    gap

Ga                     gallium

$GeO_2$                germanium oxide

GRIN                   graded-index

HC-MOF                 hollow-core microstructured optical fiber

Hg                     mercury

Ho                     holmium

HWG                    hollow waveguide

*in vivo*              Latin: (within the living)

ID                     inner diameter

In                     indium

IPA                    isopropyl alcohol

IR                     infrared

J                      joule

| | |
|---|---|
| K | potassium |
| KrF | krypton fluoride |
| KTP | potassium titanyl phosphate |
| λ | wavelength |
| $l$ | angular mode number |
| LASIK | laser-assisted in situ keratomileusis |
| LED | light emitting diode |
| μ | permeability |
| μ- | micro- $(10^{-6})$ |
| f- | femto- $(10^{-15})$ |
| $m$ | azimuthal mode number |
| m | meter |
| m- | milli- $(10^{-3})$ |
| min | minute |
| MBE | molecular beam epitaxy |
| MCW | microsphere-chain waveguide |
| MWIR | mid-wave infrared |
| NCW | nanosphere-chain waveguide |
| NIM | nanojet-induced mode |
| NSOM | near-field scanning optical microscope |
| N | number |
| $n$ | index of refraction |
| n- | nano- $(10^{-9})$ |
| NA | numerical aperture |
| OD | outer diameter |
| PDMS | polydimethylsiloxane |
| PDR | proliferative diabetic retinopathy |
| PFM | periodically-focused mode |
| PM | photonic molecule |
| PML | perfectly matched layer |
| PSF | point spread function |
| $q$ | radial mode number |
| $Q$ | quality factor |
| QDIP | quantum dot infrared photodetector |
| $R$ | radius of curvature |
| r | radius of sphere |
| S- | source |
| Sb | antimony |
| SEM | scanning electron micoscope |
| s *or* sec | second |
| SRH | Shockley-Read-Hall |
| SLS | strained-layer superlattice |
| $SiO_2$ | silicon dioxide |
| Te | tellurium |
| TE | transverse electric |
| TM | transverse magnetic |
| $\theta_i$ | incident angle |
| $\theta_r$ | refracted angle |
| $\theta_B$ | Brewster's angle |
| 3-D | three dimensional |
| UV | ultraviolet |
| W | watt |
| WGM | whispering-gallery mode |
| YAG | yttrium aluminum garnet |
| X | size parameter |

# CHAPTER 1: INTRODUCTION

## 1.1: Outline and Overview of the Dissertation

In this dissertation, we investigate optical waves interacting with dielectric spheres from a single element to complex configurations of multiple spheres. These structures exhibit strong lensing effects which allow for the relay of light within the arranged spheres [1-24]. Also, due to the circular symmetry, the light waves can be confined within the cavities manifesting photonic states [25-50]. A variety of optical effects can be observed in these structures such as optical transport phenomena [1-10], enhanced imaging capabilities [51-70], and highly concentrated optical beams [71-100].

Microspherical photonics is a term that recently emerged [49-50, 113] in the context of developing novel ways of sorting "photonic atoms" with overlapping positions of uncoupled whispering-gallery mode (WGM) resonances. The idea of this technology is based on using the phenomenon of resonant radiative pressure for sorting microspheres with nearly indistinguishable positions of WGMs called "photonic atoms". Classical photonic atoms can behave similar to quantum mechanical atoms in photonic applications in the sense that the tight-binding approximation can be implemented to engineer photonic dispersions in the structures [114-116]. In this dissertation, we will use the terminology of microspherical photonics in a broader sense, to encompass structures with non-resonant phenomena, as well as WGMs. Such structures would still possess photonic nanojets [73] and nanojet-induced modes [13]. The coupling between WGMs can be weakened in such structures [117], however coupling can still be provided with smaller efficiency when compared to resonant coupling. Basically, the use of the term "microspherical photonics" encompasses any structures formed by microspheres including resonant and non-resonant cases, regular and high-index spheres, and arrays of high-index spheres embedded in elastomeric transparent materials.

Chapter 1 is a literature review on relevant material existing in the field of study. The introductory chapter is delineated so that the sub-sections are congruent with the dissertation chapters.

In Chapter 2, we investigate light focusing and transport properties in microsphere-chain waveguides and photonic molecules. First, a non-resonant transport mechanism is studied where it is observed that polystyrene microsphere-chain waveguides (MCWs) coupled to a multimodal source form periodically focused beams with progressively smaller beam waists along the chain [1-10]. It is shown that a "beam tapering" effect occurs in MCWs formed by spheres with diameters ranging from $4\lambda$ to $10\lambda$, where $\lambda$ is the wavelength of light, but it does not occur for larger spheres ($D>10\lambda$). It is demonstrated that the propagation losses (~0.1 dB/sphere) in mesoscale MCWs are far below any estimations based on geometrical optics [112]. The ability of wavelength-scale MCWs to filter certain polarization eigenstates is described, demonstrating their potential use as polarization filters [5]. Second, a resonant transport mechanism is investigated using finite-difference time-domain (FDTD) simulations, where clusters of circular resonators with degenerate whispering gallery mode (WGM) resonances are coupled [25]. These arrangements of multiple spheres, acting as photonic atoms, form photonic molecules which exhibit interesting spectral and optical transport properties [25, 33, 101-110]. Using evanescent couplers, we show that various configurations of cavities display distinct spectral features such as mode splitting, in the vicinity of their WGM resonances [25]. These results can be used for understanding mechanisms of light transport in complex optical networks, and for developing spectral filters, delay lines, microspectrometers, and spectral markers.

In Chapter 3, we designed and tested the concept for a novel single-mode precision laser scalpel [71]. We developed single-mode designs of such devices providing higher efficiency and significantly smaller focal spot sizes compared to multimodal systems [195]. The proposed single-mode systems include: i) diode-pumped Er:YAG laser source operating at the wavelength corresponding to the maximal water absorption peak in the tissue ($\lambda$=2.94 $\mu$m), ii) low-loss hollow-core microstructured fiber delivery, and iii) high-index ($n$~1.8) focusing barium-titanate glass microsphere integrated at the distal tip of the fiber. By testing the system in air we demonstrated the focal spot diameters to be less than ~4$\lambda$. Our numerical modeling shows that, in principal, it is possible to achieve diffraction-limited spot sizes on the order of ~$\lambda$/2. Since the photonic jet is produced at the back surface of the spheres it is not strongly affected by the presence of fluid, which can exist in medical applications such as ultraprecise contact intraocular, brain or cellular microsurgeries.

In Chapter 4, we developed a novel microscopy component for microsphere-assisted optical super-resolution imaging, for practical applications [70]. We fabricated thin-films made from polydimethylsiloxane (PDMS) with embedded high-index ($n$~1.9-2.2) microspheres for super-resolution imaging applications. To precisely control the position of microspheres, the films can be translated along the surface of the nanoplasmonic structure to be imaged, with proper lubrication provided by a thin lubricating layer of isopropyl alcohol (IPA). The lateral magnification is observed to be larger prior to the evaporation of the lubricating layer. As the lubricating layer evaporates, there is a boost in the resolution of the images produced through the microspheres. Microsphere-assisted imaging, through these matrices, provided lateral resolution of ~$\lambda$/7 of nanoplasmonic dimer arrays with an illuminating wavelength $\lambda$=405 nm. These thin films can be used as contact optical components to boost the resolution capability of conventional microscope systems.

In Chapter 5, we developed a method for enhancing the photocurrent of photodetectors by coupling photonic jets into their light sensitive regions [72]. Photonic jets are light beams focused by dielectric microspheres down to subwavelength dimensions [73]. In this dissertation, we show that they can be used for enhancing the performance of strained-layer superlattice (SLS) infrared (IR) photodiodes in the midwave-infrared spectral band ($\lambda$=3-5 $\mu$m). We optimized the design of these structures using FDTD simulations and experimentally demonstrated the increased sensitivity compared to conventional SLS photodetectors.

In Chapter 6, we draw conclusions and propose future work that would expand upon the content in this dissertation.

## 1.2: Non-Resonant Optical Effects of Dielectric Spheres

In this section, we discuss non-resonant mechanisms of light interacting with dielectric spheres. First, we will review the light interaction under the considerations of geometrical optics. Secondly, we discuss wave optics effects emerging from a single wavelength scale sphere. Following this discussion, we describe the optical properties emerging from chains of dielectric spheres.

### 1.2.1: Dielectric Sphere as a Thick Lens

This section is dedicated to the analysis of the focusing properties of a spherical thick lens with a geometrical optics approach. Assuming the wavelength of the illuminating source is much smaller than the size of the object, this is a valid approximation. The sphere is a special case of the thick lens, where the front and back radii of curvature, $R_1$ and $R_2$ respectively, are equal to half of the separation distance between the principal planes, $d$. For the case of spherical geometry, the refractive index contrast between the lens and the surrounding medium, $n'$, and the diameter, $D$, are the only variables of the lens determining the effective focal length, $f$, of the system. Considering an object at infinity, the effective focal length can be mathematically expressed with the equation below.

$$f = \frac{n'D}{4(n'-1)} \tag{1.1}$$

It can be seen that if $n'>2$ the $f<D/2$, hence the light will be focused inside the lens. Similarly, if $n'<2$ the $f>D/2$, the light will be focused outside the lens. For many applications it is advantageous to provide focusing of light at the back surface of the sphere. By substituting $f=D/2$, it can be seen that this condition is satisfied when $n'=2$.

### 1.2.2: Photonic Nanojets

In the previous sub-section, discussing the sphere as a thick lens, the geometric properties were expressed. However, this is only valid when the wavelength of the illuminating light wave is significantly smaller than the sphere diameter. If we consider a sphere within the limits where geometrical optics are valid, and apply wave optics assumptions, the incident plane wave will form an Airy pattern irradiance distribution with a FWHM~$\lambda$/2, ignoring aberration. However, spherical aberration will cause the FWHM to exceed this value. This section of the chapter explores interesting optical effects that occur when the lens diameter is comparable to the wavelength of the illuminating light wave.

This topic of wavelength-scale spheres and cylinders has sparked a significant body of literature since the pioneering paper was published in 2004, by Z. G. Chen, A. Taflove, and V. Backman [73]. It should be noted, that the possibility of achieving focused beams with sizes less than diffraction-limited dimensions has been pointed out by ref. [111], however it was only after the comprehensive study of this effect in ref. [73] that this area attracted great attention from the photonics community. The topic of the origins of the photonic nanojet is addressed in further detail in Appendix C.

In ref. [73], these optical effects were studied by two dimensional (2-D) FDTD computational solutions to Maxwell's equations, where they coined the term "photonic nanojets". The physical mechanism that manifests this strong optical confinement is the result of a complex interplay of scattering and interference effects. For a lossless dielectric cylinder ($D$=5 $\mu$m) in free space ($n$=1) the photonic nanojet can be seen to evolve, Fig. 1.1, for different refractive indices of cylinders. As the refractive index is reduced from $n'$=3.5, 2.5, to 1.7, the photonic nanojet emerges out to the back surface of the lossless dielectric cylinder with a narrow beamwaist which is elongated. The photonic nanojet produced by the cylinder of $n'$=1.7 illuminated by a plane wave $\lambda$=500 nm, in Fig. 1.1(c), propagates along the optical axis with subdiffraction-limited lateral dimensions, FWHM~200nm, for a distance of ~900 nm.

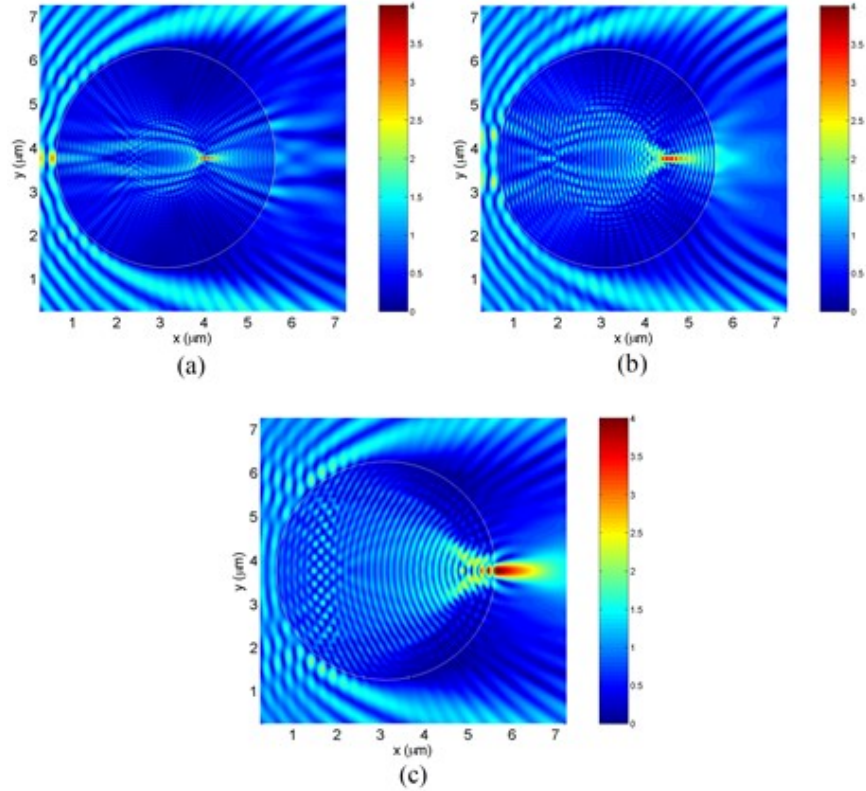

Figure 1.1: Evolution of the formation of photonic nanojet for different combinations of refractive index of the microcylinder ($D$=5 $\mu$m) in free space ($n$=1) illuminated by a planewave ($\lambda$=500 nm), (a) $n'$=3.5, (b) $n'$=2.5, and (c) $n'$=1.7 [73].

Expanding upon the research performed on 2-D microcylinders, dielectric microspheres were studied by FDTD computations of Maxwell's equations for the 3-D case [74], illustrated in Fig. 1.2. It is observed that dielectric spheres illuminated with plane waves $\lambda$=400 nm, generate photonic nanojets on the shadow side with beamwaists as small as FWHM~130 nm, shown in Fig. 1.2(a). The $D$=3.5 $\mu$m sphere produced a photonic jet with a peak intensity ~160x greater than the illuminating wave, along with a subdiffraction-limited FWHM~190 nm. Under these conditions, spheres with $D$<10$\lambda$ provide subdiffraction-limited beamwaists. It was also observed that the backscattered field intensity from a nanoparticle can be enhanced up to 8 orders of magnitude.

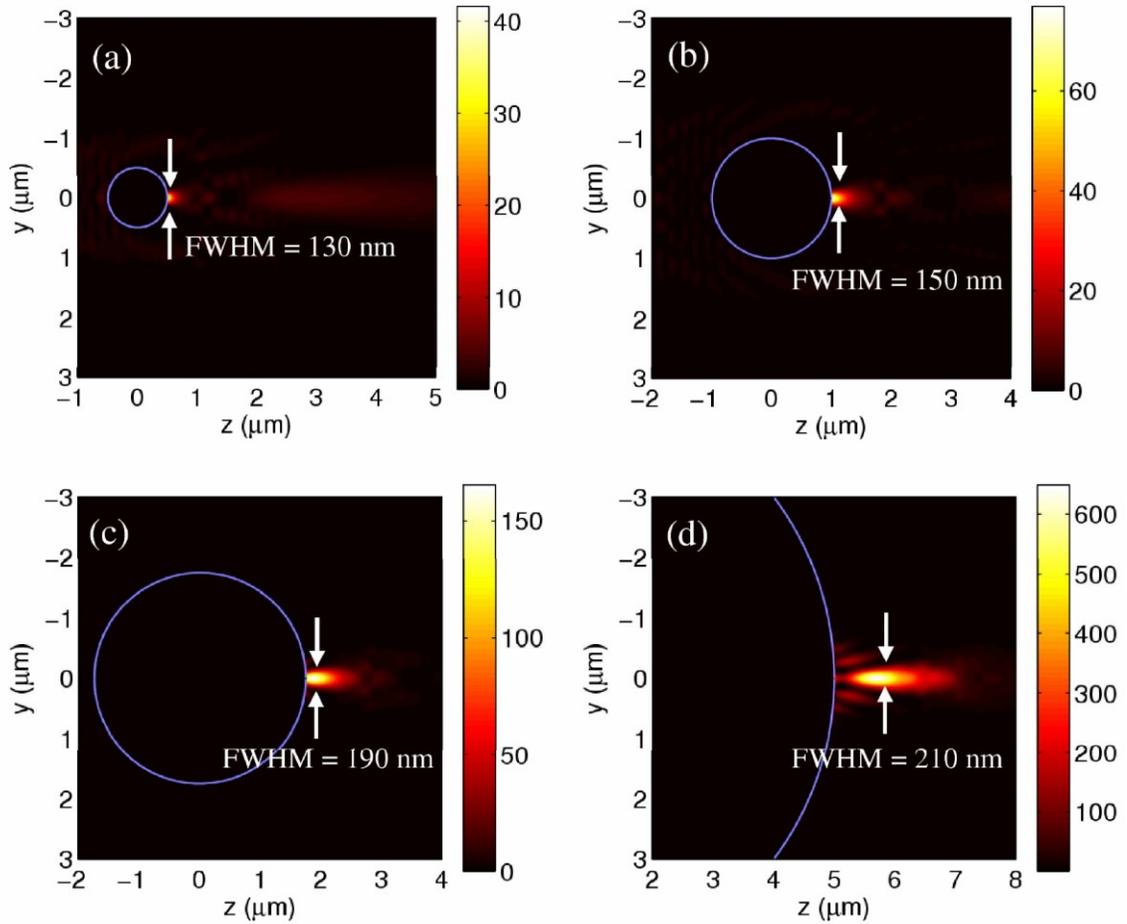

Figure 1.2: Photonic nanojets generated by dielectric spheres ($n'$=1.59) illuminated by a plane wave in free-space ($\lambda$=400 nm), with the electric field vector oscillating along the $x$-axis and propagating in the $z$-axis. (a) Sphere diameter, $D$=1 $\mu$m. (b) $D$=2 $\mu$m. (c) $D$=3.5 $\mu$m. (d) $D$=8 $\mu$m [74].

After FDTD computational studies were performed of photonic nanojets produced by spheres and cylinders [73, 74], experimental investigations were conducted in an attempt to directly observe such photonic nanojets [79]. Latex microspheres ($n'$=1.6) with diameters ($D$=1, 3, and 5 $\mu$m) were deposited on a coverslip (borosilicate glass, refractive index of 1.51, with a thickness of 150 $\mu$m). In order to visualize photonic nanojets, they used a custom laser scanning confocal microscopy system with an illuminating source of

*λ*=520 nm. The FWHM of the nanojets produced were 300, 270, and 320 nm, with enhancement factors of 3, 29, and 59x, for spheres with the *D*=1, 3, and 5 *μ*m, respectively. Numerical simulations performed using quasi-exact Mie theory are in good agreement with the experimentally obtained results, which are valid due to the wavelength-scale size of these spheres [79].

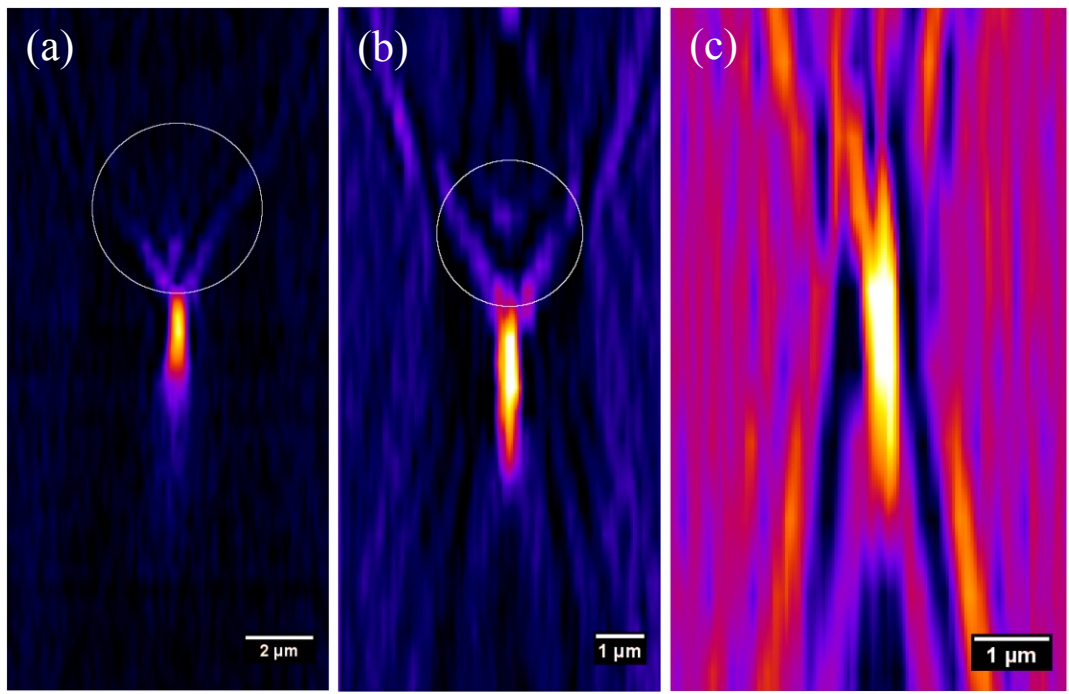

Figure 1.3: Direct observation of photonic jets produced by latex microspheres illuminated by plane waves of *λ*=520 nm. (a) Sphere diameter, *D*=5 *μ*m. (b) *D*=3 *μ*m. (c) *D*=1 *μ*m [79].

Photonic nanojets are narrow, high-intensity electromagnetic output beams that emerge from a lossless dielectric cylinder or sphere. There are certain optical properties that distinguish this effect from standard focusing:

(1) Extraordinarily small full-width at half-maximum (FWHM) can be obtained, down to values of ~$\lambda$/3 at the minimal waist.

(2) The transverse beamwidth along the direction of propagation can maintain dimensions smaller than the classical diffraction limit.

(3) As a non-resonant phenomena, this optical effect is present for diameters spanning ~$2\lambda$ to ~$20\lambda$, under the condition that the refractive index contrast is less than 2:1.

(4) Due to the field concentration, the intensity can substantially surpass that of the illuminating wave by more than 100 times.

(5) Far-field backscattering of the field decays as a third power law of the diameter, versus the significantly faster decay of the sixth power law of Rayleigh scattering.

Photonic jets can be engineered by adjusting the illumination conditions of the input beam such as polarization, phase, and wave front [87]. This was studied by measuring the nanojet's amplitude and phase with a high resolution interference microscope. Ref. [87] studied spatial beam profiles, along with polarization and wavelength of illumination producing photonic nanojets.

For the experiments in [87] a cylindrical vector beam is used, forming doughnut-shaped and two-half-circles spatial beam profiles when focused depending on the polarization state. This azimuthally polarized cylindrical vector beam illuminated a $D$=1, 3, and 10 $\mu$m borosilicate glass spheres. For the smallest sphere, $D$=1 $\mu$m sphere, as shown in Fig. 1.4(b), only one portion of the illumination, either a bright or a dark part, is exposed to the sphere. Thus, a nanojet with one spot is observed when the sphere is in the bright intensity area because the sphere is essentially illuminated by quasi-plane waves. Figure 1.4(c) shows the $D$=3 $\mu$m sphere, which is covered by both dark and bright parts of the spatial profile of the illuminating beam. Since the sphere is smaller than the illumination spot, a portion of the direct illumination is still observed at the focal plane of the nanojet as shown in Fig. 1.4(c). In the case that the sphere $D$ extends beyond the illumination spot, such as a 10 $\mu$m sphere, the nanojet takes the form of the illumination as presented in Fig. 1.4(d). Beam shape control produces the most advantageous results as the sphere extends beyond, or is equal to, the illuminating spot size.

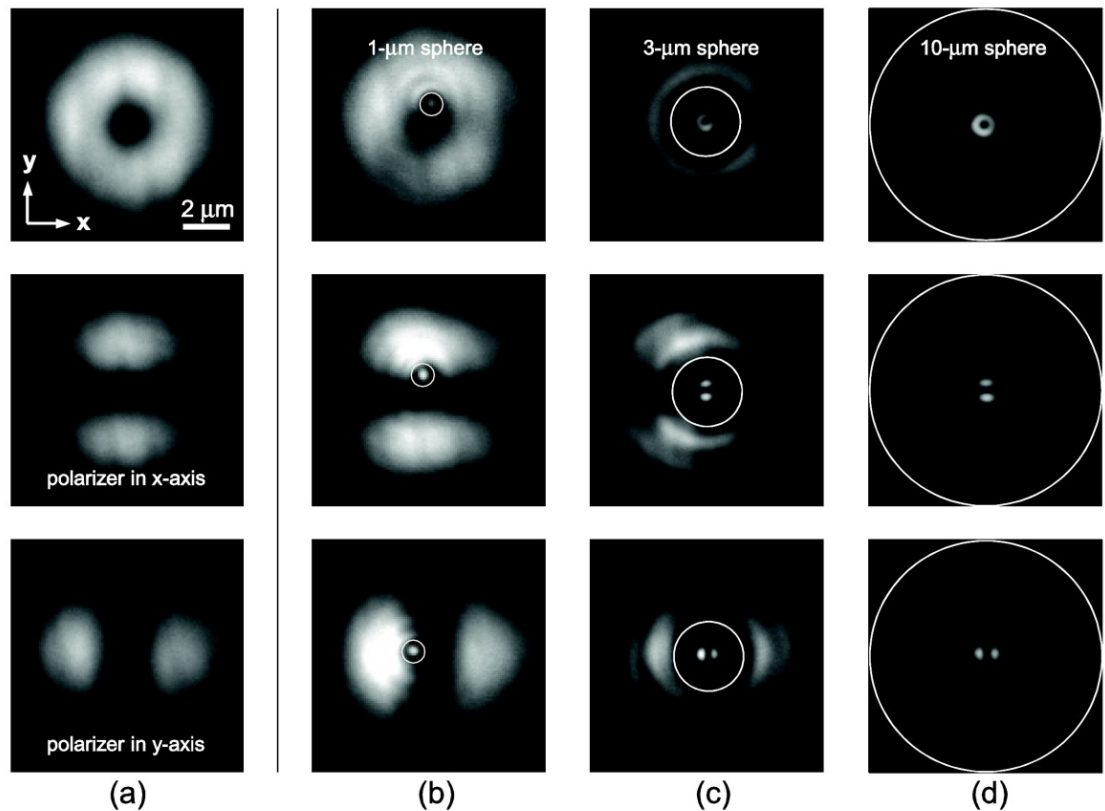

Figure 1.4: Grayscale CCD images of particular illumination conditions and corresponding nanojet by spheres of different size: (a) illumination spots at the entrance plane without a sphere, (b) $D$=1 $\mu$m, (c) $D$=3 $\mu$m, and (d) $D$=10 $\mu$m sphere. The white circle indicates microspheres. Image size is 10 x 10 $\mu$m$^2$ [87].

### 1.2.3: Nanojet Induced Modes and Periodically Focused Modes

Previous research showed that the photonic nanojet can be engineered by altering the beam properties of the optical wave illuminating the sphere. However, in the case that there are multimodal illumination conditions the light transport is more complex. A single sphere is no longer sufficient to provide the tightly focused output beams associated with the photonic nanojet. However, for a linear chain of wavelength-scale spheres, nanojet-induced modes (NIMs) emerge as the light propagates through the linear chain of spheres. The consequence of this filtering process leads to progressively smaller spatial beam profiles at the interface between spheres, as shown in Fig. 1.5. This concept was observed in a linear chain of touching 2.9 $\mu$m polystyrene spheres, integrated with a fluorescent ($\lambda$=530 nm) dye-doped sphere as the source of light. [13]

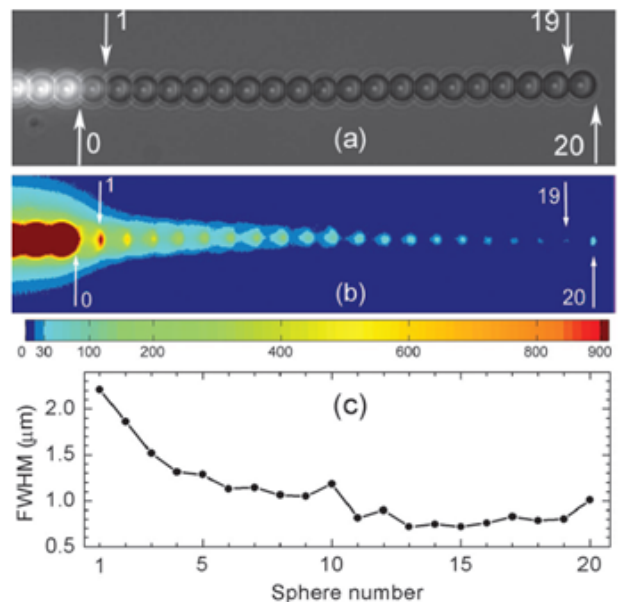

Figure 1.5: (a) Optical micrograph of a linear chain of 2.9 $\mu$m polystyrene microspheres, integrated with fluorescently dye-doped spheres as the source of light. (b) Pseudo-color plot of illustration the beam reduction as the light propagates through the spheres. (c) Cross-sectional FWHM of the bright spots at the interfaces between the undoped spheres [13].

It should be noted that the optical transport in chains of coupled microspheres was theoretically considered in ref. [12] by generalized multiparticle Mie theory. This work introduced two mechanisms of optical transport, *i*) coupling between WGMs and *ii*) periodical beam relaying, as shown in Fig. 1.6. This work played an important role in stimulating further experiments in chains of spheres. However, the concept of nanojet-induced modes with 2*D* period and the minimal losses has not been introduced in this work.

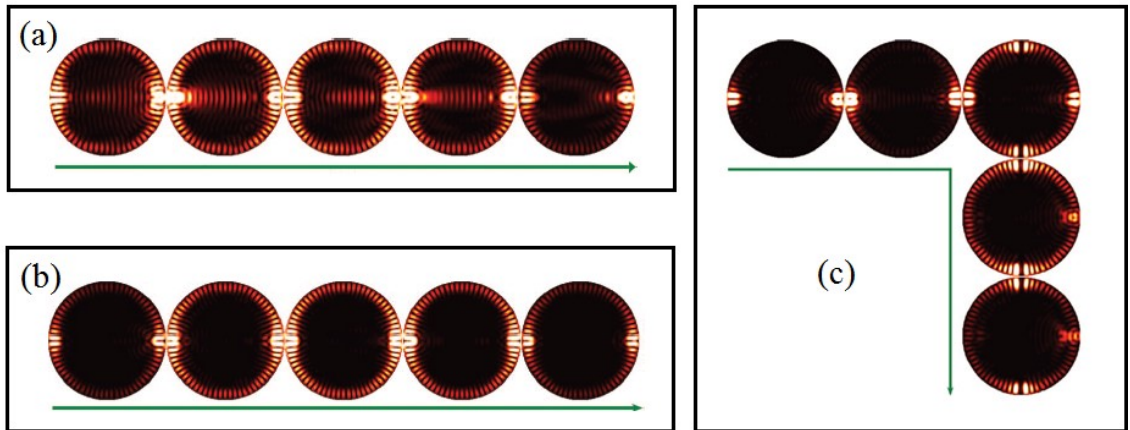

Figure 1.6: Illustration of resonant and non-resonant transport through chains of *D*=3 μm dielectric spheres [12].

Further investigations of these NIMs lead to spatially resolved spectroscopy measurements of long straight chains of touching 5 μm polystyrene microspheres [14]. Two cases were studied (*i*) the source sphere was touching the chain so that multiple modes are coupled, and (*ii*) the source sphere was separated so that only quasi-plane waves are illuminating the chain, as illustrated in Fig. 1.7. The intensity distributions, in Fig. 1.7, illustrate that in case (*i*) there are two NIMs excited in the chain offset by one sphere. This comes from the fact that due to 2*D* periodicity the field maxima are expected after every other sphere. However, experimentally, the maxima are found to present after each sphere.

This interpretation is supported by Fig. 1.7 illustrating the case (*ii*) with the separated light source. In this case, the illumination is provided by quasi-plane waves which are coupled to only one fundamental NIM. As a result, the dominant peaks in Fig. 1.7 have 2*D* period.

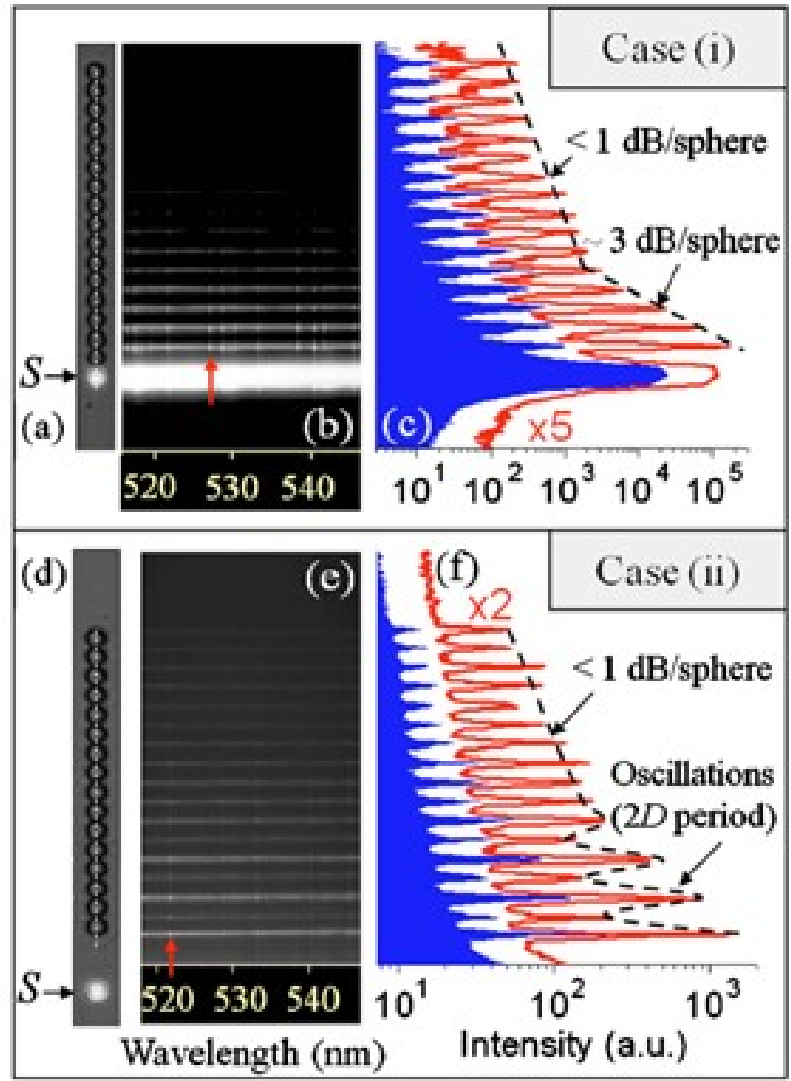

Figure 1.7: Scattering spectroscopy: (a) image of the chain in contact with the dye-doped source sphere with the central section indicated by the dashed line, (b) scattering spectral image at the central section of the equatorial plane, and (c) the red line indicates that the spheres are at the resonance wavelength while the blue curve is off resonance. [(d)-(f)] Similar plots where the source sphere is located ~12 $\mu$m from the chain [14].

Direct measurements of the transmission spectra were performed for longer chains of spheres, 20≤$N$≤88 where $N$ is the number of spheres [14]. These measurements have been performed by using a double-pass geometry for the light; the excitation wavelength propagates through the chain and excites the source sphere, which then emits light back through the chain of spheres. Sequentially, spheres were removed and the evolution of the transmission spectra was measured, as shown in Fig. 1.8. Extraordinarily low attenuation values of 0.08 dB per sphere were observed for these NIMs. The combination of focusing properties and extraordinarily low attenuation values make structures that support NIMs attractive for applications [14].

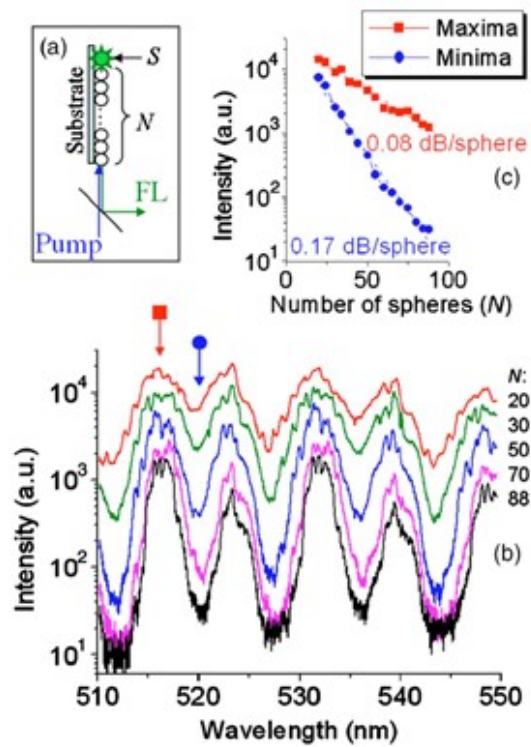

Figure 1.8: Transmission spectroscopy: (a) experimental setup, (b) double pass spectra measured for chains with 20≤$N$≤88, (c) intensities at the maximum at 516.3 nm (squares), and at the minimum of transmission at 520.0 nm (circles) plotted as a function of $N$. The straight lines represent an exponential fit to the experimental data [14].

Some fundamental optical properties of NIMs have been understood in the limit of geometrical optics [112]. In this limit, the phase properties of the light beam are not considered; however, the periodical focusing effects can be easily evaluated by performing numerical ray tracing. Clearly this approach is applicable only to sufficiently large spheres ($D/\lambda >> 20$). In the limit of geometrical optics, the corresponding modes with $2D$ period and minimal propagation losses have been termed "periodically focused modes" (PFMs) [112]. These modes can be considered the geometrical optics analog of the NIMs introduced previously for smaller spheres.

If the conditions are met, such that the refractive index contrast between the spheres and the background is $\sqrt{3}$, and the axial displacement of the rays are at a height so that the incident rays are at Brewster's angle ($\theta_B$), there will be a spatially reproducing pattern of the trajectory of the rays with a period of $2D$. Considering the symmetry about the optical axis, when the PFMs are defocused, they produce a radially polarized cylindrical vector beam and a longitudinally polarized focused beam. For a multimodal source coupled to chains of spheres, it is observed that the chains work as filters of such modes at $1.72 < n < 1.85$, leading to tapering of the focused beams and simultaneously reducing the attenuation of the light propagating within the chain, as shown in Fig. 1.9. The local minima in the attenuation curve presented in Fig. 1.9(d) indicates a mode which is supported by the chain of dielectric spheres. Also, in Fig. 1.9(c) the most drastic beam focusing occurs at the same index as the local minima in attenuation. This is a direct indicator that the periodically focused modes are supported by the chain of geometrical optics scale spheres.

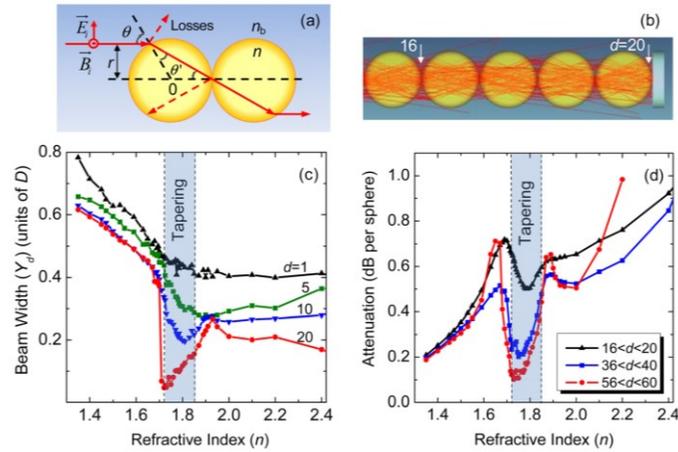

Figure 1.9: (a) TM polarized ray with the directions of incident electric and magnetic vectors indicated, (b) ray tracing from 16th to 20th sphere for $n$=1.785, (c) FWHM of the central peak calculated as a function of $n$ at different distances from the source, (d) attenuation per sphere as a function of $n$ calculated for different segments of the chain [112].

### 1.2.4: Whispering Gallery Modes

Optical waves can be trapped in cavities with circular symmetry, such as microrings [37], microcylinders [38], and microspheres [26-36], due to total internal reflection of light and interference phenomena. This subsection of the dissertation is dedicated to whispering gallery modes in spherical resonators.

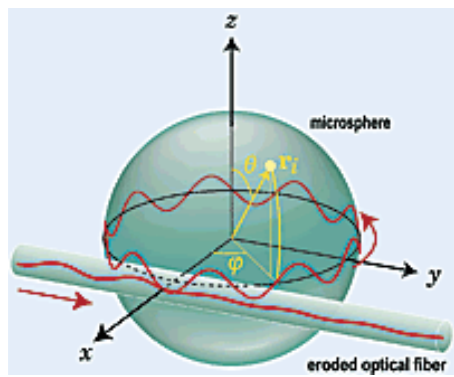

Figure 1.10: Whispering gallery mode resonance inside of a spherical cavity, excited by side-coupling provided by an eroded optical fiber [36].

Due to spherical symmetry, whispering gallery modes (WGMs) in microspheres are characterized [43, 44] by radial $q$, angular $l$, and azimuthal $m$ mode numbers. The radial number, $q$, indicates the number of WGM intensity maxima along the radial direction, whereas the angular number, $l$, represents the number of modal wavelengths that fit into the circumference of the equatorial plane of the dielectric sphere.

These electromagnetic waves propagate inside the dielectric sphere close to its surface so that their trajectory traverse a distance of $\sim 2\pi a$ in a round trip, where $a$ is the sphere radius. The condition of constructive interference of the waves in the cavity with circular symmetry can be approximated as $2\pi a \approx l(\lambda/n)$, where $\lambda/n$ is the wavelength in the medium with refractive index $n$ [33]. When this constructive interference condition is satisfied the light undergoes harmonious trajectory and standing WGMs are formed [118]. This condition can be expressed in terms of the size parameter, $X = 2\pi a/\lambda \approx l/n$.

In an ideal free-standing sphere, without any degree of ellipticity or surface degradation, the azimuthal modes represented by $m$ numbers are degenerate. However, in practical experimental conditions, this degeneracy is broken by deformations of the microspheres causing deviations from a spherical shape. A single sphere on a substrate is rotationally invariant around the normal vector from the contact point, $z$-axis, defining the polar axis. Thus, the fundamental whispering gallery modes with $m=l$ are defined [119] by the equatorial plane of the sphere, that is parallel to the supporting substrate. These fundamental modes have the highest $Q$ factors, due to the separation from the substrate by the radius of the sphere [33]. In contrast, the modes with $m \ll l$ are dampened, due to the fact that they have a spatial distribution with maximum intensity approaching the substrate of high index of refraction for the whispering gallery modes with $q=2$ [33].

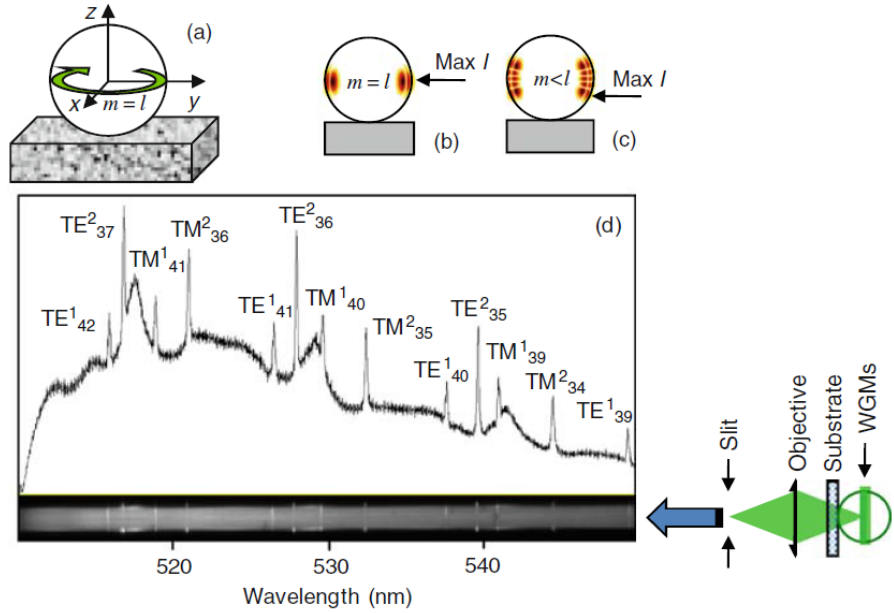

Figure 1.11: (a–c) Schematics of a single sphere on a substrate illustrating intensity maxima distributions for azimuthal modes with different m numbers. (d) Fluorescence spectral image and emission spectrum of a single green fluorescent $D$=5.0 $\mu$m sphere with WGM peaks measured through the transparent substrate [120, 121].

A typical fluorescence (FL) spectrum of a single dye-doped 5 $\mu$m polystyrene (index of refraction $n$=1.59) sphere on a glass substrate, is illustrated in Fig. 1.11 [120, 121]. The whispering gallery mode peaks with orthogonal polarizations are labeled TE$^q_{\ l}$ and TM$^q_{\ l}$, respectively. The peaks are characterized to be $Q$=4×10$^3$ for the modes with $q$ =1, which is well below the theoretical limit ($Q$ >10$^5$) for perfect spheres with a diameter $D$ =5 $\mu$m, as illustrated in Fig. 1.11.

The positions of fundamental WGM peaks can be derived from Maxwell's equations solved in spherical coordinates by using the Mie scattering formalism [27, 122]:

$$P\frac{J'_{l+\frac{1}{2}}(NX)}{J_{l+\frac{1}{2}}(NX)} = \frac{N'_{l+\frac{1}{2}}(NX)}{N_{l+\frac{1}{2}}(NX)} \tag{1.2}$$

where $P = N$ for TE polarization ($P = N^{-1}$ for TM), and the ($l + 1/2$) term appears, due to translating the spherical Bessel and Neumann functions to their cylindrical counterparts [33]. By expanding the quantities in (1.2) as an asymptotic series in powers of $(l + 1/2)^{-1/3}$, it is possible to express first terms of the whispering gallery mode resonances [27] in terms of the size parameter:

$$nX q_{,l} = l + \frac{1}{2} - \left(\frac{l + 1/2}{2}\right)^{1/3}\alpha_n - \frac{P}{\sqrt{n^2 - 1}} + ...,$$ (1.3)

where $q$ is the radial number, and $\alpha_n$ are the roots of the Airy function Ai($-z$).

The spectrum is composed of a periodic distribution of WGM resonances, versus $l$ with the optical frequency separation between the peaks, with the same polarization states, represented by a free spectral range (FSR) [33]:

$$\Delta\omega = \frac{c}{2\pi a}\frac{tan^{-1}\sqrt{n^2-1}}{\sqrt{n^2-1}}$$ (1.4)

The optical frequency spacing between WGMs having the same $q$ and $l$ but different polarization states can be approximated [123] by the following formula,

$$\Delta\omega_{q,l}^{\text{TE−TM}} = \frac{c}{2\pi a n}\sqrt{n^2 - \frac{1}{n}}$$ (1.5)

For studies of cavity quantum electrodynamics, the figure-of-merit for the photonic cavities is typically estimated to be $\sim Q/\sqrt{V}$, where $V$ is the modal volume [33]. One fundamental property of WGMs is that the electromagnetic fields of the mode are confined, so that they occupy a relatively small volume in comparison to the total volume of the cavity [64, 73]:

$$V = 3.4\pi^{3/2}\left(\frac{\lambda}{2\pi n}\right)^3 l^{11/6}\sqrt{l - m + 1}$$ (1.6)

The degeneracy of azimuthal ($m$) modes that exist in perfect spheres can be broken by deformations or deviations from the nominal spherical shape [17, 30], leading to the appearance of quasi-equidistant closely spaced peaks [33]. The WGM has a broadening effect caused by the interaction with the substrate [116]. It should be noted, that the deformations and deviations from the nominal spherical shapes not only breaks the symmetry around the *z*-axis but can also induce mixing of the modes of different *m*, thus broadening all of these modes, due to optical tunneling to the substrate [119]. These effects can be proposed as a reasonable explanation to whispering gallery linewidth broadening observed in the experimental spectra obtained for microspheres [33].

So far, only the case of a single dielectric sphere has been described. However, if the spheres are in close proximity, subwavelength separation, whispering gallery modes in individual spheres can be strongly coupled. Due to the subwavelength separation, the evanescent field couples directly into the adjacent sphere. Drawing an analogy with electronic states, in atoms and molecules, WGMs can act as photonic states in photonic atoms, and photonic molecules can be built from these building blocks [108].

Furthermore, this resonant transport can potentially be used for developing delay lines [109], and ultranarrow spectral filters [33]. However, size uniformity presents a problem since all spheres diameters deviate from the nominal design, due to manufacturing errors which corresponds to shifting of the positions for WGM resonances [113]. Experiments have been performed with spectroscopically selected spheres that showed very efficient WGM coupling [120]. However, in the case of commercially available suspensions of microspheres with the size variations ~1% the propagation losses for WGM based transport are ~2-3 dB per sphere [14]. These losses are too high for developing applications [33]. In terms of biosensing, such WGM-based sensors could be used for label-free detection down to the level of single molecules, due to their extreme sensitivity [39].

## 1.3: Infrared Waveguides

In this section of the introductory chapter, infrared waveguides will be discussed in order to provide the proper background for chapter 3 on precision laser scalpels that use hollow core waveguides to transmit the radiation. Hollow waveguides are appealing in terms of transmission properties in comparison to solid-core infrared fibers, because of the low propagation losses and coupling losses [124]. Traditional hollow waveguides are fabricated from plastic, metal, or glass with highly reflective coatings deposited on the inner surface and operate by attenuated total reflectance and total internal reflectance [124]. However, there are other classes of hollow waveguides based on the photonic bandgap and low density of states guidance [125-145]. These fibers can be used in a variety of applications, spanning from high-power laser delivery to chemical sensors.

### 1.3.1: Attenuated Total Reflectance and Total Internal Reflectance Infrared Waveguides

This section addresses mid-IR waveguide systems. In this dissertation, we develop a focusing microprobe for laser surgery applications. In Chapter 3, we illustrate the application of photonic nanojets for laser surgery [71]. In this regard, we use the focusing properties of dielectric microspheres as described in section 1.2. However, the design of this device goes beyond the focusing spherical tip of the probe, to include a flexible optical delivery system and a compact single-mode source of the electromagnetic power. For this reason, in this section we review some properties of mid-IR waveguides.

These fibers consists of three main groups: glass, crystal, and hollow waveguides [124]. Glass fibers can be divided into subcategories of heavy metal fluoride, germanate, and chalcogenide. Crystal fibers have subcategories of polycrystalline or single crystal such as sapphire. Hollow waveguides consists of the subcategories of metal/dielectric film, and refractive index <1 such as sapphire at $\lambda$=10.6 $\mu$m [124].

The subcategory of metal/dielectric films consist of an inner core material with $n$>1, act as leaky waveguides [124]. In the case, that $n$<1 for the inner core material the guiding is similar to total internal reflection (TIR) [124]. However, the evanescent field extends into the outer core material and is attenuated. This effect is called attenuated total reflectance (ATR). ATR guides can be used for applications in IR spectroscopy [210-215].

The spatial beam profile of the output beams transmitted through these hollow-core waveguides are strongly dependent on the coupling parameters of the input beam from the source. Due to the strongly attenuated higher-order modes, these hollow-core waveguides are low-order mode waveguides. Although these hollow-core waveguides are favorable as low-order mode waveguides, they can support higher-order modes under certain launch conditions and if radial forces are applied to the waveguide.

There are two types of losses generally associated with these hollow-core waveguides. First, the propagation loss that follows an inverse third power law with respect to the hollow-core bore radius. Second, the bend loss that is inversely proportional to the radius of curvature of the bend.

### 1.3.2: Hollow-Core Microstructured Optical Fibers

Recently, there have been developments in novel waveguiding technology in the infrared, namely hollow-core microstructured optical fibers (HC-MOFs) [136]. This new form of optical fiber waveguide has unique properties that can be utilized for many applications such as high power and ultra-short pulse delivery, light-gas interactions and terahertz radiation, along with flexible delivery for surgical purposes [135-145]. The technology of drawing fibers is well developed for a variety of glasses. However, silica technology has reached high levels of fidelity in terms of the quality of materials, and purity of interfaces, leading to extremely small losses in the telecom regime. Unfortunately, in the range of interest for surgical applications (~3 $\mu$m) silica is strongly absorbing, therefore this well-developed technology cannot be implemented. Other materials sometimes used in mid-IR regime have their own limitations. For example, some of the materials are not considered to be biocompatible like chalcogenide glasses. Other technologies have not reached the level of quality and reproducibility required for single-mode applications. The advantage of the hollow-core microstructured fibers is that they are built on the well-developed foundation of silica technology. For particular geometries there can be an extremely small overlap of the optical fields with the silica sladding which greatly reduces the absorption. In this regard, the design is not perfect because it requires certain trade-offs. However, if the waveguide system does not require substantial lengths, these hollow-core microstructured optical fibers can be a viable solution for such mid-IR applications.

Electromagnetic fields are localized within the hollow core surrounded by microstructured cladding, providing lower absorption losses as the light propagates in air. There are two waveguiding mechanisms which support the propagation of electromagnetic waves confined within hollow cores surrounded by microstructured cladding. First is the photonic band gap mechanism, where the cladding rigorously forbids modes for a certain range of wavelengths and propagation constants [125-133]. The rejection from the cladding modes forces the electromagnetic waves to propagate in the hollow core with low-losses because of the absence of material absorption. The second mechanism, is termed "low density of states guidance" whereas the photonic bandgap is not supported in the HC-MOF and the core modes are weakly coupled with the cladding modes [136]. Consequently, the transmission losses are higher but the HC-MOF waveguides for a significantly larger bandwidth. The HC-MOFs also have a large diameter air core which significantly exceeds the core diameters in single-mode fibers. As an example, for mid-IR range, an ultralow loss single-mode Fluoride fiber diameter is ~15 microns [216-218]. Whereas, the core diameter of the hollow-core microstructured optical fiber can be up to ~100 microns [136]. The concept of single-mode guiding is difficult to apply to cores with such large diameters. As already stated, these waveguides are fundamentally leaky. However, for applications using sufficiently short fiber lengths in their delivery systems, such waveguides can be very advantages compared to conventional single-mode waveguides due to the simplified coupling of light from the source and simplified integration with the focusing element.

The HC-MOFs, used in Chapter 3 of this dissertation, are formed by a hollow-core with a cladding which consists of 8 contiguous silica fibers within a single encompassing capillary for support, as shown in Fig. 1.12. The HC-MOFs presented in this section of the introduction have different size effective air core diameters; however, the mechanism of electromagnetic wave transport is the same.

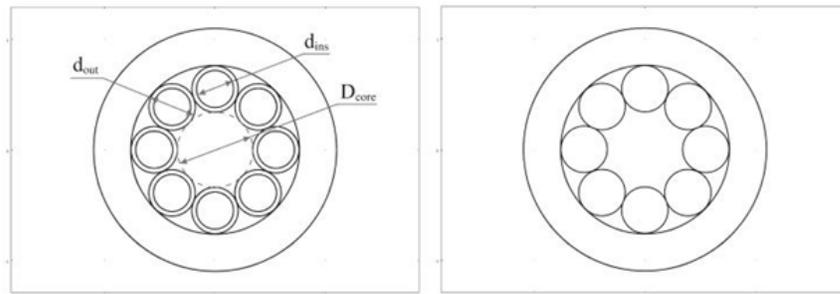

Figure 1.12: HC-MOF with a negative curvature of the core with cladding consisting of one row of capillaries (left) and with rods (right) [136].

These HC-MOFs were studied both numerically, using the finite element method (FEM), and experimentally, by the Fiber Optics Research Center of the Russian Academy of Sciences [136]. In the FEM simulation, a comparison between this technology with different geometry capillaries and a standard tube were considered. The FEM simulations provided calculations of the total loss for the fundamental mode (FM) that was transmitted through the waveguides.

The HC-MOFs have an effective air core diameter ($D_{air}$) of 36 $\mu$m. The outer diameter of the capillary and rod ($d_{out}$) is 22.5 $\mu$m [136]. The inner diameter of the capillary is $d_{in}$=0.76*$d_{out}$. The waveguiding of the FM in these structures are compared with a simple dielectric tube of silica, with a bore diameter equal to the effective air core diameter of the HC-MOFs, 36 $\mu$m [136]. The FM is well confined within the HC-MOF with cladding consisting of capillaries, as shown in Fig. 1.13.

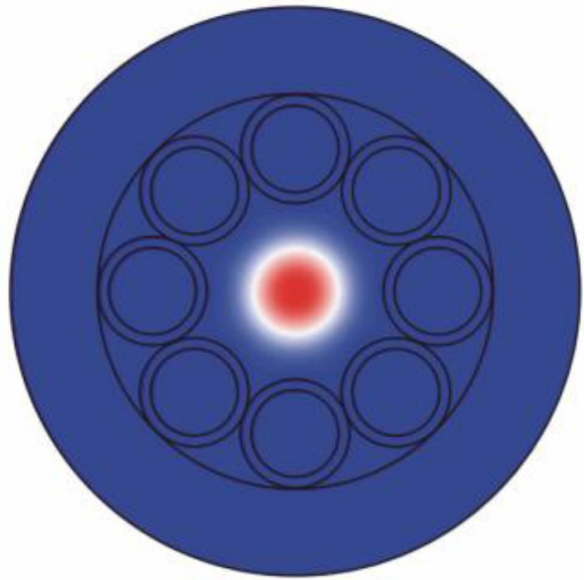

Figure 1.13: Fundamental mode confined within the HC-MOF formed by a cladding consisting of capillaries [136].

Simulations indicate that the HC-MOFs consisting of capillary cladding provide high throughput at certain wavelengths. Whereas, when the microstructure was formed by rods the transmission performance was slightly worse. The standard dielectric fibers have significantly worse transmission properties in comparison with either case. The ineffectiveness of the microstructued rods can be explained by their poor scattering properties in the regime where $\sim \lambda / d_{out}$. Whereas, the rod has a larger number of its own resonance wavelengths and a high density of states for leaky modes [136], the low-loss transmission bands do not emerge [136]. These results are demonstrated in Fig. 1.14.

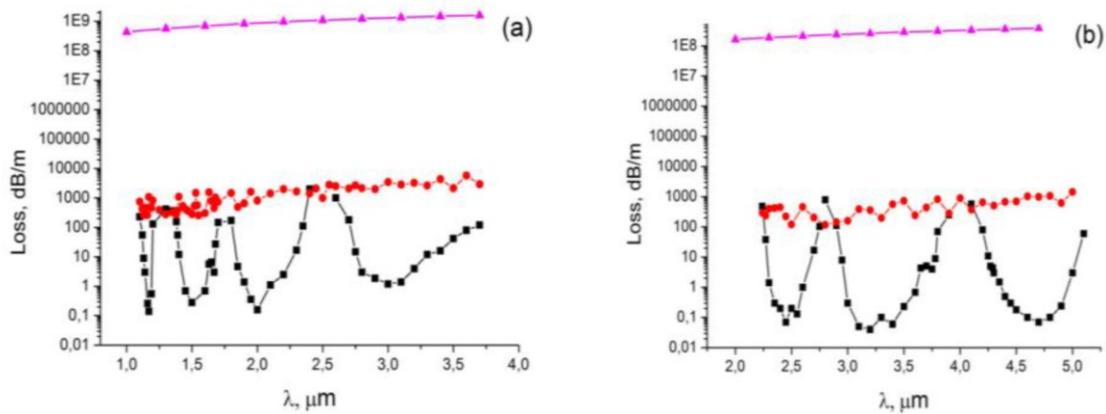

Figure 1.14: (a) Total loss for HC-MOFs with $D$core = 36 $\mu$m in the case of cladding consisting of capillaries (square), rods (circle) and simple dielectric tube (triangular) (b) the total loss for HC-MOFs with $D_{\text{core}}$ = 68 $\mu$m and all notations is the same as in (a) [136].

This work demonstrated the existence of several transmission bands, with a total loss significantly lower than the material losses which are associated with silica used as the cladding. The mechanism responsible for this waveguiding effect in HC-MOFs can be attributed to the negative curvature of the core boundary and the lower density of states of scattering elements that comprise the cladding [136]. These HC-MOFs can be advantageous for applications in the mid-infrared regime, since the transmission bands of the waveguide can been engineered to overlap with the source of light for the task providing a high throughput delivery system.

## 1.4: Super-Resolution Microscopy

The direction of propagation of light through the spherical lens can be reversed, so that instead of a focusing element such as for surgical applications, it can be used for imaging applications. Some of these properties can be understood by using geometrical optics [54, 55, 67]. It typically leads to a conclusion that such spheres in contact with objects tend to produce virtual images [51-70]. These properties are much more intriguing because they do not have a simple explanation based on geometrical or even wave optics assumptions [55]. The most important and critical issue is the resolution produced by such magnifying lenses, in contact with the objects [51-70]. There have been recent reports that this resolution exceeds the diffraction limit and the mechanisms of this super-resolution imaging are debated in the literature [51-55, 58-61, 63, 65-70]. This area is the subject of Chapter 4 in this dissertation. For this reason, in this section, we review the basic concepts and text book definitions associated with defining the resolution of an optical system.

The spatial resolution in the far-field of a lens based optical imaging system is limited by several factors, namely, the wavelength of light illuminating the object, and the lenses ability to collect light which is described by the numerical aperture. The numerical aperture of the objective lens can be mathematically expressed as

$$NA = n\sin\theta, \tag{1.7}$$

where $n$ is the refractive index in the object space and $\theta$ is the half-angle representing the maximum angle of light that can be collected by the objective lens, as illustrated in Fig. 1.15. For a standard circular lens in the far-field, with diameter ($D$) and focal length ($f$) the numerical aperture can be expressed as

$$NA = \frac{D}{2f} \tag{1.8}$$

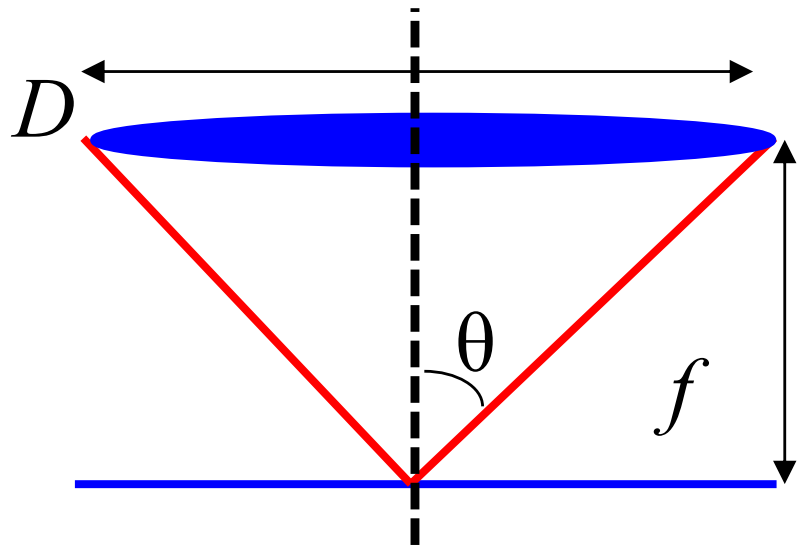

Figure 1.15: Illustration of the numerical aperture of an objective lens.

By optimizing these parameters the spatial resolution of the optical imaging system can be improved. For example the refractive index in the object-space can be increased with solid immersion lenses [146-158], either hemispherical (h-SIL) or super-spherical (s-SIL) geometries. The collection angle can be increased with the s-SIL. Finally, shorter wavelengths can be used for the illumination as in the case of deep-UV microscopes [159-161]. Super-resolution imaging occurs when the resolution of the imaging system exceeds the values predicted by these classical far-field diffraction considerations.

### 1.4.1: Diffraction Limit and Resolution Criteria

Ultimately the spatial resolution of a far-field imaging system is going to be restricted by the diffraction of light. A point in the object plane will result in a finite-sized spot in the image plane, as a manifestation of the diffraction limit. If the point in the object plane is considered as a Dirac delta function, the result in the image plane will be the consequence of the convolution of the delta function with the point spread function (PSF) of the imaging system [168].

An aberration-free system is still limited by diffraction phenomena. In order to quantify the smallest discernable feature of the imaging system, various criteria have been proposed such as Abbe [162], Rayleigh [163], Sparrow [164], and Houston [165]. The criteria are built around the minimum separation distance between two point sources at which the objects can be distinguished in the resulting irradiance pattern in the image plane. For this consideration we will assume incoherent image formation.

For optical imaging systems, the object features can be considered as a diffraction grating. The grating's Fourier transform forms at the back focal plane of the objective lens.

The superposition of these spatial frequencies at the imaging plane forms the image. However, due to the finite size of the lens, not all of these spatial frequencies pass through the lens. The limited collection efficiency of the imaging system operates as a low-pass filter; therefore, the highest spatial frequency is a measure of the spatial resolution [167].

In 1873 Ernst Abbe developed a theory [162] of image formation. According to the theory, images are formed by light diffracted by the objects. If an array of objects, such as a diffraction grating, is illuminated only some of the diffraction orders will be imaged due to the system's limited collection efficiency. This enables the use of the grating equation,

$$d \sin \theta_m = m\lambda,$$

(1.9)

where $\theta_m$ is the angle between the $m^{th}$ diffraction order and the normal vector to the surface of the diffraction grating. This expression is in agreement with the concept that the $d$ has an inversely proportional relationship to $\theta_m$ for any single $m^{th}$ diffraction order. A solution for the expression under the constraint that the first order diffracted beam passes through the lens gives the Abbe diffraction limit. This was simultaneously mathematically formulized by Helmholtz [166], although credit is generally attributed to Abbe:

$$d = \frac{\lambda}{2NA},$$

(1.10)

where $\lambda$ is considered the free-space wavelength. The Abbe limit suggests that objects separated less than roughly half the wavelength of illumination cannot be resolved, a modest ~200-250 nm for conventional white-light microscope systems.

Rayleigh's resolution criterion is a heuristic estimation, whereas Abbe's theory is derived by a physical model. Rayleigh states that two, equal intensity, point sources are just resolved under the condition that the central maxima of the irradiance diffraction patterns coincides with the counterpart's first zero. The summation of the irradiance patterns produced from diffraction by square and circular shape apertures produce saddle points of 81.1% and 73.6%, respectively, applying Rayleigh's resolution criteria. We consider a circular shaped aperture for the remainder of this section. The point spread function (PSF) for the circular aperture is in the form of a *Jinc²*, where *Jinc(x)=J₁(x)/x* in which $J_1$ is the Bessel function of first kind of order 1. The appearance of the irradiance profile for the circular aperture's Airy disk is [167]

$$I(r) = I_0\left(\frac{2J_1(k_0 rNA)}{k_0 rNA}\right)^2,$$   (1.11)

where $k_0$ is the free-space wave number of the illuminating light, $2\pi/\lambda_0$.

The first zero of the *J₁(x)* occurs at *x*=1.22π, and Rayleigh's criteria is satisfied when the distance between the point sources is equal to the radius of the mutual Airy disks. The diffraction-limited lateral resolution as defined by Rayleigh is

$$\Delta_x = \frac{0.61\lambda_0}{NA}.$$   (1.12)

The Sparrow resolution criteria [164] due to the "undulation condition" is very sensitive to irradiance differences, making the criterion appealing in the case that there are no well-defined zeros in the diffraction pattern [167]. The criterion claims that two equal intensity point sources, separated by a distance *d*, are just resolved under the condition that the composite irradiance distribution has a minimum on the line joining their centers [167], i.e. $\frac{\partial^2 I}{\partial x^2} = 0$ at *x*=0 [164]. Therefore, Sparrow's diffraction-limited resolution is

$$\Delta_x = \frac{0.473\lambda_0}{NA}. \tag{1.13}$$

Houston proposed, in 1927, a resolution criteria stating that two equal intensity point sources are resolved if the distance between the central maxima of the summation of their irradiance profiles is equal to the full-width at half maximum (FWHM) of the irradiance profile of either individual point source [165]. The FWHM of the PSF from a single circular aperture with a NA, related to the $D$ by equation (1.8), is

$$\text{FWHM} = \frac{0.515\lambda_0}{NA} = \Delta_x. \tag{1.14}$$

It should be noted, that these criterion neglect to consider signal-to-noise (SNR), which is always present during experimental measurements. Also, the experimental PSF will not be identical to the model because its profile will be perturbed due to aberrations, physical deviations from nominal designs of the objective lens, and other irregularities.

### 1.4.2: Super-Resolution Techniques

After achieving the ultimate resolution by exhausting the optimization of the far-field diffraction consideration, researchers pursued alternative means to advance the capabilities of imaging system's resolution. This was accomplished by obtaining information from evanescently coupling with the objects near-field. The parallel components of the wave vector of these standing surface waves, which exponentially decay away from the surface, are generally larger than the free-space propagation wave vector [168]. This gives access to finer structural information of the objects of interest.

$$\mathbf{K}_e = \mathbf{k}_{//} + i\mathbf{k}_\perp; \ = |\mathbf{K}_e|^2 = \left|\mathbf{k}_{//}\right|^2 + |\mathbf{k}_\perp|^2 = \left(\frac{2\pi}{\lambda}\right)^2 \tag{1.15}$$

In 1972, the first system utilizing this phenomena was the near-field scanning optical microscope (NSOM) [169]. The probe simultaneously scattered and collected evanescent waves in the near-field to extract information of higher spatial frequencies [168]. The NSOM performed with a resolution exceeding the far-field diffraction limit [169]. The resolution depends on two components, illustrated in Fig. 1.16. First, the resolution depends on the size of the aperture at the tip of the probe. Second, resolution also depends on the separation distance between the tip of the probe and the sample surface. Consequently, an inherent tradeoff exist between scanning time and resolution. The NSOMs enhanced performance stimulated the development of scanning probe techniques, such as scanning tunneling microscopy [170], and atomic force microscopy [171].

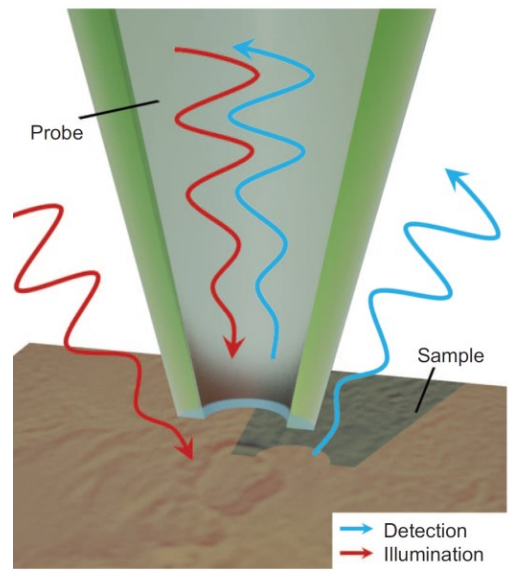

Figure 1.16: Schematic illustration of the near-field scanning optical microscope. The resolution depends two components: the size of the probes aperture and the separation distance of the tip of the probe from the object. Generally speaking such probes can be used as the source for near-field illumination, the detection probe of the near-field's evanescent waves, or both [168].

Recently, another branch of super-resolution techniques has been pursued based on the amplification of evanescent waves. In 2000, Pendry published a galvanizing paper [172] entailing a "perfect lens" made of a slab of negative index material. This work demonstrated that such a perfect lens had the power to focus all Fourier components of a 2-D image, including the propagating and evanescent waves. However, there are practical issues such as natural negative index materials do not exist on the visible scale [168].

Under the illumination conditions that the light is TM polarized, the dependence of the transition on the coefficient $\mu$ is eliminated [168]. This allows for a portion of the function to be realized for a TM polarized beam and a thin metallic film [168]. In 2005, Zhang and others [173, 174], confirmed this with a silver superlens achieving super-resolution. Later, this was extended by using the concept of frequency shift [175], allowing for the far-field superlens (FSL) [176]. Introducing curvature to these lenses provided additional magnification, and was termed hyperlens [177]. Although these lenses, in Fig. 1.17, work in principal, there are "real-world" limitations which restrict the systems [178].

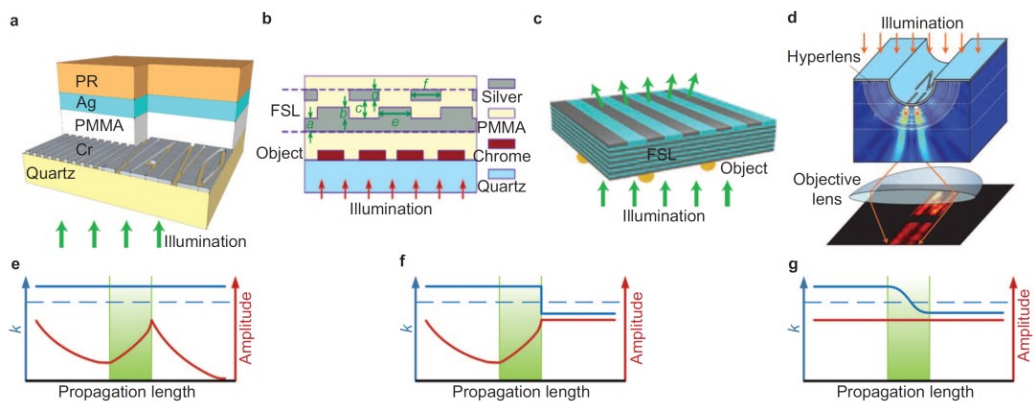

Figure 1.17: Schematic of (a) superlens, (b) FSL, (c) 2-D FSL and (d) hyperlens. (e–g) Theoretical comparison [168, 173, 176, 177, 193].

Illuminating with evanescent waves is another path that leads to optical super-resolution [168]. After the illumination has been provided by the evanescent waves, surface scattering or other mechanics can project these waves into the far-field as propagating waves. For example, metallic samples support surface plasmon polaritons (SPPs) [179]. Such SPPs have larger lateral wave vectors than the free-space propagating wave that would excite them, resulting in a local field enhancement at the metal/dielectric interface [168]. These SPPs wave vectors can be expressed as:

$$\boldsymbol{k}_{sp} = \boldsymbol{k}_0 \left(\frac{\varepsilon_d \varepsilon_m}{\varepsilon_d + \varepsilon_m}\right)^{\frac{1}{2}} \gg \boldsymbol{k}_0 \tag{1.16}$$

where $\boldsymbol{k}_0$ is the wave vector in free-space; $\varepsilon_d$ and $\varepsilon_m$ are the dielectric functions of the dielectric and metal, respectively.

Such techniques have demonstrated optical super-resolution imaging [180, 181]. However there is an inherent flaw in the technique, requiring the sample to be coated with a metallic film. A design based on illuminating the sample with evanescent waves, Fig. 1.18, from a fiber led to resolution values on the order of ~75 nm [182].

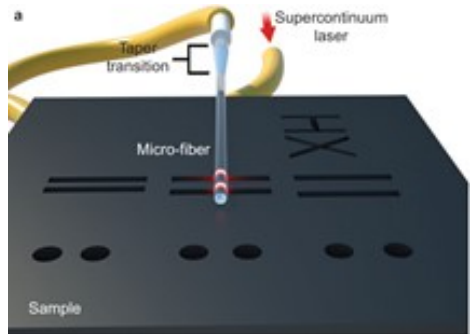

Figure 1.18: Schematic of illuminating a sample with evanescent waves by a microfiber to achieve optical super-resolution imaging [182].

Another approach stems from solid immersion lens (SIL) technology, the nanometric scale SIL (nSIL), shown in Fig. 1.19 [183]. Such nSILs have demonstrated the capability to improve lateral resolution by as much as 25%, in comparison to its macroscopic counterpart. The nSIL acts as an aperture which narrows side lobes from the diffraction pattern, and blocks subcritical rays which typically degrade the contrast [184].

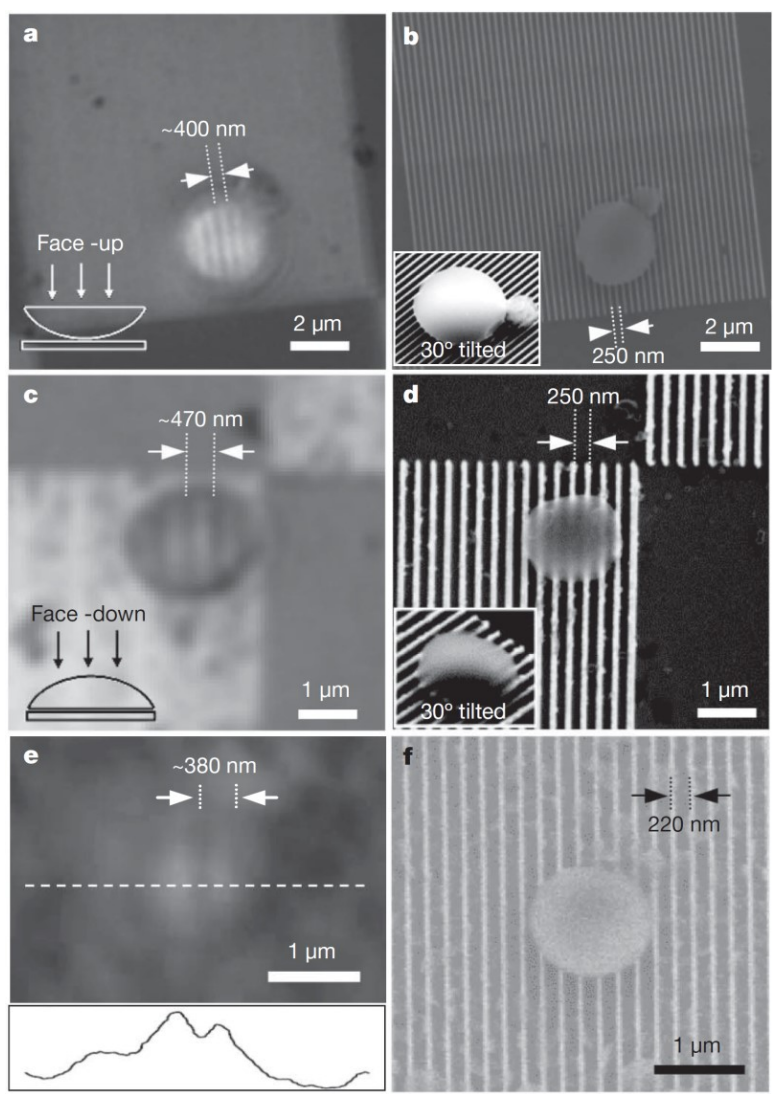

Figure 1.19: Optical images accompanied by SEM images for nSILs, (a)-(b) face up, and (c)-(f) face down [183].

In 2011, Wang, et. al, demonstrated the white light nanoscope [51]. Silica (*n*~1.46) microspheres 2<*D*<9 $\mu$m were deposited onto a sample, and the virtual image was captured by imaging through the sphere [51]. In order to show the power of this technique images of the stripes of a Blu-ray® disk were obtained, where the stripes have a width of 170 nm and are separated by 130 nm., along with a nanometric scale star fabricated with a ~90 nm corner [51], as shown in Fig. 1.20.

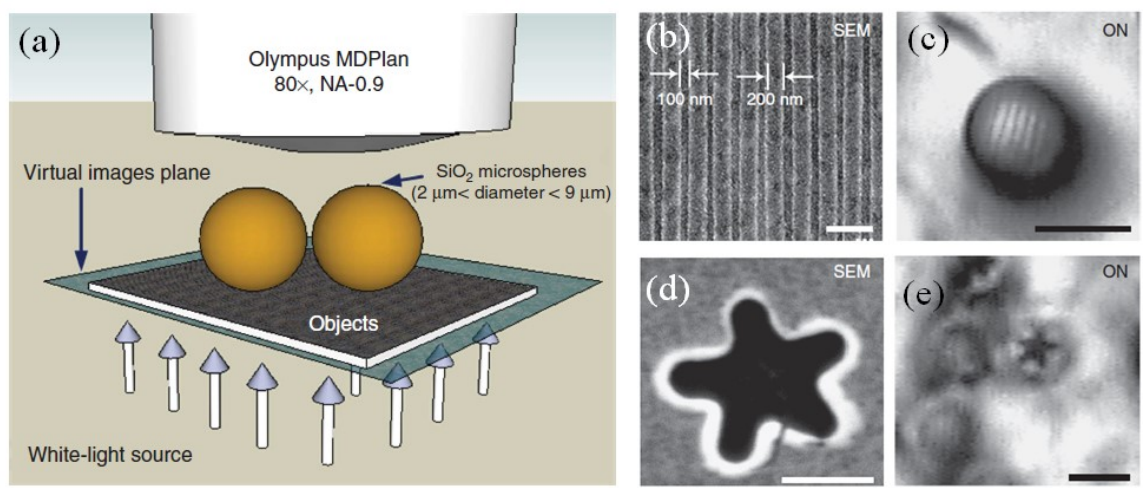

Figure 1.20: (a) Illustration of microsphere nanoscopy technique. SEM and optical images of (b)-(c) Blu-ray® disk, and (d)-(e) star with ~90 nm corners [51].

It was proposed that the microsphere's ability to resolve fine features is a consequence of optical reciprocity, and the spheres ability to produce photonic nanojets as discussed in the second subsection of the introductory chapter. Also, it was argued that this technique does not work under conditions of full-immersion and that the effect is present for spheres with refractive indices less than 1.8 [51].

More recently, research from Professor Astratov's Mesophotonics Laboratory has given more insight into this microsphere-assisted optical super-resolution technique [53]. The experimental studies were performed imaging an array of Au nanoplasmonic dimers and commercially available Blu-ray® disks. It was demonstrated that full liquid-immersion of barium titanate glass microspheres with index, $n \sim$1.9-2.1, produce super-resolution images, as shown in Fig. 1.21. It was also observed that minimal feature sizes on the order of $\sim \lambda/7$ could be discerned, for wavelength-scale spheres [53]. This value increases to $\sim \lambda/4$ for larger spheres; however, the field-of-view increases linearly as a function of $D$. Semi-quantitative arguments describing the field-of-view can be found in ref. [55]. They are based on the idea that not only the contact point, but a certain area of the spherical surface near the contact point are involved in the formation of super-resolved images. Within this area on the sample surface, the spherical surface is separated from the object less than $\lambda$ [55]. The size of this area increases with the sphere radius. Lateral magnification values were observed to be as large as $\sim$4.5, over a range of $2 < D < 200$ $\mu$m [53], as shown in Fig. 1.21.

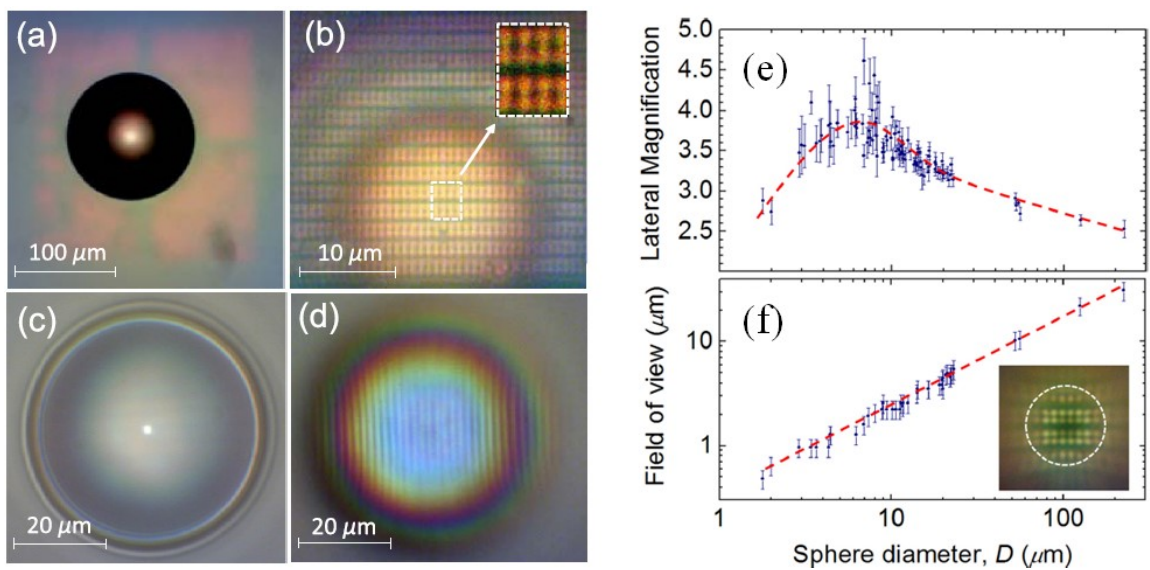

Figure 1.21: (a)-(b) Optical images obtained at the sample plane and virtual image plane through the sphere, of Au nanoplasmonic dimers. (c)-(d) Similar to [(a)-(b)] of a Blu-ray® disk. (e), (f), lateral magnification and field of view as a function of sphere diameter, respectively [53].

These studies were followed by direct comparisons of the microsphere-assisted optical super-resolution imaging technique with existing technologies [65]. It can be seen, in Fig. 1.22 (a)-(d), that microsphere-assisted imaging (*D*~15 μm, *n*~1.9) provide better image quality in comparison with standard microscopy, and confocal microscopy. All of the images in this comparison were performed through a 100x (NA=0.95) lens while imaging an array of Au nanoplasmonic posts forming a square where the posts are 100 nm in diameter and separated by ~50-60 nm [65]. Furthermore, the microsphere-assisted imaging technique (*n*~1.9) provides superior resolution in direct comparison with SIL lens (*n*~2), while imaging nanoplasmonic objects with ~50-60 nm gaps as shown in Fig. 1.22 (e)-(f), imaged through an objective lens with 20x (NA=0.60) with an illuminating wavelength *λ*=405 nm [65].

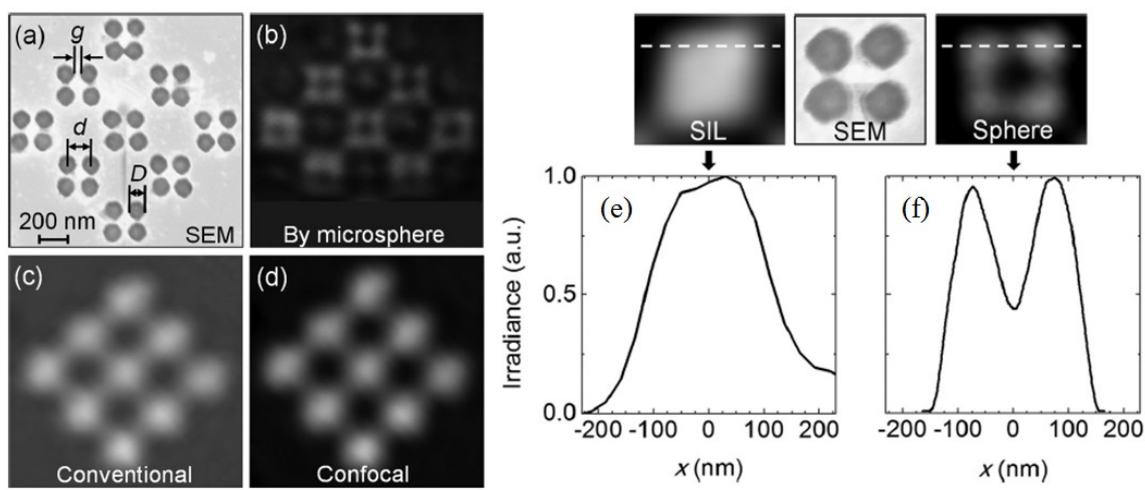

Figure 1.22: (a) SEM of the Au nanoplasmonic structure with gaps of ~50-60 nm and posts of 100 nm in diameter. (b) Image through the microsphere demonstrating the microsphere-assisted super-resolution technique. (c) Conventional microscopy image of the same object. (d) Confocal microscopy image of the same object. All of these images (b)-(d) were obtained through a 100x (NA=0.95). (e) and (f) are irradiance line profiles through the cross-section of the object imaged through a 20x (NA=0.6) objective obtained by a SIL and microsphere, respectively [65].

A standard white light microscope was used to investigate the role of the numerical aperture of the objective lens. Microsphere-assisted imaging was used with a $D$~20 $\mu$m barium titanate glass sphere to image features of ~200 nm. The numerical aperture of the objective lenses with NA=0.4 and 0.9, was used in combination with the microsphere-assisted imaging technique, Fig. 1.23 (b)-(c). Although the change in NA was drastic, there was very little change in the image because the collection of light is provided by the sphere itself. However, it can be seen in Fig. 1.23 (d)-(f), that the image changes as expected by the classical far-field diffraction formula as the NA is changed. Images were obtained through an objective lens with NA labeled in the figure with an illuminating central wavelength of $\lambda_c$=550 nm.

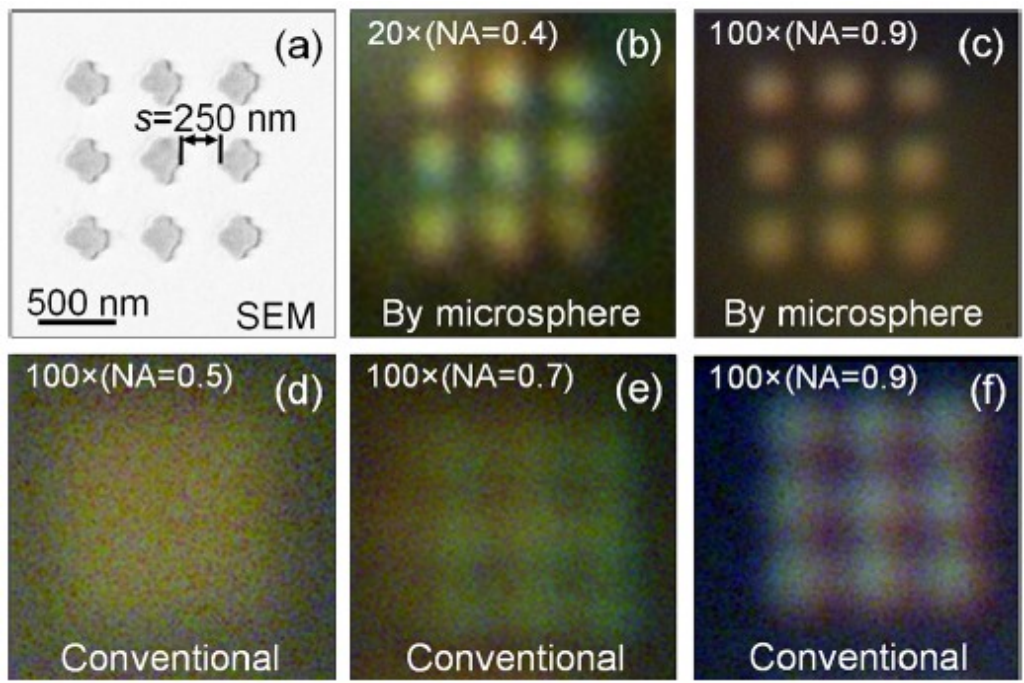

Figure 1.23: (a) SEM of the sample. (b and c) Images through a 20 $\mu$m liquid immersed barium titanate glass ($n$=1.9) sphere. (d-f) Conventional optical micrographs using different numerical apertures [65].

Pushing the limits of optical imaging opens up opportunities in a broad range of disciplines. For example, potentially allowing dynamic studies of biological samples, on the scale of viruses [60, 63]. These techniques could also be implemented in microelectronics for on-chip level inspection. In terms of fundamental research, these studies give insight to light-matter interactions on a nanometric-scale and its influences on image formation.

## 1.5: Sensitivity Enhancement of Infrared Photodetectors

Another application of microspheres is associated with their ability to channel and concentrate light beams which can be very useful for improving sensitivity of the light detectors, especially detectors with limited photosensitive area or a device mesa [71, 185]. Since 1970 there has been significant interest in semiconductor low-dimensional solids, provoked by the proposal of the advent of molecular beam epitaxy (MBE) by Esaki and Tsu [186]. This galvanized interest led to the development of new physical concepts and phenomena with promising applications [187]. These continual efforts have resulted in significant improvements in infrared photodetector technology. One example is with type-II strained-layer superlattice (SLS) detectors which have several advantages that have stimulated research interests. SLS detectors operate as a photodiode. These SLS materials can be utilized for two- or three-color detectors. Also, these detectors are tunable over a wide spectral band from 2-30 $\mu$m. Recently, it was theoretically proposed to couple photonic nanojets with single-channel germanium photodiodes for enhanced photocurrent and speed [185]. This topic is the focus of Chapter 5 where we will study coupling of photonic jets to SLS infrared photodetectors for increased sensitivity.

However, there is an alternative approach for increasing the sensitivity of the detectors which is based on integrating a resonant component, such as a plasmonic grating or nano-antenna, to produce local field enhancement. An illustration of this was provided by ref. [188] where a nanometer-scale germanium photodetector was enhanced by a near-infrared dipole antenna. Based on a common concept used in radio frequencies ref. [188] used a half-wave Hertz dipole antenna (length of ~380 nm) to concentrate the near-infrared radiation (~1300 nm) into the nanometer-scale photodetector. The strong confinement of light enabled ref. [188] to fabricate, as illustrated in Figs. 1.24(a)–1.24(c), the smallest semiconductor active region to be experimentally demonstrated at the time, on the scale ~150 x 60 x 80 nm. This device resulted in ~20 times photocurrent enhancement at ~1390 nm and with such a small capacitance from the reduction in the active volume it is estimated to have a cutoff frequency over 100 GHz [188].

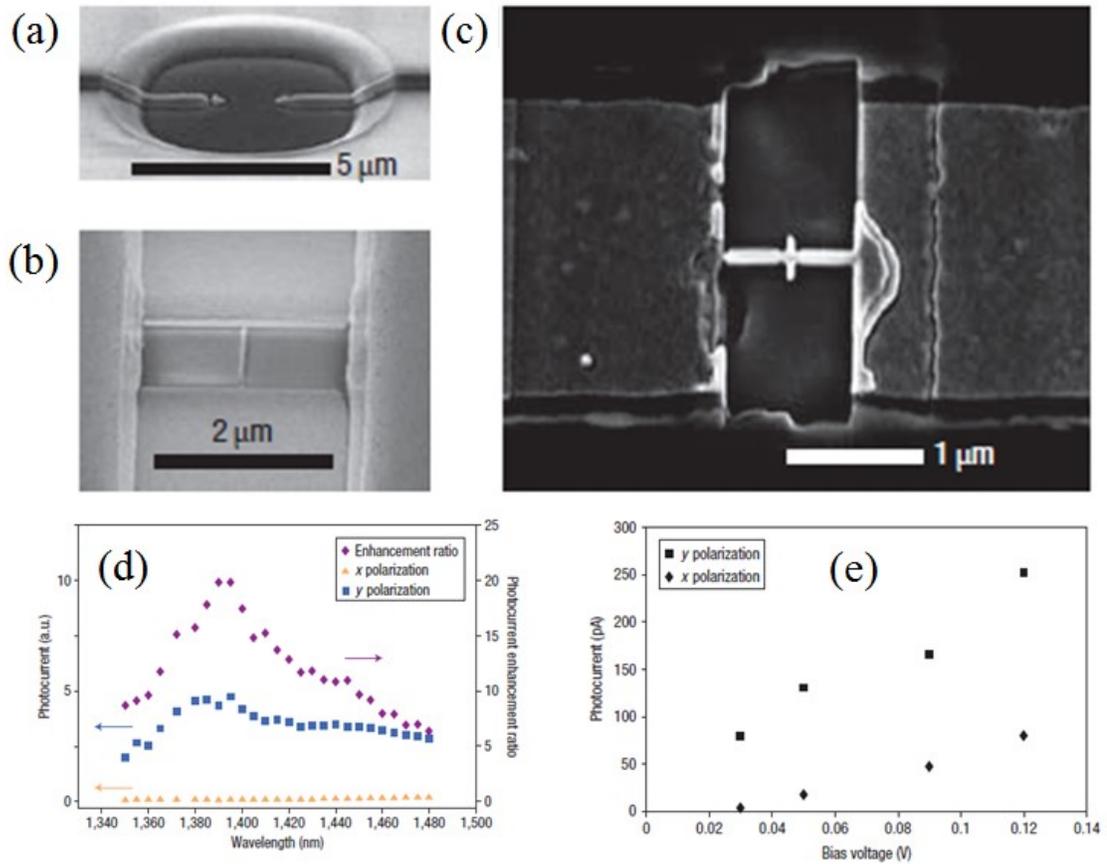

Figure 1.24: (a)-(c) Fabrication of the half-wave Hertz dipole antenna. (d)-(e) Improvement of the photocurrent and photoresponse, respectively [188].

More recently, in 2011, a photodetector enhanced by Fano-type interference was reported in ref. [189]. A metallic film perforated with two-dimensional subwavelength holes forming an array was integrated on top of the quantum dot infrared photodetector (QDIP) mesa, as illustrated in Fig. 1.25. The photocurrent was enhanced by 100% for the Fano dip of the first order plasmonic mode at $\lambda$=8.3 $\mu$m [189]. This option is suitable for QDIPs because of their poor quantum efficiency. For QDIPs the enhancement is not needed for their speed, therefore the detector mesa area is typically larger.

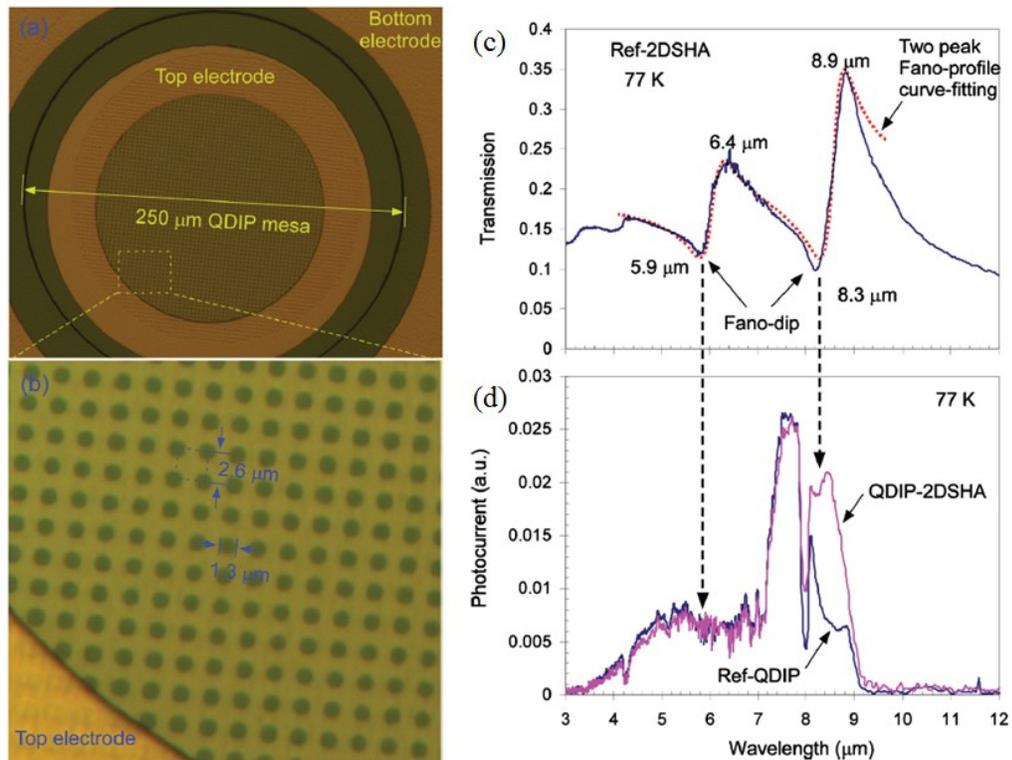

Figure 1.25: (a)-(b) Detector mesa with 2-dimensional subwavelength metallic hole arrays integrated on top of the active region of the device. (c)-(d) Transmission and photocurrent enhancement, respectively [189].

Three-dimensional FDTD simulations were performed for near-infrared wavelengths to demonstrate advantages provided by photonic nanojets produced by a 6.5 $\mu$m polystyrene sphere directing the beam into the active regions of the detector [185], as illustrated in Fig. 1.26. Since the resonant enhancement only occurs over a narrow spectral bandwidth [183,184], the photonic jet-enhancement technique is an excellent broadband alternative [51, 185].

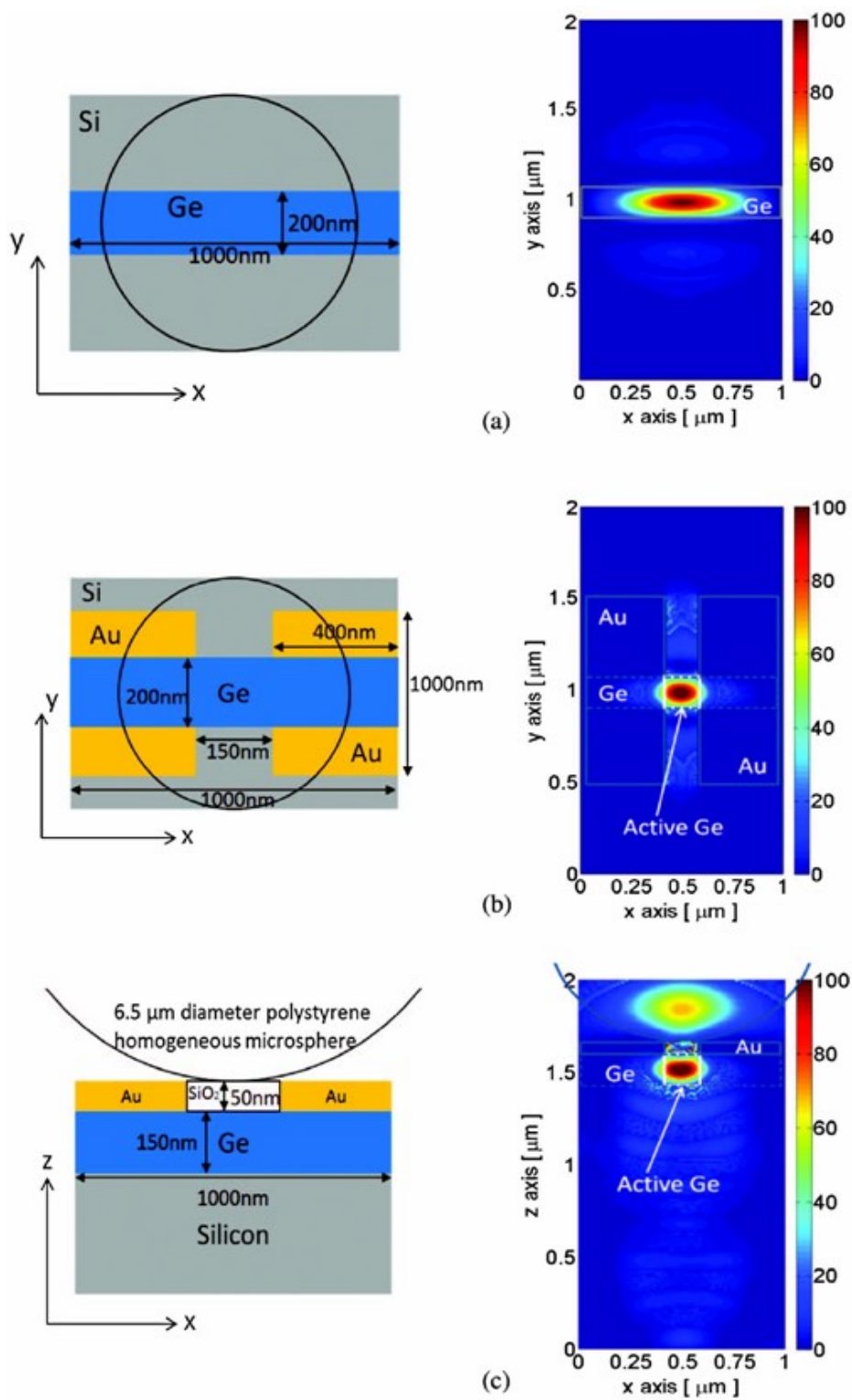

Figure 1.26: FDTD-calculated optical near-field intensity in the *x-y* plane 30 nm above the substrate. (a) and (b) are with and without gold electrodes, respectively. (c) *x-z* plane with gold electrodes. The color scale bar indicates the enhancement factor relative to the square of the incident fields' magnitude [185].

Ref. [185] showed with FDTD-calculations that a polystyrene sphere with a diameter of 5 times the wavelength can provide 26 times enhancement of energy directed into the active volume of the photodetector. This technique provides advantages over its resonant counterpart seeing as it is a broadband effect, and is polarization insensitive.

## 1.6: Summary

This dissertation work is devoted to the optical properties of structures and devices formed by dielectric spheres. The subject of focusing light by a single dielectric sphere seems to be well-known, however, even in this simplest case, there were recently new developments represented by the terminology of "photonic nanojets" introduced in 2004 by the group of Allen Taflove [73]. In this work, we considered all types of optical phenomena which could take place in microspherical arrays, taking into account new achievements in this area of microspherical photonics such as photonic nanojets [1-24], ultrahigh quality WGMs in individual spheres [26-50], coupling effects between WGMs in neighboring spheres [25, 101-110, 113, 114, 117, 120, 121], tight-binding theory [115, 116], and near-field imaging properties of individual dielectric spheres [51-70]. Our motivation was to develop a deeper understanding of novel physical properties, such as nanojet-induced modes in chains of spheres, develop novel optical devices such as laser surgical scalpels, to design structures with coupled WGMs for potential sensor applications, to enhance the sensitivity to light of photodetectors, and to increase the capabilities of imaging techniques based on using microspheres operating in contact with the nanoscale objects.

In Chapter 2 we consider microsphere-chain waveguides and photonic molecules. There have been many observations and results previously obtained in this area, however the holistic picture of the optical properties were largely missing. Historically, the nanojet-induced modes were first observed in polystyrene chains formed by small ($D<20\lambda$) spheres [13, 14]. The two main properties of these linear chains are (*i*) the beam tapering effect and (*ii*) small propagation losses that have been observed experimentally [1-4, 13, 14]. After the initial studies in [13, 14] attention gradually shifted to the opposite limit of geometrical optics ($D>>20\lambda$), where similar effects have been reproduced by numerical ray tracing [1, 5-11, 112]. Still, some questions remained unanswered. Geometrical optics theory scales with the sphere diameter for sufficiently large spheres, however it predicts, both effects at the indices $1.72<n<1.85$ while experiments with small polystyrene spheres revealed these effects for much smaller spheres at a significantly smaller index $n$=1.59. In this dissertation, we decided to study the entire evolution of the optical properties of these chains as a function of the spheres diameters, varied by small increments. The goal was to study the evolution from geometrical optics to wave optics in such chains by small steps to understand how and where this transition occurs for polystyrene spheres. We found that this transition indeed occurs approximately around $D\sim20\lambda$. We supported our experimental observation by numerical finite element modeling for small spheres ($D$=3 $\mu$m). We show how dramatic the transition is between the two regimes of wave optics and geometrical optics in chains of dielectric spheres. The optical losses at $n$=1.59 for wavelength-scale spheres becomes significantly smaller than losses calculated using geometrical optics for the index of 1.59, corresponding to the index of polystyrene which was used for the experimental work. Furthermore, the attenuation values were lower than any attenuation values predicted by geometrical optics, including the regime of $1.72<n<1.85$. Our results can stimulate the development of a more complete wave theory of light propagation in such structures. They can also stimulate the development of periodically focusing waveguides which can be used as radial polarization mode filters and as focusing optical multimodal microprobes.

In Chapter 3, a single-mode system for these precision laser scalpels are presented. Previously, there was an attempt to focus multi-modal beams for a precision laser scalpel for surgical applications using an Er:YAG laser diode at $\lambda$=2.94 $\mu$m. However, more recently there have been developments in regards to single-mode delivery systems such as the HC-MOFs described in section 1.3. Laser scalpels operating at $\lambda$=2.94 $\mu$m with high optical throughput are good candidates for vitreorentinal surgery, with the goal of dissecting or removing fibrotic membranes developed on the retina's surface, due to onset proliferative vitreoretinopathy. First, the single-mode HC-MOFs were characterized in terms of their waveguiding capabilities. A suitable HC-MOF candidate was selected then integrated with a dielectric microsphere. It was demonstrated that we can achieve wavelength-scale dimensions for the output beam ~4$\lambda$ while simultaneously providing high optical throughput. These experimental results were in qualitative agreement with numerical simulations performed in the frequency domain. This proof of concept could be a promising direction for precision laser scalpels for surgical action.

In Chapter 4, we consider the field of super-resolution microscopy which has made significant progress since the 1990's. One particular direction, which emerged in 2011, is the microsphere-assisted optical super-resolution technique [51-70]. The appeal of this technique is connected with its simplicity by not requiring piezoelectric stages, bulky equipment, low-signal detectors, or special illumination conditions. It was demonstrated that high-index spheres provide superior resolution on the order of ~$\lambda$/7 during full liquid immersion in comparison to low-index spheres in air [53]. Also, it was shown that this technique's resolution capabilities exceed that of conventional microscopies, confocal microscopies, and SIL technology [65]. This effect also exists when poor quality numerical aperture objective lenses are used NA~0.4 [65], making it appealing for applications. However, the current stage of microsphere-assisted imaging with high-index spheres requires uncontrollable deposition of spheres onto the sample. This process is random, not allowing you to target objects of interest, and has the potential to contaminate the sample. In Chapter 4, thin-films with high-index sphere embedded inside of transparent solids are investigated for practical applications [70]. It is observed that this gives similar resolution as previously observed by the microsphere-assisted imaging technique with high-index spheres [70]. However, it offers an advantage of being able to controllably translate the sphere across the sample's surface to locate objects of interest. Also, the thin-film can be removed without leaving significant residue on the sample. These films are fabricated out of polydimethysiloxane (PDMS), which is bio-compatible, flexible, and mechanically robust, the films demonstrate reversible adhesions properties, and are optically transparent and reusable. The role of the gap between the back surface of the sphere and the object is studied, showing that magnification follows closely to geometrical optics predictions, while showing enhanced resolution in the regime of nanometric gaps. Due to the adhesive properties of PDMS to rigid samples, we were able to ensure close proximity to the object, in order to perform ultimate resolution studies, showing values on the order of ~$\lambda$/7.

In Chapter 5, we investigated an enhancement technique based on microspherical photonics for infrared photodetectors [71]. Such devices have been an area of significant interest in recent years, due to applications ranging from the medical field to military imaging systems. Numerous developments have occurred in the area of infrared materials, leading to devices with engineered bandgaps. Such materials with engineered bandgaps, like type-II strained-layer superlattices, also have their limitations in design architecture, battling effects like Shockley-Read-Hall (SRH). Further advancements, such as $n$B$n$ devices tackle these important issues. However, in terms of hierarchical components, there is room for improvement. For example, 2-D plasmonic meshes have been successfully integrated on the detector mesa to boost the performance of photocurrent [189]. Although the resonant techniques have had some degree of success [188], they only operate over a narrow spectral bandwidth, and are often polarization dependent. In Chapter 5, we integrate dielectric spheres on top of type-II strained-layer superlattice detector mesas to couple photonic jets into the light sensitive regions of the photodetector. We directly observe ~100x enhancement of the photocurrent due to the photonic jet. FDTD simulations are performed with reasonable qualitative agreement with the experimentally obtained results. Also, self-assembly methods are performed on arrays of prefabricated dents in order to have a proof-of-concept of extending this technology past the single-channel to focal plane arrays (FPAs).

# CHAPTER 2: MICROSPHERE-CHAIN WAVEGUIDES AND PHOTONIC MOLECULES

## 2.1: Introduction to Microsphere-Chain Waveguides

In recent years the microsphere-chain waveguides (MCWs) emerged as a paradigm for photonic applications, because of their ability to periodically relay light [1-20]. This interest was stimulated by two main factors. The first factor was connected with developing techniques of self-assembly, which made it possible to fabricate extremely long chains on a substrate without any structural defects or significant sphere misalignments, as illustrated in Fig. 2.1. The second factor was connected with an interest in novel optical modes and mechanisms of optical transport in these mesophotonic structures [5, 13-16, 112].

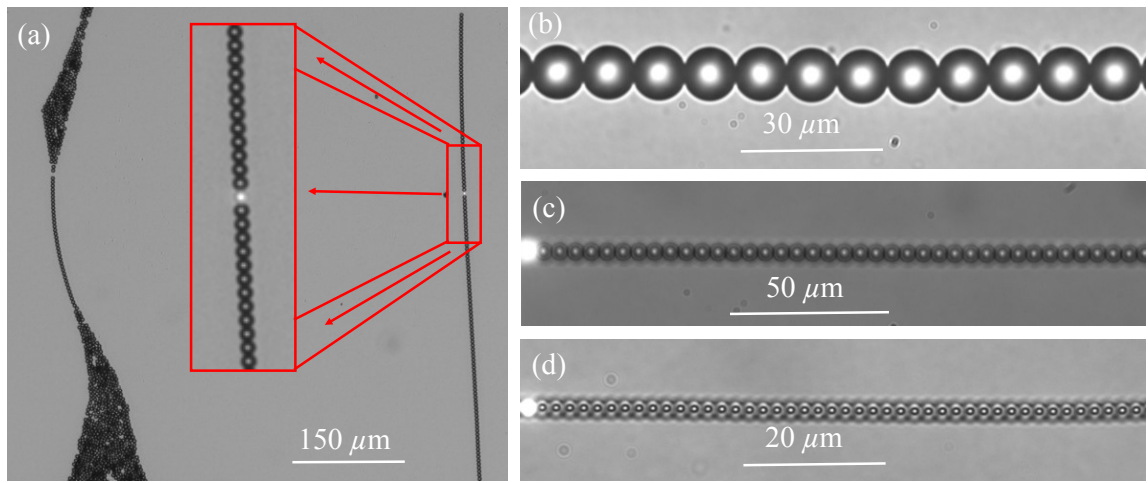

Figure 2.1: Optical micrograph of linear chains of touching dielectric spheres formed by self-assembly. The sphere diameters ranged from 5 $\mu$m in (a) and (c), 10 $\mu$m in (b) and 2 $\mu$m in (d).

Initially, these modes have been experimentally observed [13] in mesoscale MCWs formed by dielectric spheres made of polystyrene with $D<10\lambda$, where $\lambda$ is the wavelength of light. These modes have been termed nanojet-induced modes (NIMs) based on an analogy with tightly focused beams, "photonic nanojets" [33, 71-100] produced by individual spheres, shown in Fig. 2.2, which were discussed in Chapter 1. It was shown by performing spatially resolved spectroscopy and applying a Fabry-Perot model to the transmission spectra that NIMs are the dominate mode in size disordered chains of spheres, and WGM peaks are attenuated strongly at the rate of ~3 dB per sphere [14]. It was explained that these NIMs couple more efficiently to the MCWs than WGMs because of their tolerance to size disorder. Later, it was shown by FDTD simulations that the micro-joints which form during the self-assembly process of fabricating these chains, can drastically increase the NIM coupling between spheres [20]. These focusing effects occur over a broad spectral range, which makes them attractive for device applications.

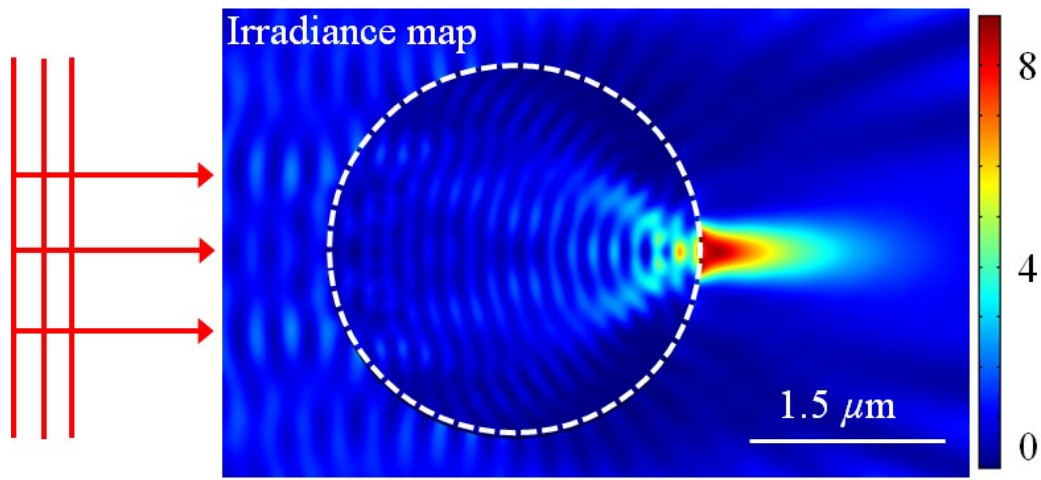

Figure 2.2: Simulation illustrating a photonic nanojet produced by a lossless, perfect electric conductor, dielectric cylinder with a diameter of 3 $\mu$m, index of 1.6, and illuminated with a plane wave with $\lambda$=0.53 $\mu$m.

These studies, on wavelength-scale structures, have shed light onto two effects in the optical properties of linear chains of dielectric spheres which are still not adequately understood. The first effect was connected with gradual tapering of the periodically focused beams in chains of dielectric spheres [13]. The second effect was connected with extremely small losses, on the order of 0.1 dB/sphere, observed in these photonic structures at distant portions of the linear chain of dielectric spheres which are well separated from the locally integrated light source [14, 19]. This level of losses was found to be significantly smaller than the attenuation due to reflection, ~0.27 dB/sphere, as estimated for incident plane waves with random polarization eigenstates. These transport and focusing effects were observed under multimodal illumination conditions, given by a fluorescent dye-doped source sphere, locally integrated into the chain of passive dielectric polystyrene spheres in contact position.

In recent years there has been extensive numerical modeling of these effects, within the limit of geometrical optics $D>>10\lambda$. Therefore these results are applicable for sufficiently large spheres, in comparison to the wavelength of light. This numerical modeling showed that both of the transport and focusing effects mentioned previously for wavelength-scale spheres, $4<D/\lambda<10$, present themselves in the geometrical optics regime, $D/\lambda>>10$. However it should be noted, that in contrast to the wavelength-scale spheres with $n=1.59$ where these effects were initially observed, the calculations predict that in the limit of geometrical optics these transport and focusing effects should take place at much higher indices within the range of $1.72<n<1.85$. Furthermore, it was shown that these effects are manifested from a condition that satisfies periodically focused modes (PFMs) which can propagate through the chains of spheres *without attenuation* under the Brewster angle condition for TM polarized rays [112]. If the azimuthal symmetry of the optical axis is considered, this PFM condition gives rise to radially polarized cylindrical vector beams within the chain of dielectric spheres [5]. It was suggested that PFMs can be considered as the geometrical optics analog to NIMs [112].

In this dissertation, we experimentally study the transition from geometrical to mesoscale optical transport in MCWs formed by polystyrene spheres ($n$=1.59), with gradually increasing diameters from 2 $\mu$m to 30 $\mu$m (~4$\lambda$ to ~60$\lambda$), with a locally integrated multimodal fluorescent dye-doped light source. Also, NIMs formation and polarization effects in a MCW comprised of 3 $\mu$m and 5 $\mu$m spheres is observed by full-wave simulations with COMSOL Multiphysics in the frequency domain.

We show that NIMs lead to an extremely well pronounced beam tapering effect for wavelength scale sphere, $D \leq 10\lambda$. However, we show this phenomena is absent for spheres with $D \geq 20\lambda$, as expected from geometrical optics. In terms of transport in such MCWs, we demonstrate that the propagation losses in mesoscale chains are too small (~0.1 dB/sphere) to be explained by geometrical optics theory. We supported some of our NIMs observations by comparison with full-wave numerical modeling, in the frequency domain, which also demonstrated that MCWs act as a polarization filter with periodic focusing capabilities.

### 2.1.1: MCW Experimental Setup

The light focusing and transport properties were investigated in chains of dielectric microspheres made from polystyrene (Thermo Fisher Sci., $n$=1.59 at $\lambda$=0.53$\mu$m). These spheres were monodispersive, and the diameters varied no more than 1-2%. These MCWs were formed from spheres diameters ranging from 1.9-30 $\mu$m, as shown in Fig. 2.4. Such MCWs were self-assembled on glass substrates. This self-assembly method is based on the evaporation of a liquid film containing spheres, which is sandwiched between two hydrophilic glass slides. As the liquid evaporates the liquid film forms pipelike flows stretching from massive deposits of microspheres to "pinning" points nucleated by spheres which have fallen into defects on the substrate, as shown in Fig. 2.3. Evaporation of microflows in the lateral direction leads to the formation of extremely straight and long (up to ~100 spheres) chains of touching spheres, as shown in Figs. 2.1(a) and 2.3(f).

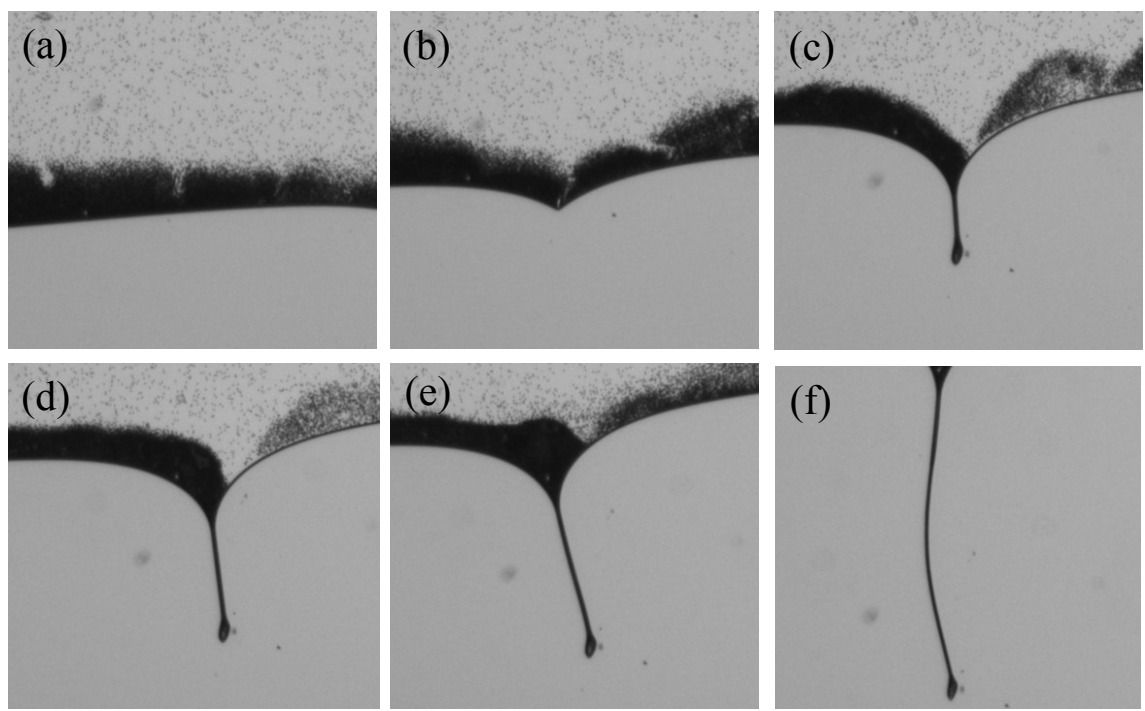

Figure 2.3: Optical micrographs demonstrating the self-assembly process. (a) Shows the evaporating front which contains a large concentration of polystyrene spheres. (b) Shows the wave front passing a defect, where spheres will tend to be trapped. (c)-(e) After the cluster of spheres nucleate in the defect the front of evaporating liquid continues to move due to evaporation. (f) The sides of the long liquid channel are still subject to evaporation, hence the straightening.

The chains of spheres are straight because of the minimization of the surface energy. For a straight chain the surface is minimal. Spheres tend to come into a contact position due to microfluidic forces at the interface between spheres. The study of material science aspects of such self-assembly processes goes beyond the scope of this dissertation. However, experimentally, we found that the chains contain practically no missing spheres or structural defects as shown in Figs. 2.4(b-d). Sometimes the chain can be broken in several parts, however in each section all spheres were found to be in a contact position along a straight line. The source of light, the fluorescently dye-doped polystyrene microsphere which is size-matched, was placed at one end of the chain in a contact position. This micromanipulation control was provided by a hydraulic micromanipulator connected with either a tapered metallic probe or fiber.

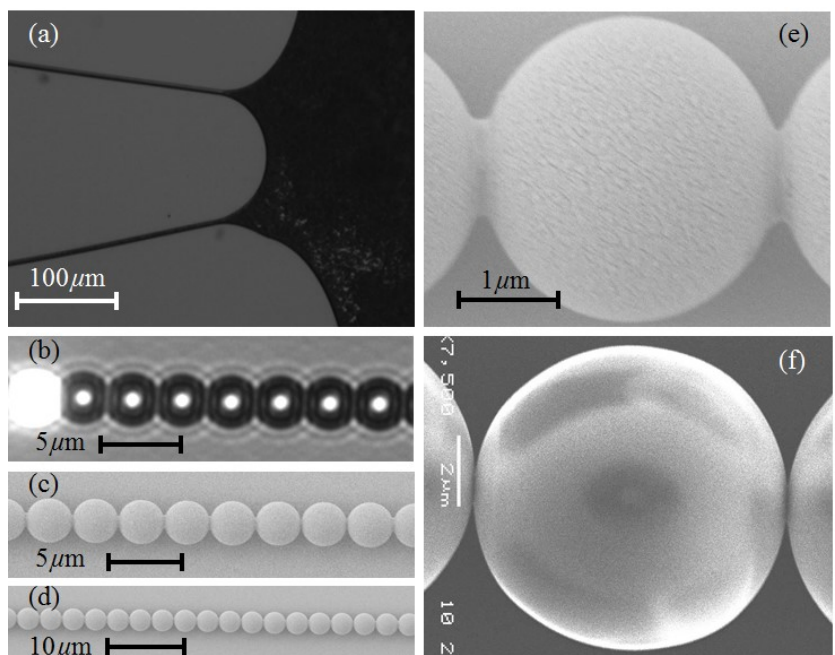

Figure 2.4: (a) Optical micrograph demonstrating the self-assembly process. (b) Optical and (c), (d) Scanning electron microscope (SEM) images of a chain of $3\mu m$ spheres. (e), (f) SEM images of micro-joints at the interface between spheres with $D$=3 and $10\mu m$, respectively.

The fluorescence excitation was provided at 460-500 nm by a mercury lamp integrated with an inverted IX-71 Olympus microscope. Fluorescence emission occurred in a 500-570 nm band having a maximum at 530 nm. A small fraction of the total emitted power was coupled into the propagating modes of the MCW. These propagation effects were visualized with the inverted IX-71 Olympus microscope, schematically illustrated below in Fig. 2.5. The images illustrating propagation effects were captured by a CCD video camera with variable exposure time and 12-bit dynamic range.

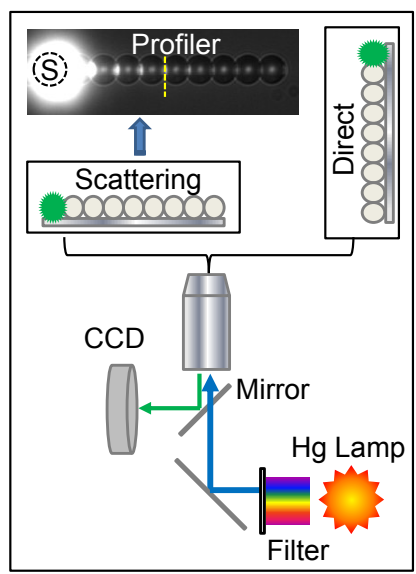

Figure 2.5: Schematic of experimental setup of direct and scattered imaging techniques with an optical micrograph showing a scattered image. Also, the schematic illustrates the mercury lamp exciting the fluorescently doped source sphere, through the objective lens then the emitted light is collected for imaging through the same objective.

The beam tapering effect is illustrated in Fig. 2.7 for MCWs formed by 5 $\mu$m spheres by two techniques. The first method is based on using scattered light with the imaging perpendicular to the MCW axis. We assume that due to surface scattering the transverse intensity profile represents the width of the beam guided within the MCW along the axial direction, see the vertical line termed "profiler" in Fig. 2.5. This allows for the determination of the beam full width at half maximum (FWHM) at different distance from the source, as illustrated in Fig. 2.5. The second method is based on using the light which is transmitted through the MCW and directly collected by the objective along the optical axis.

By removing spheres sequentially and reattaching the source sphere, as illustrated in Fig. 2.6, we were able to control the length of the MCW and establish a good agreement of the beam FWHM measurements by both methods, as shown in Fig. 2.7 for 5 $\mu$m spheres. Using the first method based on the scattered light imaging is advantageous, in comparison with transmitted light, due to the fact that it simultaneously captures several spatial beam profiles under the exact same imaging conditions rather than just one spatial beam profile. Therefore, all of the data collected for characterization of the transport and focusing properties of these MCW in this chapter of the dissertation were obtained by the scattered light imaging technique.

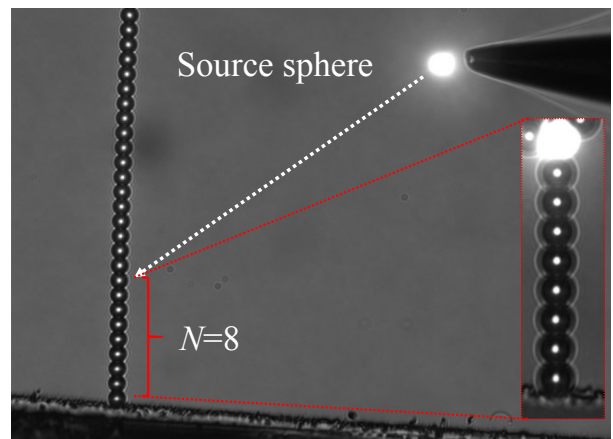

Figure 2.6: Optical micrograph illustrating the source sphere being integrated with the linear chain of $D$=5 $\mu$m polystyrene microspheres.

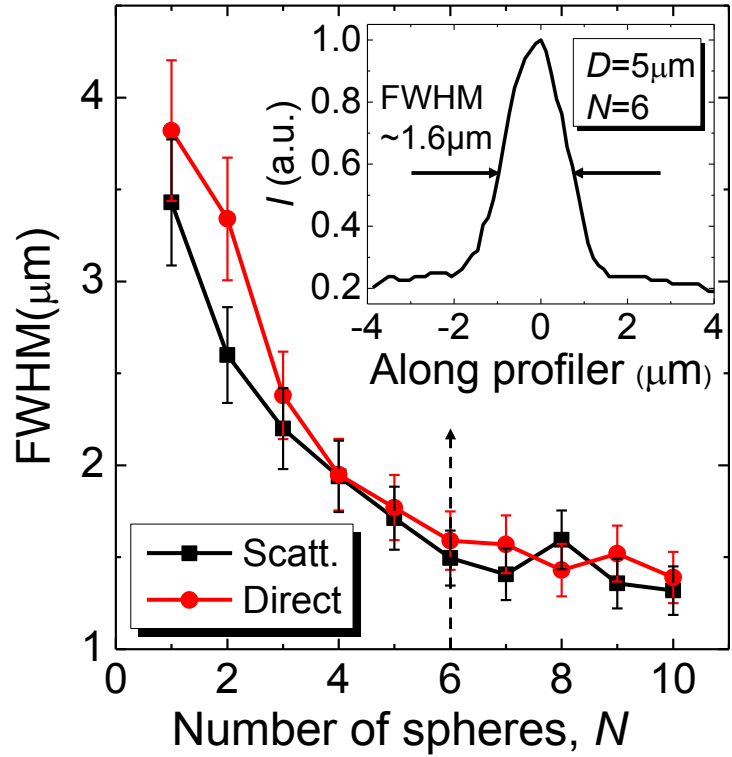

Figure 2.7: Illustration of beam tapering effect by two imaging techniques: i) scattered light perpendicular to the chain axis (squares) and ii) direct imaging along the axis of the chain (circles). Inset in (b) shows the FWHM scattering measurement after 6[th] sphere with $D$=5 $\mu$m.

### 2.1.2: MCW Focusing Properties

To accurately compare the results of the optical transport and focusing properties in MCWs with $D$ varying from 2 to 30 $\mu$m, it is essential to present the results in dimensionless units, as shown in Fig. 2.8. The FWHM measured after the $N$[th] sphere was normalized by the FWHM after the first sphere, $Y_N/Y_1$. Due to the multimodal nature of the source sphere emission, we typically observe a broad beam waist after the first sphere, $Y_1 \sim D/3$-$D/2$. As seen in Fig. 2.8, MCWs comprised of polystyrene spheres ($n$=1.59) with $D/\lambda \geq 20$ ($D \geq 10$ $\mu$m at $\lambda$=0.53$\mu$m) do not demonstrate a pronounced beam tapering effect. However, the mesoscale MCWs comprised of spheres with $4 \leq D/\lambda \leq 10$ ($2 \leq D \leq 5$ $\mu$m at $\lambda$=0.53$\mu$m) demonstrate a significant FWHM reduction with $Y_{10}/Y_1$<0.4. For MCWs formed by the smallest spheres, $D$=2 and 3 $\mu$m, it takes approximately 10 spheres for the FWHM to reach the dimensions $Y_{10}$~0.4-0.5 $\mu$m determined by the diffraction limit of our imaging system (~440 nm for NA=0.6).

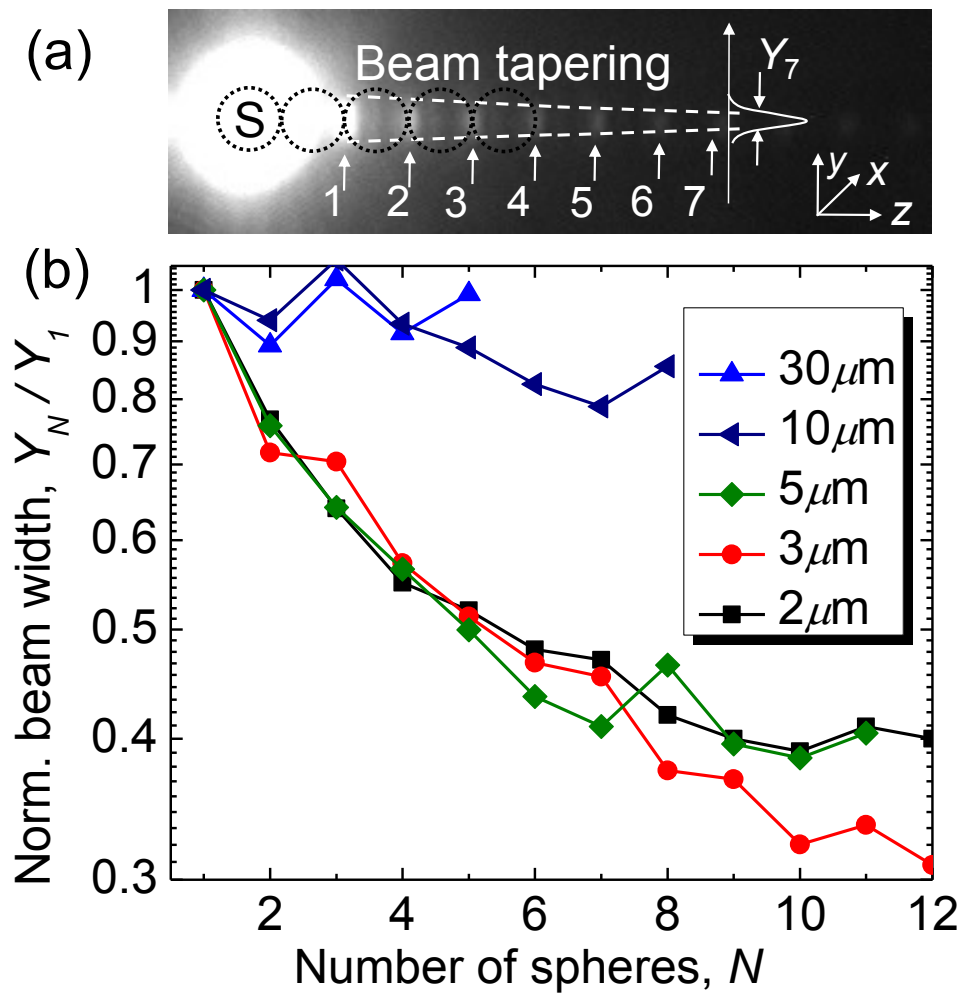

Figure 2.8: (a) Optical micrograph of experimental structures showing the near contact regions where the spatial beam profiling was performed, (b) Experimental results of the beam profiling at different distances from the source (S) sphere for spheres from 2 to 30 $\mu$m in diameter.

The MCWs beam tapering properties were modeled within the geometrical optics limit, by former Mesophotonic's Laboratory member Dr. Arash Darafsheh, using ZEMAX-EE ray tracing software, as illustrated in Fig. 2.9. Previously, this effect has been studied in a broad range of sphere indices [112], where it has been shown that this effect takes place in a relatively narrow range $1.72<n<1.82$. In our work, this effect was illustrated by calculations performed for chains of $N$ identical spheres with $n$=1.59, 1.77, 1.8, and 1.9. The emission of the fluorescence dye-doped sphere was modeled by using point ray sources with random polarization states, which were distributed randomly throughout the volume of the source sphere.

The evolution of the spatial beam profile was investigated at the interface between spheres using a flat square detector with the size equal to the diameter of the spheres in the MCW, $D$. Irradiance distributions were calculated from the density of rays at the detector plane. It can be seen from Fig. 2.9 that MCWs comprised of spheres with $n$=1.77 and 1.80 demonstrate marked FWHM reduction, $Y_{12}/Y_1{\sim}0.4$. This beam tapering effect is attributed to gradual filtering of periodically focused modes, PFMs, in such structures [112]. According to numerical modeling [112], PFMs are sharply focused in the regions where the spheres touch. On the other hand, the other modes tend to be focused away from the axis of the chain. For this reason, the modes different from PFMs cause the broadening of the beam at the sphere's contact regions. These modes, however, have larger propagation losses compared to PFMs. As a result, only PFMs survive in sufficiently long chains and, at the same time, the sizes of the focused beams are reduced along the chain. Significantly smaller beam tapering, $Y_{12}/Y_1{\sim}0.7$, was observed for $n$=1.9. For our experimental case of $n$=1.59, such beam tapering effects were practically absent, which was consistent with our measurements for spheres with $D/\lambda \geq 20$ ($D \geq 10$ $\mu$m at $\lambda$=0.53 $\mu$m), as shown in Fig. 2.8.

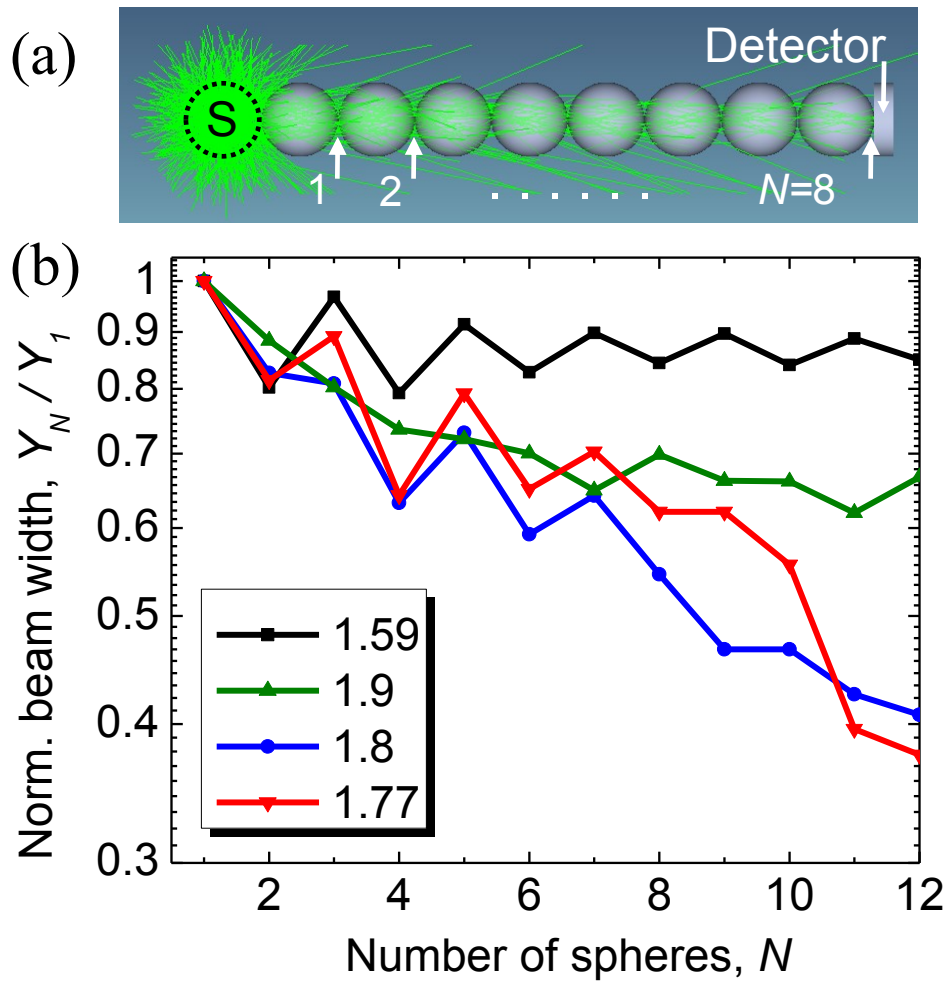

Figure 2.9: (a) Illustration of numerical ray tracing performed to study the evolution of the spatial beam profile through the sphere chain waveguide. (b) Geometrical optics numerical modeling results for the normalized beam width.

In order to shed some light into properties of mesoscale MCWs, we performed full-wave modeling by finite element method (FEM) in the frequency domain, as illustrated in

Fig. 2.10. The simulation was performed using COMSOL in a 2-D geometry for a chain of touching losses dielectric cylinders with $D$=5 $\mu$m and $n$=1.6. I initiated this modeling work in order to mimic our experiments with 5 $\mu$m polystyrene microspheres. Clearly, 2-D modeling can be used only for qualitative or semi-quantitative comparison with the experiments. However, our computational resources allowed us to perform such calculations only in 2-D case. Later, another student in the Mesophotonics Lab, Farzaneh Abolmaali, joined this project and carried out some of these calculations in collaboration with me. As a light source, we used a source cylinder containing 1400 dipole resonators which emit light at $\lambda$=0.53 $\mu$m, schematically shown in Fig. 2.10. The dipoles with various oscillation directions, to mimic random polarization, were randomly placed along several radial construction lines inside the source cylinder. In order to reduce the coupling to whispering gallery modes, the dipoles were removed from 0.5 $\mu$m near-surface region inside the source cylinder. The calculations were performed on an adjustable triangle mesh with minimal $\lambda$/180 (~3 nm) and maximal $\lambda$/18 (~30 nm) sizes.

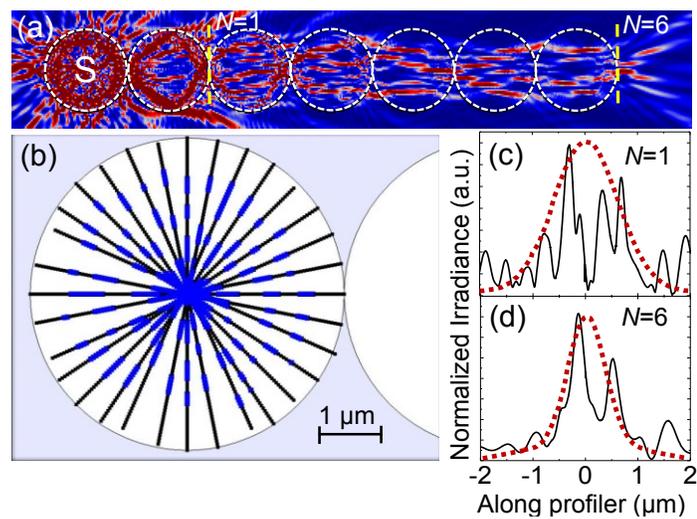

Figure 2.10: (a) 2-D irradiance map. (b) S-cylinder, dipoles placed along radial directions. Line profiles of (c) the first and (d) the sixth cylinder.

In the FEM simulations, a plethora of dipole oscillators are emitting at the same frequency, and have a fixed phase relationship which is established by the optical path difference separating them, and therefore are coherent. This strong coherence due to correlated frequency results in very pronounced interference effects, is evident in the irradiance map presented in Fig. 2.10(a). These effects manifest themselves in the multiple narrow beams in the 2-D irradiance map, shown in Fig. 2.10(a). Interference effects can also be seen in the transverse irradiance profiles, calculated at the interface between cylinders, as sharp peaks as illustrated in Fig. 2.10(c)-(d). We found that although the positions of the sharp peaks are very sensitive to the spatial distribution of the dipole oscillators, the envelope function is much less sensitive to such configuration effects. We hypothesize that the envelopes are representative of the spatial beam profiles expected from an incoherent multimodal source similar to that provided by the fluorescent dye-doped spheres used in our experimental studies shown throughout the chapter.

Such envelopes become narrower as light propagates through the MCW, reminiscent of our experimentally observed beam tapering effect. The envelope FWHM is rather broad after the first cylinder, $Y_1 \sim 1.7 \, \mu$m, as illustrated in Fig. 2.10(c). However, after the light has propagated through six dielectric cylinders, a large fraction of the modes have been filtered out of the cylinder-chain waveguide, resulting in a very narrow and tightly focused beam waist with $Y_6 \sim 1 \, \mu$m, as illustrated by the envelope in Fig. 2.10(d). It can also be seen in Fig. 2.10(a) that the calculated mode patterns are reminiscent of $2D$ periodic focusing properties of NIMs [13] or PFMs [112]. Thus, this modeling generally confirms that in mesoscale MCWs beam tapering can take place at smaller indices of refraction (~1.6), in contradiction to its geometrical counterpart which requires higher indices of refraction for such effects, $1.77 \leq n \leq 1.9$ [112].

### 2.1.3: MCW Transport Properties

Gradual filtering of extremely low-loss NIMs in such chains can be a key factor for the explanation of the MCW attenuation properties which are presented in Fig. 2.12. Power attenuation was studied by integrating the irradiance distributions observed at the interface region between the spheres in the scattering images acquired perpendicular to the MCWs axis, as shown in Fig. 2.12(a). The power estimation after $N^{th}$ sphere was normalized by the power after the first sphere, $P_N/P_1$. Significant coupling loss was observed in the section of the chain adjacent to the source sphere, $N<10$, as illustrated by a steep slope of dependencies in Fig. 2.12(a) and in the optical micrographs in Fig. 2.11. However, as the light propagates away from the source sphere, the losses are gradually reduced to extremely low levels. As illustrated in Fig. 2.12(b), for a distant section of the chain with $35 \leq N \leq 39$ the measured losses are found to be 0.076, 0.12, and 0.15 dB/sphere for 2 $\mu$m (red star), 3 $\mu$m (green circle), and 5 $\mu$m (blue triangle) spheres, respectively.

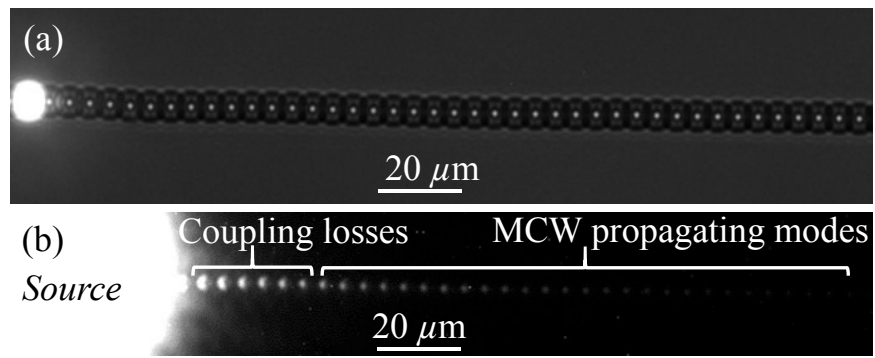

Figure 2.11: Optical micrographs of (a) a linear chain of ~40 polystyrene spheres with diameters of 5 $\mu$m and (b) light coupling into the propagating modes of the MCW.

67

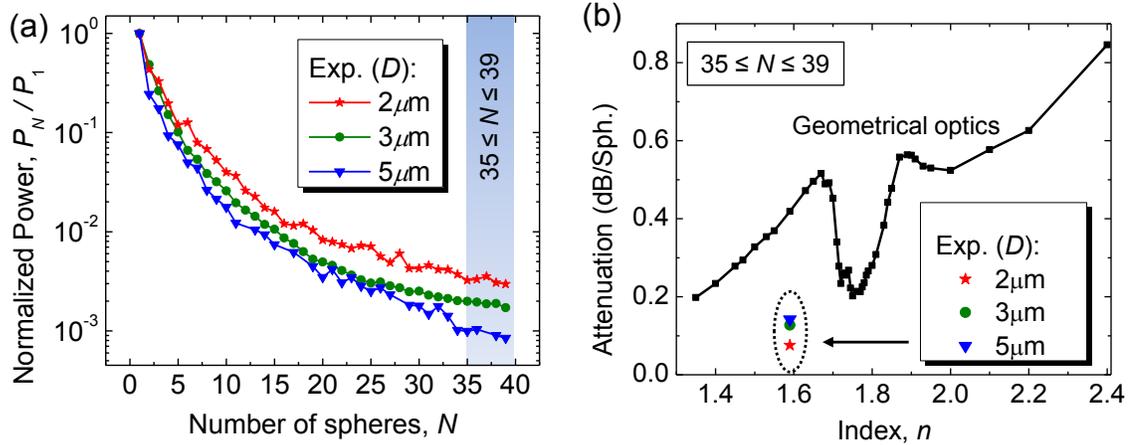

Figure 2.12: (a) Normalized power loss measured in MCWs formed by the polystyrene microspheres with $D$=2, 3, and 5 $\mu$m. (b) A comparison of the experimental data with the geometrical optics model.

These losses are far less than any estimation based on geometrical optics theory [1-11, 112]. Previously, these geometrical optics calculations have been performed for a broad range of indices 1.3<$n$<2.4 [3, 112]. In our work, we repeated such calculations for a section of the chain (35≤$N$≤39) which has been studied experimentally. The calculations were performed by a member of Mesophotonics Lab, Arash Darafsheh. The simulation results are represented by the curve in Fig. 2.12(b). The calculations are performed in a broad range of indices 1.4≤$n$≤2.4. The trend for increased attenuation with $n$ is due to enhanced Fresnel reflections by the spherical interfaces. The geometrical optics predicts a local minimum of attenuation (~0.2 dB/sphere) in a narrow range of indices around $n$=1.77 due to filtering of PFMs with the 2$D$ period and radial state of polarization [5]. For the experimentally studied case of $n$=1.59 the geometrical optics predicts attenuation ~0.4 dB/sphere. These results are expected to scale with $D$ for sufficiently large spheres where the geometrical optics is applicable.

As seen in Fig. 2.12(b), the experimentally observed attenuation in such chains is significantly smaller than the smallest values of attenuation predicted by geometrical optics at any $n$ from 1.4 to 2.4 (for $35 \leq N \leq 39$). It should be noted that such dramatically reduced attenuation takes place despite the random ~1-2% variations of the sphere diameters.

### 2.1.4: MCWs as Polarization Filters

The result of two-dimensional FEM modeling also sheds light into the polarization dependence of transmission. Light launched with TM polarization is propagated with less attenuation than TE polarized beams, as they are periodically focused within the chain. The operation of such structures is based on an analogy with larger scale structures, within geometrical optics, where their photonic eigenstates are represented by PFMs with orthogonal polarizations, which have dramatically different propagation losses [5].

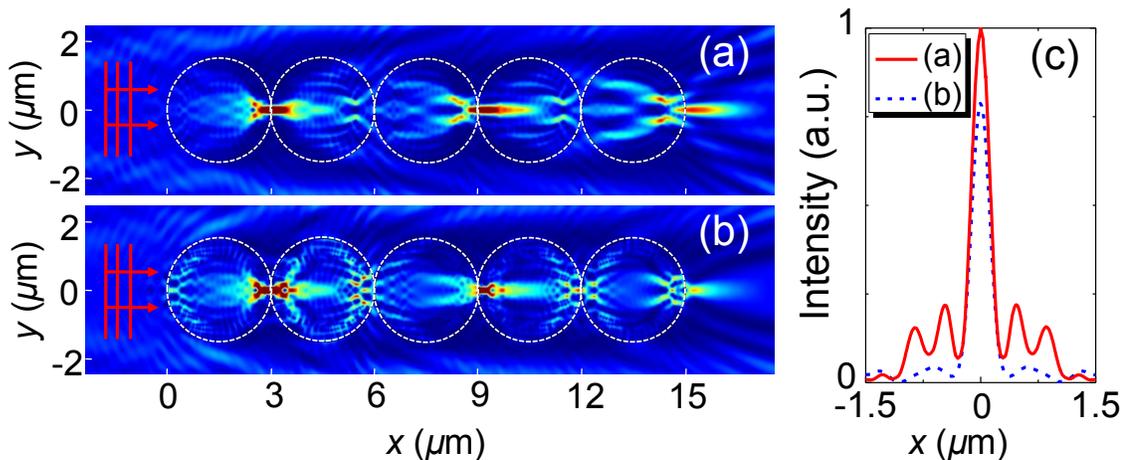

Figure 2.13: 2-D electric field intensity maps calculated for touching cylinders with $D$=3 $\mu$m and $n=\sqrt{3}$ at $\lambda$=0.53 $\mu$m for the incident plane waves with electric field oscillating along the (a) $y$-axis and (b) $z$-axis. (c) The intensity profiles of the focused beams calculated 0.62 $\mu$m away from the tip of the end cylinder.

To illustrate this analogy between PFMs and NIMs, simulations are performed for coupled cylinders where the eigenstates of the input beam are linearly polarized. The lossless dielectric, perfectly electric conducting, cylinders of 3 $\mu$m and index of refraction $\sqrt{3}$ are illuminated by plane waves propagating along the x-axis, polarized along the *y*-axis and *z*-axis. Plane waves, $\lambda$=0.53 $\mu$m, which were launched with their electric field vector oscillating along the y-axis (TM) led to higher transmission in comparison to plane waves launched with their electric field vectors oscillating along the z-axis (TE), as illustrated in Fig. 2.13.

Higher transmission for incident polarization along the *xy* plane, illustrated in Fig. 2.13(c), can be explained by the role of the Brewster angle for this polarization. Since only a small fraction of the wavefront is incident on the cylinder surface at the Brewster angle, this model cannot be used for quantitative analysis of such mesoscale structures; however it provides a physical explanation for this observation.

### 2.1.5: Conclusions for MCWs

The theoretical understanding of these NIMs requires an in-depth modal analysis. A simple limiting case is represented by the nanosphere-chain waveguide (NCW) formed by nanospheres with $D \ll \lambda$. In this case, the losses per nanosphere can be extremely small due to the fact that the NCW can be modeled as a single-mode low-loss nanofiber with effective index guiding [15]. It is interesting to note that for the high-index ($n$>1.9) NCWs the losses can be completely eliminated in certain frequency intervals $\omega(k) \leq ck < c\pi/D$, where $k$ is the quasiwave vector of the polariton modes traveling along the chain and $c$ is the vacuum speed of light [15]. For larger ($D \sim \lambda$) spheres corresponding MCWs becomes multimodal. Due to symmetry, the modes of the structure should be represented by periodically focused Laguerre-Gaussian beams which should be characterized by the indexes *l* and *m*, characterizing the azimuthal and radial distributions, respectively. Wave optics calculation of attenuation of these modes goes beyond the scope of this dissertation.

An additional factor which can play an important role in our experiments can be connected with microjoints [16, 18, 20] inevitably occurring between the polystyrene spheres, as shown in Fig. 2.14. Such microjoints can significantly reduce scattering losses in periodically relaying structures [20]. A detailed study of the influence of microjoints on the optical transport for MCWs is beyond the scope of this dissertation.

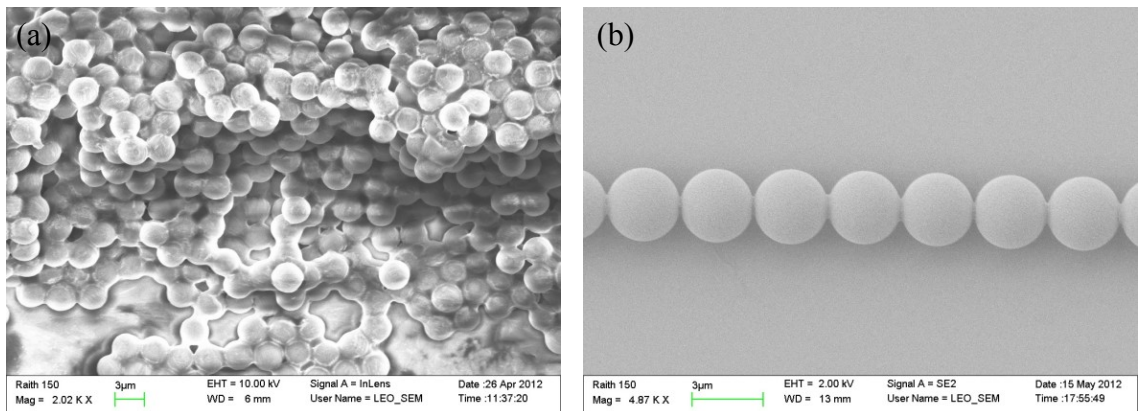

Figure 2.14: Scanning electron microscope images of microjoints formed from material reflow which occurs during the self-assembly process used in order to form organized arrangements of spheres such as (a) clusters of spheres, and (b) linear chains of spheres.

In conclusion, it was demonstrated in the first part of this chapter dedicated to the non-resonant transport of light through chains of microspheres, that mesoscale ($4 \leq D/\lambda \leq 10$) self-assembled polystyrene sphere-chain waveguides, coupled to a multimodal source, gradually filter periodically focused beams with the diffraction-limited beam waists. We showed that such "beam tapering" takes place at much smaller indices ($n$=1.59) than is predicted by geometrical optics. The propagation losses deeply inside such chains were also found to be smaller than any prediction based on geometrical optics. The results of this work are important for developing waveguiding structures with focusing capability which can be used as local microprobes, ultraprecise laser scalpels, and polarization filters. They can also stimulate development of the full wave theory of such structures and devices.

## 2.2: Introduction to Photonic Molecules

In addition to the non-resonant light transport of light, presented in the previous sections of this chapter, there has also been interest stimulated in the resonant transport of light in coupled resonant cavities [12, 14, 20, 25, 33, 101-110, 113, 120-123]. More specifically there has been a large effort focused on coupling between whispering-gallery modes in spherical cavities [25-36, 38-50, 113, 120-123]. The transport of these confined photonic states present striking similarities to confined electron states in an atom [190]. Initially this phenomena was observed in the context of a microsphere maintaining WGM resonances while in contact with microspheres of random sizes. Although the system follows classical electrodynamics this result in [190] demonstrates the analogy of tight-binding from quantum mechanical systems [114-116]. Naturally the term "photonic atom" has emerged from this analogy. Extrapolating on this terminology, a chain or cluster of so called photonic atoms form a photonic molecule (PM). Controlling the coupling parameters and spatial configurations of the photonic atoms allows for the spectral engineering of PMs [108]. Furthermore, this opens an avenue to engineer electromagnetic structures to expand the frontier of novel physical regimes in atomic physics and quantum optics [108].

In this dissertation, we studied chains and clusters formed by individual dielectric microspheres with degenerate WGM resonances termed photonic atoms, Fig. 2.16. These studies consisted of FDTD simulations and in parallel experimental work was done by a Mesophotonic Laboratory member Yangcheng Li. However, those extensive experimental results go beyond the scope of this work and will be presented in his dissertation.

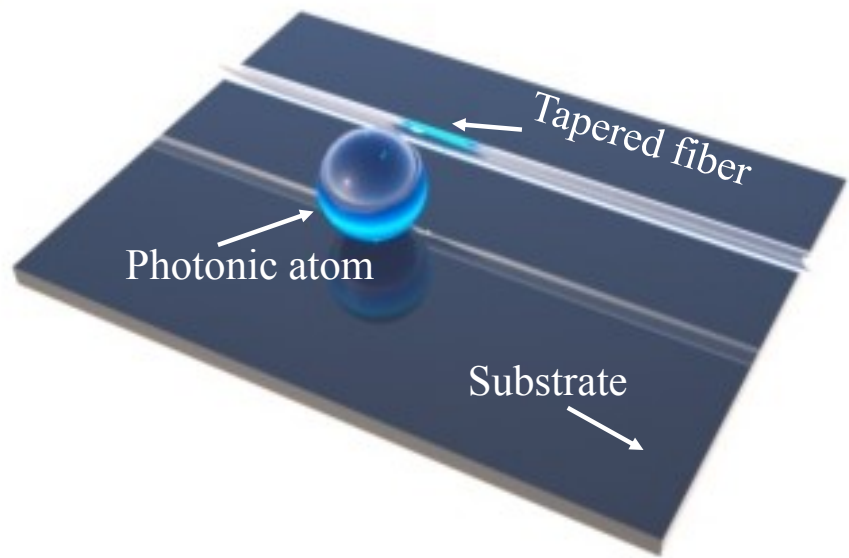

Figure 2.15: Illustration of the sphere, photonic atom, on a substrate evanescently coupled to a tapered fiber.

The photonic states of such chains and clusters are excited using side-coupled waveguides. The numerical modeling is performed for a 2-D cross section, through the 3-D geometry at the equatorial plane of the spheres, by FDTD calculations. However the results are qualitatively applicable to various 3-D structures which have the same cross sectional geometry such as clusters of dielectric microspheres, cylinders, disks, toroids or rings.

In this work, 25 $\mu$m diameter microspheres with a refractive index of 1.59 were studied in computational space with refractive index of 1.33. This was done intentionally to match the experimental conditions, not presented in this dissertation, where polystyrene ($n$=1.59) spheres were sorted with overlapping resonant WGMs and a variety of photonic molecules were arranged while immersed in water ($n$=1.33).

## 2.2.1: Simulation Parameters of Photonic Molecules

Two-dimensional numerical simulation using finite-difference time-domain (FDTD) method was performed, using Lumerical FDTD solutions, to study the optical coupling properties of WGMs between spherical resonant cavities [21-23]. The theoretical model is illustrated in Fig. 2.16 for the case of a single sphere ($n_s$=1.59, $D$=25 $\mu$m) side-coupled to a fiber waveguide ($n_{wg}$=1.45, $w$=1 $\mu$m) in air, shown in Fig. 2.15(a). A Gaussian modulated pulse of ~10 femtosecond (fs) length with central wavelength at 1310 nm was injected into the waveguide. The polarization eigenstates of the light injected into the waveguides was composed of the fundamental TM and TE modes. Demonstrating the electric field of the injected wave, Fig. 2.15(a) and the corresponding spectra Fig. 2.15(b), show the electric field response in the time domain and the irradiance spectra in the frequency domain represented by wavelength, respectively. At the output of the waveguide there was a time domain monitor placed in order to record the transmitted electric field response, Fig. 2.16(e). The simulation time was usually on the order of ~20,000 fs, which is necessary for a sufficient amount of field decay, $10^{-6}$ of the initial amplitude measured within the perfectly matched layers confining the computational grid. Coupling of light to the microsphere is illustrated by the electric field map in Fig. 2.16(d). After the electric field has been recorded in time, Fig. 2.16(e), a discrete Fourier transform of the signal was then performed to obtain the frequency domain response, as presented in Fig. 2.16(f). The spectrum shows periodic dips due to coupling of light to WGMs in sphere.

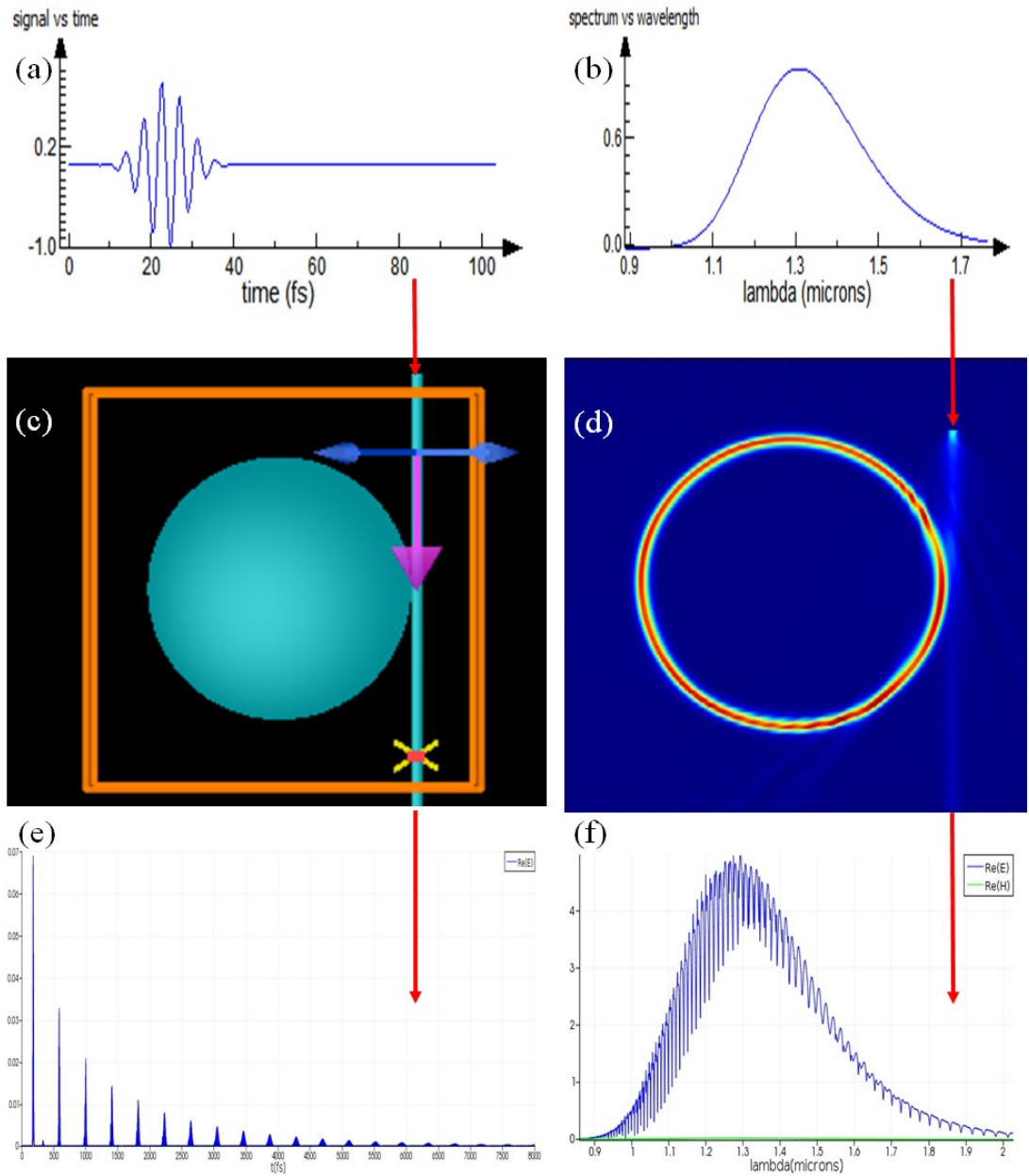

Figure 2.16: Results from FDTD simulation of single sphere side coupled to tapered fiber waveguide (a) Input pulse of the electric field; (b) Corresponding spectra; (c) geometry (d) Power map. Transmitted E-field in (e) time domain and (f) frequency domain, in terms of wavelength.

2.2.2: Simulated Spectral Finger-Prints of Photonic Molecules

It has been well established, as discussed in the introduction to this chapter, that an analogy can be drawn between strongly coupled resonant cavities and strongly driven two-level quantum mechanical systems [25, 108, 190]. First, let's consider the well-known case of the biatomic photonic molecule [25, 120, 121]. Biatomic photonic molecules have one coupling region which can be termed bonding domain to be consistent with the analogy and a 180 degree boding angle. The photonic atoms, acting as the building blocks for this photonic molecule, have $D$=25 $\mu$m, $n$=1.59, in water. There are red dotted construction lines passing through the uncoupled resonance in Fig. 2.17(b), aligning with the coupled modes obtained by simulations in Figs. 2.17(c), 2.17(d), 2.17(f), 2.17(h). The FDTD simulations show that uncoupled resonances split into two coupled resonances, as shown in Fig. 2.17(d), analogous to the classical bonding and antibonding molecular states for the bisphere case.

FDTD simulations were also performed for 3 photonic atoms ($D$=25 $\mu$m, $n$=1.59, in a computational grid $n$=1.33, to mimic water) forming a linear chain with 2 bonding domains with 180 degree bonding angles, as shown in Fig. 2.17(e). The red construction lines are indicative of the uncoupled WGM resonance for TE and TM polarization states. It can be seen in Fig. 2.17(f) that for the FDTD simulations the uncoupled resonance splits into 5 coupled modes. It should also be noted, that there could potentially be a degenerate coupled mode, of the TE and TM polarization eigenstates.

Extending the molecule to 4 photonic atoms ($D$=25 $\mu$m, $n$=1.59, in a computational grid $n$=1.33, to mimic water) formed a linear chain with three bonding domains with 180 degree bonding angles, as shown in Fig. 2.17(g). The red construction lines are indicative of the uncoupled WGM resonance for TE and TM polarization states. It can be seen in Fig. 2.17(h) that for the FDTD simulations the uncoupled resonance splits into 7 coupled modes. Again it should also be noted, that there could potentially be a degenerate coupled mode, from the two polarization states.

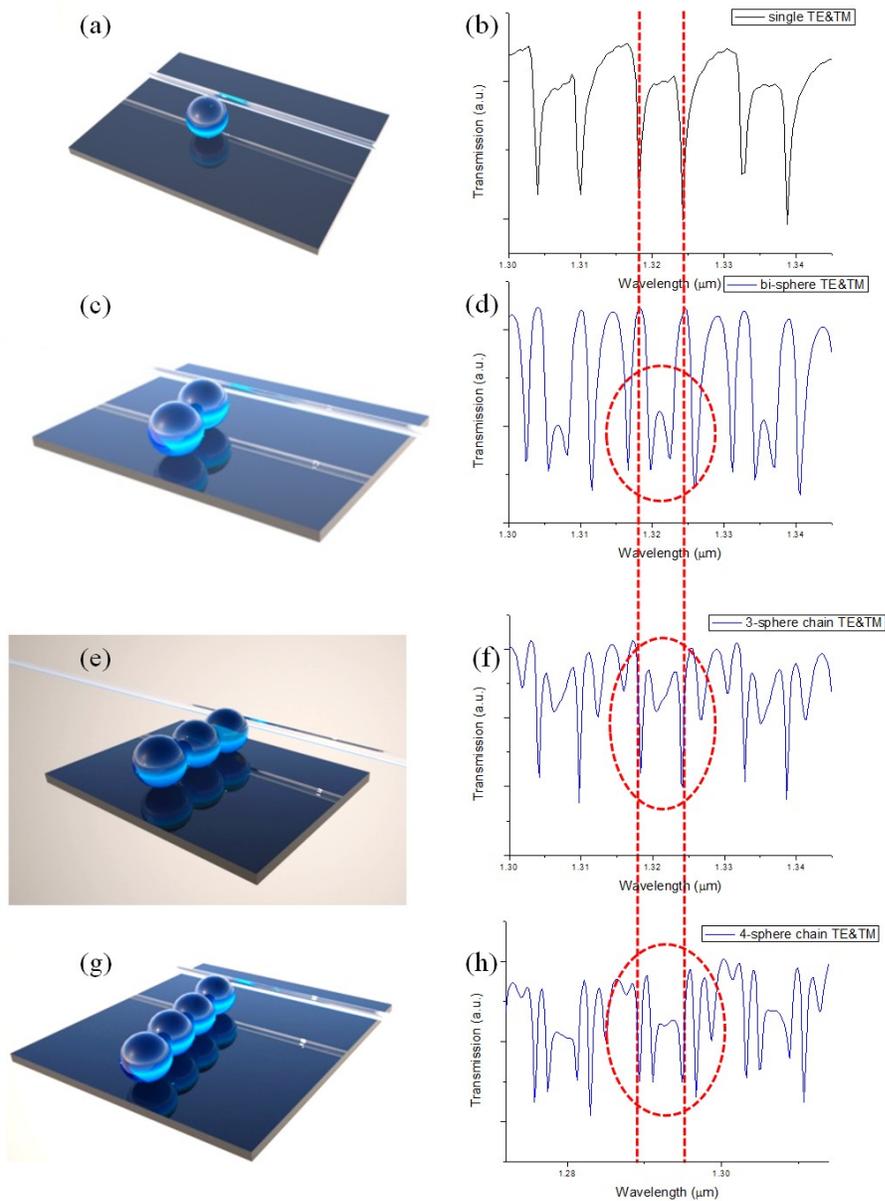

Figure 2.17: Transmission spectra produced by linear chains of photonic atoms (a)-(b), of a bisphere (c)-(d), 3-photonic atom chain (e)-(f) and 4-photonic atom chain (g)-(h).

Another arrangement for 3 photonic atoms ($D$=25 $\mu$m, n=1.59, in water with n=1.33) is trigonal planar, which has an additional bonding domain when comparing with the linear chain, with three bonding domains, as shown in Figs. 2.17(e) and 2.18(c). Another key distinction between the linear chain of 3 photonic atoms and the trigonal planar is that the structure extends into 2-D causing the bond angle to decrease to a value of 120 degrees. With a difference of 60 degrees for the bond angles, and one additional bonding domain, we can anticipate different optical path and photonic states. The red construction lines are indicative of the uncoupled WGM resonance for TE and TM polarization states. It can be seen that the uncoupled TM resonance splits into 3 coupled modes, according to FDTD simulations and is confirmed experimentally.

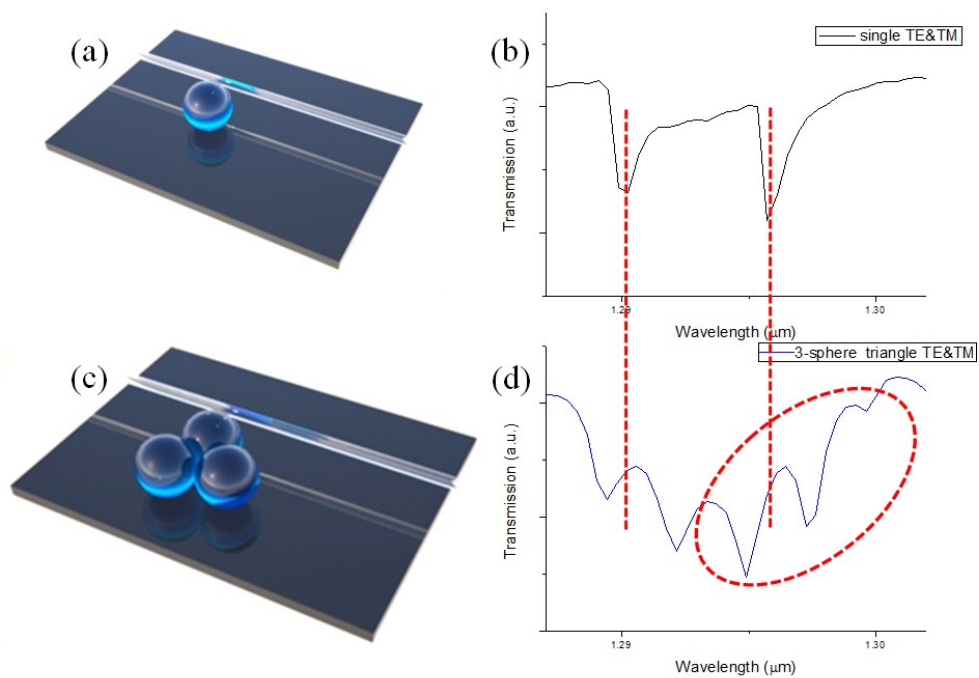

Figure 2.18: FDTD simulation results for the triagonal planar photonic molecule (c)-(f). Directly compared with the uncoupled resonance of the photonic atom (a)-(b)

### 2.2.3: Conclusions on Photonic Molecules

We show that photonic molecules have certain spectral properties which are characteristic for a given spatial configuration; depending on the number of bonding domains, and the bond angle between domains of their constituting photonic atoms. We called these features spectral finger-prints of photonic molecules. The spectra of photonic molecules also depend on the coupling strength between the atoms and their specific parameters. We observed specific splitting in the spectra of biatomic, 3-atom chain, 4-atom chain, and planar triganol photonic molecules which can be related to symmetry properties and spatial configurations. In some cases, $Q$-factors of such molecular photonic states were found to exceed the corresponding uncoupled resonance in a single sphere. This observation can be related to super-modes which have been known [108] to emerge from coupled resonators. These FDTD simulations were in excellent agreement when directly compared with experimental results obtained by Mesophotonics Laboratory member Yangcheng.

# CHAPTER 3: PHOTONIC JETS PRODUCED BY MICROSPHERES INTEGRATED WITH HOLLOW-CORE FIBERS FOR ULTRAPRECISE LASER SURGERY

## 3.1: Introduction to Photonic Jets for Laser Surgery

The concept of "photonic nanojet" emerged in 2004, as discussed in previous chapters, in the context of extremely sharp focusing of light by mesoscale dielectric microspheres with diameters on the order of several wavelengths ($\lambda$) and index of refraction ($n$) about 1.6 [73]. Due to the fact that the nanojet is created near the back surface of the sphere illuminated by a plane wave, such structures are ultimately suitable for ultraprecise contact surgeries requiring very shallow and narrow cuts through the tissue [6-11]. Such optical scalpels can be used for dissecting and removing fibrotic membranes from the retina surface [191]. Our designs were developed for Erbium:YAG laser ($\lambda$=2.94 $\mu$m) with short optical penetration depth in tissue [11], as shown in Fig. 3.1.

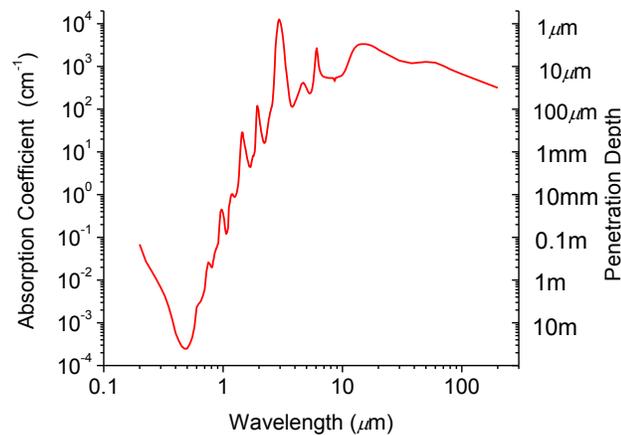

Figure: 3.1: Spectral curve for water absorption and penetration depth [192].

These optical scalpels can ablate the tissue while being operated in contact conditions. Engineering sharply focused beams at the distal tip of the device is challenging in these applications because of the multimodal structure of the beams delivered by flexible hollow waveguides (HWG) or multimode fibers [194]. Single mode fibers are not well-developed in the mid-IR range and they have limited coupling efficiency with many practical radiation sources. For these reasons previous research has been developed for surgical devices designed for multimodal input beams [6-11]. It was proposed to use chains of microspheres to filter so-called "nanojet-induced modes" (NIMs) [1-4, 13, 14] in these structures, similar to the previous chapter on microsphere-chain waveguides. It was demonstrated that periodical focusing effects lead to the gradual reduction of the beam diameters along the chains. This enabled compact beams to form in these chains to be used for ultraprecise surgery [11]. The drawback of these structures, however, was connected with their poor overall transmission efficiency on the order of $3\% - 5\%$ [195].

In this dissertation, we develop a novel single-mode approach to the design of such devices. Advantages of this approach are connected with the minimal spot sizes of photonic nanojets produced by a single microsphere under single mode illumination, and efficient delivery of the optical power by a novel mid-IR waveguide. Our approach is based on two recent developments. First, the development of a new generation of compact, powerful (~30W) and portable diode-pumped Er:YAG lasers capable of operation in a single mode regime at $\lambda$=2.94 $\mu$m [196]. Second, the development of flexible single-mode microstructured fibers with large core diameters [136]. We used novel silica hollow-core microstructured optical fibers (HC-MOF) with negative curvature of the effective air core wall, and a core diameter 53 $\mu$m [136] that significantly simplified coupling of light from the laser and integration with the focusing microsphere. We integrated HC-MOF with high-index ($n$~1.8) barium titanate glass (BaTiO$_3$) microsphere with the diameter 53 $\mu$m matching the effective air core diameter of HC-MOF, as shown in Fig. 3.2. We tested propagation and curvature losses of HC-MOF, assembled the laser scalpel system with flexible delivery based on HC-MOF, and tested the systems focusing and transmission properties. Using an objective with finite NA~0.5 we measured the focused beam diameters around ~4$\lambda$. Our numerical modeling results show that in principle that spot sizes on the order of ~$\lambda/2$ can be realized using this approach. The discrepancy between the observation and theory can be attributed to the poor collection efficiency of the mid-IR spatial characterization setup. Also, we observed much higher efficiency of the single-mode system (>20%) compared to its multimode counterpart, which will be discussed in the next section of the dissertation. Due to a combination of sharp focusing, low-loss, and flexible delivery, these systems can find broad applications in ultraprecise microsurgeries.

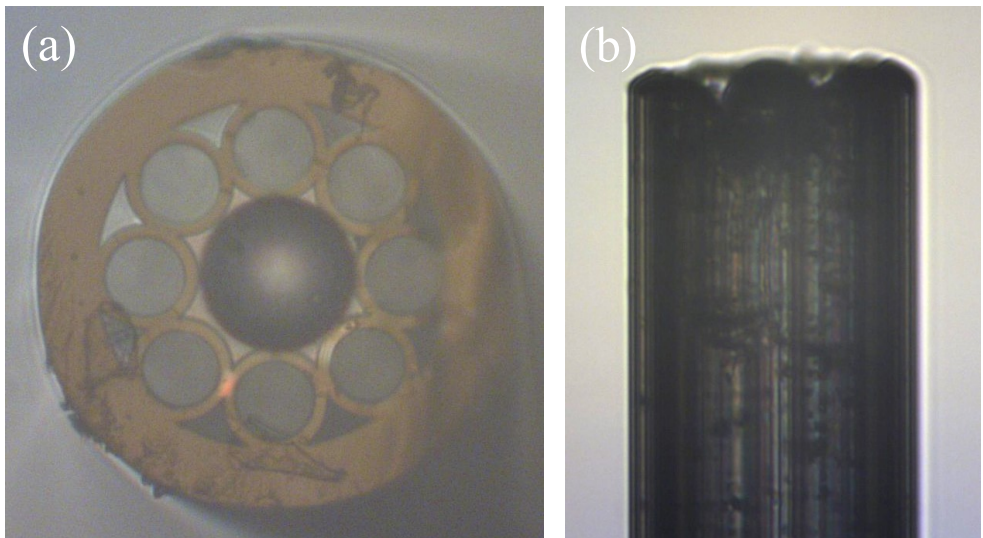

Figure 3.2: Barium titanate glass sphere ($n$~1.8) integrated into a HC-MOF, where the effective air core of the fiber is size matched with the diameter of the dielectric sphere at ~50 $\mu$m.

3.2: Multimodal Probe Design

The focusing of multimodal beams propagating through flexible optical delivery systems such as optical fibers or hollow waveguides were studied for surgical probes. In these multimodal systems each mode is focused in a separate spot, and all these spots are overlapped leading to large focused spot sizes well in excess of the diffraction limit. As an example, hollow waveguides with the diameter of internal hole around 300 $\mu$m support hundreds of modes at $\lambda$=3 $\mu$m [124]. As a result, focusing of the beams delivered by hollow waveguides would result in spot sizes ~100 $\mu$m. Such spot sizes would be considered extremely small in many types of laser surgery. However, the specific application which we are going to address in this dissertation requires smaller output beams. The fibrotic membranes developing at the surface of retina as a result of proliferative diabetic retinopathy (PDR) have a thickness of several monolayers of cells, typically on the order of 20-30 $\mu$m. The goal of the surgical procedure is to dissect and to remove these fibrotic membranes [191]. Currently, this surgery is performed by a set of steel instruments which are inserted through a hole in the sclera [197]. These procedures often results in bleeding that can lead to post-operational inflammations and complications. Laser surgery can provide a better solution for the surgery because it coagulates the tissue and stops bleeding. A natural suggestion would be to use well-established surgical tools such as LASIK (laser-assisted in situ keratomileusis) [198] for these procedures. It should be noted, however, that the femtosecond pulsed Nd:YAG laser used to create the hinged corneal flap in the LASIK procedure, penetrates to depths of ~120-180 $\mu$m. The surgical action is provided too deeply inside the tissue to be used for dissecting shallow membranes. The ultraviolet, argon fluoride excimer laser systems ($\lambda$=193 nm) that is used in LASIK to remove the tissue from the corneal bed to correct for refractive errors is well-developed. The laser is very suitable for sectioning the tissue, with extreme precision measuring ~1 $\mu$m. However, although this system could potentially dissect extremely thin membranes at the surface of retina, optical waveguides at this wavelength are not available for use in endoscopic surgery. This motivated the previous work of the Mesophotonics Laboratory.

In this dissertation, we propose a laser tool that is based on contact surgery at the strong water absorption peak wavelength of 3 $\mu$m. This tool operates when the distal tip of the device is in contact with the tissue, and the strong absorption leads to tissue ablation with very limited depth of surgical action. The limited depth of surgical action produces ablation craters on the order of ~20-30 $\mu$m [11]. For this reason, the probe can be ok for dissecting fibrotic membranes in such a way that the healthy retinal areas below these membranes are protected. As we stated above, multimodal devices tend to produce spot sizes in excess of 100 $\mu$m. These spot sizes are too large for achieving shallow surgery because it is difficult to provide the same shallow depth of surgery within larger spots. Previously, it has been demonstrated that by using chains of spheres it is possible to reduce the spot sizes down to several wavelengths in air and ~30 $\mu$m in tissue [11]. It has been shown that chains of coupled dielectric microspheres have certain advantages in comparison to single lenses in applications where multimodal sources of light are used to provide illumination. Chains of dielectric spheres filter periodically focused modes that produce sharply focused beams, as described in detail in Chapter 2. This allows for the reduction in the sizes of the focused beams in the structures, down to wavelength-scale dimensions at the cost of optical transmission.

Extensive numerical modeling and experimental work has been accomplished towards these multimodal focusing microprobes [2, 6-11]. Optimizing the parameters where $\lambda$=2.94 $\mu$m and hollow-core waveguides are used, a result of a beam waist of ~$\lambda$ can be predicted at the back surface of the distal tip after 3-5 sapphire spheres ($n$=1.71). However, this small spot size comes at the expense of the total transmitted power, so that only 3-5% of the power is coupled to the sharply focused output beam [195]. This process is illustrated in Fig. 3.3.

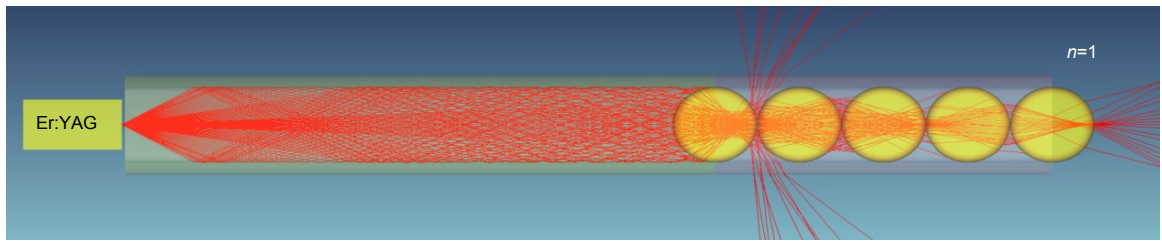

Figure 3.3: Illustration of the mode filter process in the multimodal focusing microprobe, providing a tightly focused beam waist at the back surface of the last sphere [7].

### 3.3: Single-Mode Scalpel Design

In this work, we develop a different approach to developing such laser scalpels based on an idea that we can use a single-mode Er:YAG laser source and we can design our delivery and focusing systems in such a way that the single-mode regime would be preserved throughout the system. The advantage of such an approach is that the entire system would not require any mode filtering for focusing, since we preserve a single-mode beam all the way from the source to the tissue application. This means that this design can be significantly more efficient in comparison to multimodal microprobes described previously. In addition, the focusing of the single-mode beam can be achieved by a single lens. Therefore, the single-mode design does not require a chain of microspheres for filtering certain modes. Just a single high-index sphere integrated with the flexible delivery system would provide sufficiently small spot sizes on the order of $\sim\lambda$.

As a source of light we use a portable diode-pumped Er:YAG laser capable of single-mode operation at $\lambda$=2.94 $\mu$m with nearly 1 W of power, as shown in Fig. 3.4 below. Although the Er:YAG laser has an output power of $\sim$1 W, that might be viewed as sufficient power, but in fact this level of power is relatively close to the threshold required for the tissue ablation. A CW beam cannot be used in compact laser surgery due to collateral thermal damage caused by the conduction of heat. However, heat conduction can minimized by delivery the laser beam in pulsed mode. The pulse duration of the laser can be determined by the thermal relaxation time of the tissue. This is the time that is necessary for the peak temperature to diffuse over the distance of the optical penetration depth of the laser beam. The relaxation time is proportional to the square of the penetration depth and inversely proportional to the thermal diffusivity. For micron-level penetration depth the relaxation time is in the range of tens to hundreds of microseconds.

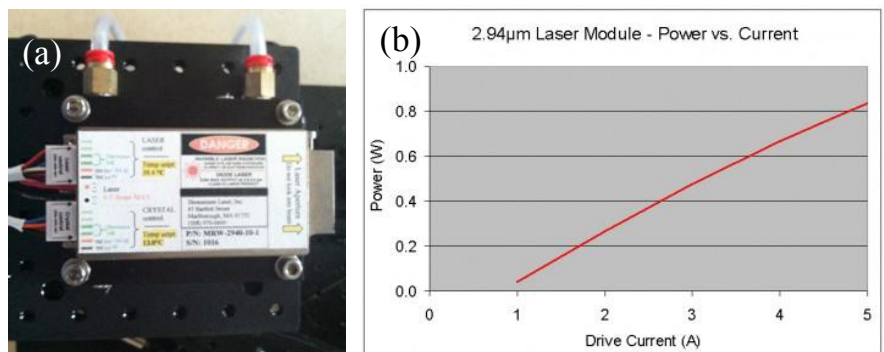

Figure 3.4: (a) Image of the portable and compact diode-pumped Er:YAG laser. (b) Output power as a function of current, for the Er:YAG laser [199].

The next component in this design is the hollow-core microstructured optical fiber (HC-MOF), provided to the Mesophotonics Laboratory by our collaborators at the Fiber Optics Research Center of the Russian Academy of Sciences. The natural suggestion would be to use a single-mode mid-IR fiber for making such flexible delivery system. It should be noted, however, that single-mode fibers are not very well developed as commercial products. Some of them have mechanical limitations such as fragility. Some of them contain materials, such as chalcogenide glasses, which are not compatible with medical applications. In addition, they usually have small-diameter cores on the order of ~15 $\mu$m [216-218]. Coupling of single-mode laser emission to small-diameter cores (~15 $\mu$m) is a feasible task, however it requires the development of very precise couplers. These couplers require periodical testing and adjustments that are difficult to provide in a hospital or in the environment of a doctor's office. For these reasons, we decided to use novel HC-MOF structures for flexible delivery of mid-IR emission. The advantage of HC-MOFs over standard single-mode mid-IR fibers is their very large core diameters ~40-70 $\mu$m that significantly simplifies coupling of the single-mode laser emission to the waveguides. Another advantage is the very small outer diameter of HC-MOF, on the order of ~300 $\mu$m, which makes them ultimately suitable for integration with current surgical procedures.

However, these HC-MOF had not been characterized previously with an Er:YAG laser system. Therefore, the first portion of the project was dedicated to studying the waveguiding properties of hollow-core microstructured optical fibers in the mid-IR regime with Er:YAG laser radiation ($\lambda$=2.94$\mu$m). The HC-MOFs studied have different effective air core diameters ($D_{core}$) and different geometries. One geometry is such that the cladding is formed by 8 contiguous capillaries, as shown in Fig. 3.5(a). There are 3 HC-MOFs with this geometry which have different effective air core diameters: 38, 53, and 70 $\mu$m. The other geometry is such that the 8 capillaries are slightly separated; as shown in Fig. 3.5(b). For this geometry there is only one HC-MOF with an effective air core diameter of 40 $\mu$m. The detailed waveguiding analysis allowed us to determine a suitable HC-MOF for the single-mode precision laser scalpel.

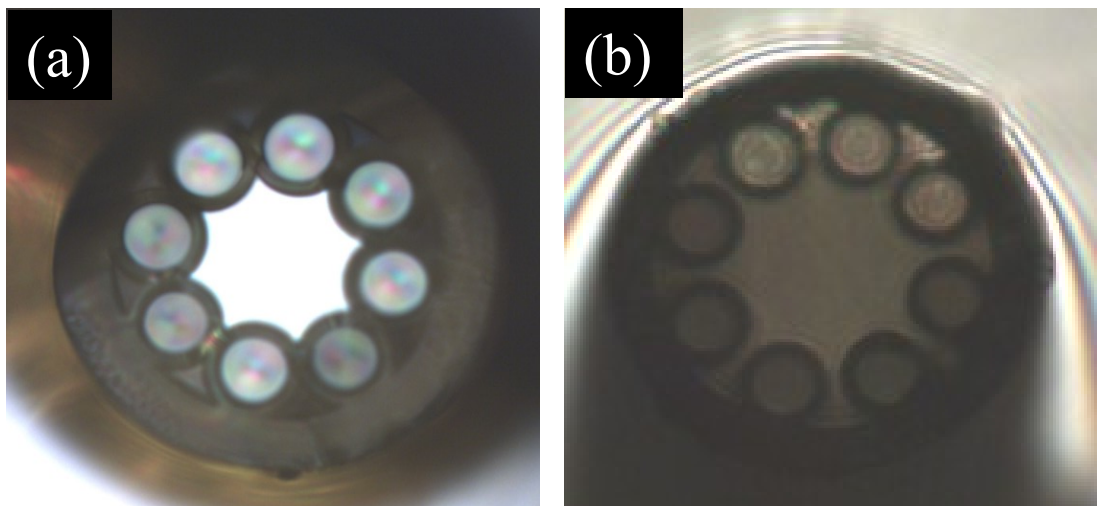

Figure 3.5: Optical micrograph of (a) contiguous geometry and (b) separated capillaries geometry for HC-MOFs.

The final component to this single-mode design is the focusing element at the distal tip of the probe. The single-mode spatial profile of the output beam from the HC-MOF allows for the use of a single dielectric microsphere. We performed FDTD simulations in order to determine the refractive index for the sphere that would provide the best focusing properties for the laser scalpel, as shown in Fig. 3.6. A FDTD computational grid of 70x90 $\mu$m$^2$ was chosen with a perfectly matched layer (PML) surrounding the computational area. A power monitor was placed over the computational area displaying the calculated fields at $\lambda$=3 $\mu$m. In order to model the single-mode beam transmitted thorough the HC-MOF, we used a Gaussian beam with a beam waist of 35 $\mu$m. Based on published results on the spatial mode profiles in HC-MOF [136], we believe that such a Gaussian beam is a reasonable approximation to the output beam of the HC-MOF with an effective air core of ~50 $\mu$m. An adjustable mesh was used throughout the FDTD computational grid with a maximum mesh cell size of $\lambda$/32.

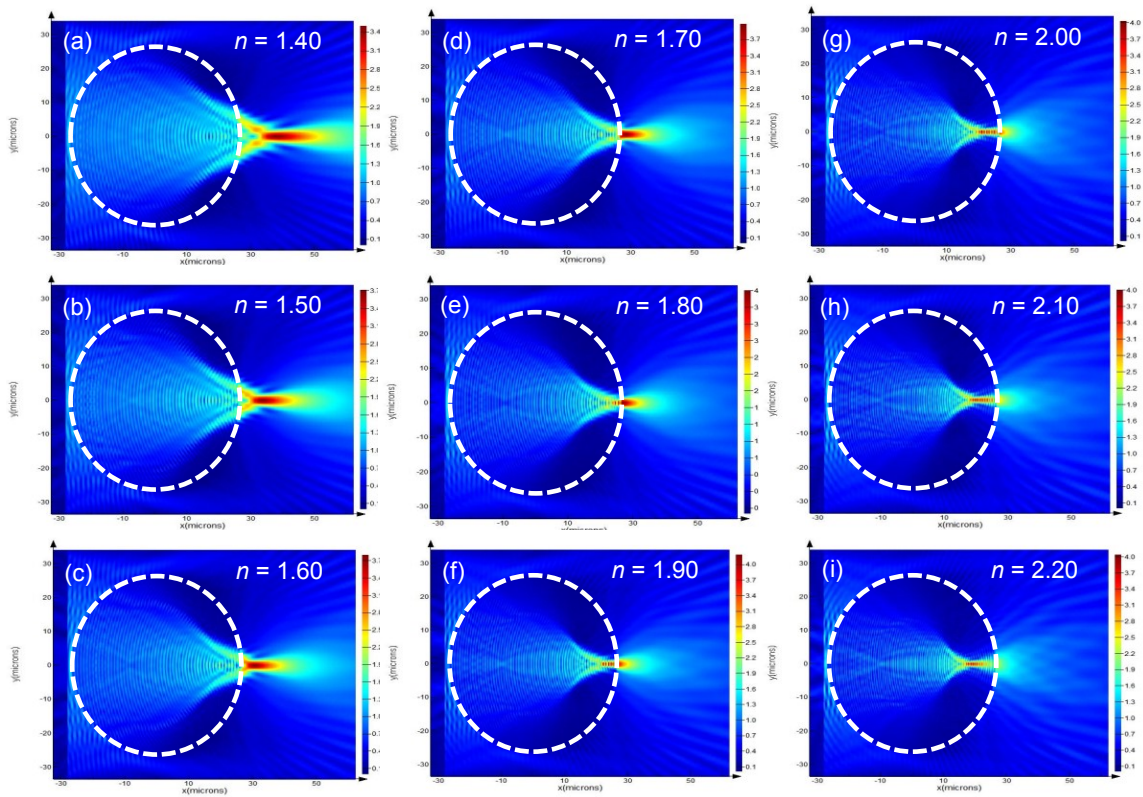

Figure 3.6: Power maps calculated by FDTD simulations for indices ranging from 1.4-2.2, (a)-(i). For the laser scalpel the results presented in the images (c)-(f) are ideal candidates due to the fact that the photonic jets is produced at the back surface of the spherical interface.

3.4: Experimental Results

This section of the chapter is two-fold: first in section 3.4.1 the HC-MOF waveguide characterization will be presented, secondly in section 3.4.2 the characterization of the focused beam of a laser scalpel in air will be shown.

### 3.4.1: HC-MOF Waveguiding Properties

Laser radiation was provided by an Er:YAG diode laser (Sheaumann Laser, inc). Imaging of the output beam's intensity spatial profile was captured by using 124x124 matrix pyroelectric IR beam analyzer (Ophir-Spiricon LLC). The physical size of each element is 85x85 $\mu m^2$. Total output power measurements were obtained by an IR power meter (Newport). The experimental set-up is illustrated in Fig. 3.7.

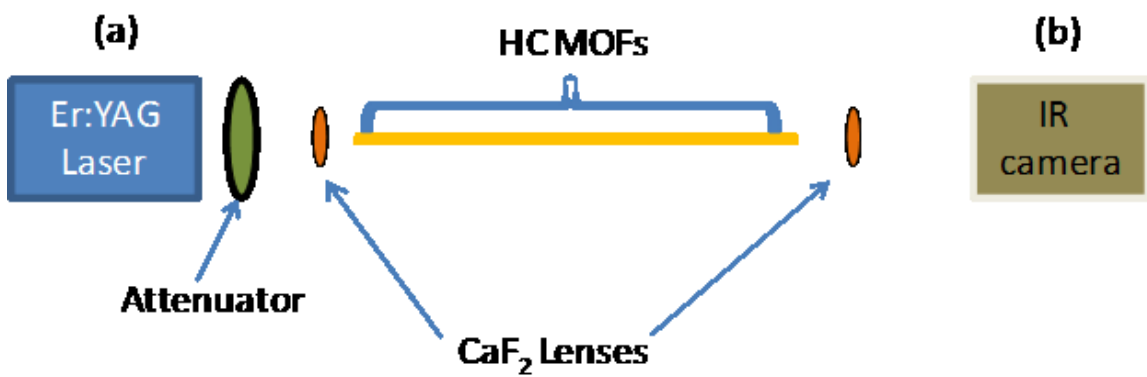

Figure 3.7: Schematic of experimental setup: (a) Er:YAG diode laser, λ=2.94μm. (b) IR camera, pyroelectric IR beam analyzer.

In order to efficiently couple the laser beam into the HC-MOF, two parameters should be considered. First, the beam-waist which is produced by the coupling lens. Second, the numerical aperture of the coupling lens. The beamwaist produced should be smaller than the effective air core diameter of the HC-MOF. In accordance with this constraint the numerical aperture produced by the lens should be less than the acceptance cone, numerical aperture, of the HC-MOF. It is a critical point to ensure that the measurements are performed when the beam is coupled into the effective air core, rather than the extremely attenuating (~44dB/m) "parasitic" modes, which occur when the beam is coupled into the cladding. This can be done by visualizing the output beam from the HC-MOF as demonstrated in Fig. 3.8.

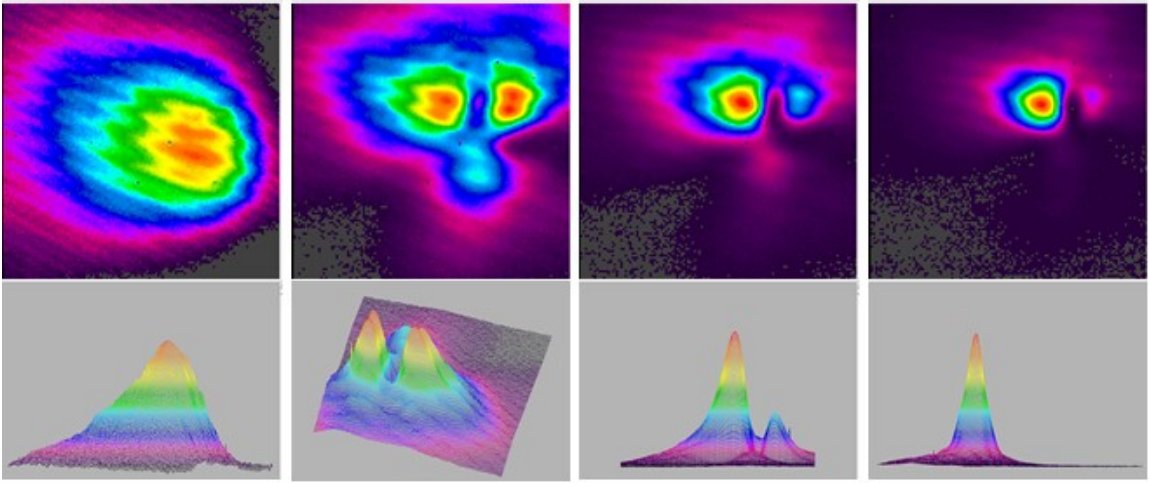

Figure 3.8: The progression, from left to right, of coupling into the cladding to the effective air core, by adjusting the coupling lens position on the optical bench. The image demonstrating coupling into the parasitic modes, far left, has two orders of magnitude longer exposure time in comparison to the image demonstrating coupling into the effective air core, far left. This was performed for a ~10 cm long piece of the HC-MOF with an effective air core diameter of ~50 $\mu$m.

The delineation for our measurement strategy is, first, adjust both parameters that are critical for coupling, $\text{FWHM}_{\text{coupling}}$ and $\text{NA}_{\text{coupling}}$ , so that coupling of the laser radiation into the HC-MOF is maximized. Second, perform the measurement of the total transmitted power and total efficiency $\eta_{total}$. Third, without perturbing the coupling stage, or the flange of the HC-MOF, assuring the coupling conditions are preserved, the HC-MOF is cleaved so that 10 cm of the fiber remains. This allows for an efficiency measurement to be performed on a shorter piece of fiber, $\eta_{0.1m}$, where the coupling conditions have not been altered. The power measurements through the long and short piece of HC-MOF were used to independently measure coupling and propagation losses ($\alpha_p$). For example, the propagation losses ($\alpha_p$) can be estimated using the idea that the transmission through the length ($l_l$-$l_s$) leads to the attenuation of the power from the value determined by $\eta_{0.1m}$ to the value determined by $\eta_{total}$. Finally, using similar ideas based on measurements on short ($l_s$) and long ($l_l$) pieces of HC-MOF the coupling loss ($\alpha_c$) can be independently determined. The total loss of the HC-MOF can be mathematically expressed by the following equation:

$$\text{Loss}_{dB/m} = [10\text{Log}_{10}(P_s/P_l)]/(\,l_l\text{-}l_s)$$

(3.1)

However, this is not the propagation loss ($\alpha_p$) of the HC-MOF, we also need to consider other losses; such as attenuation of the beam in air. Since the Mesophotonics Laboratory is located on the east coast of the United States, the air in this region can contain a significant amount of moisture. The moisture is the main cause of attenuation of laser beam with $\lambda$=3 $\mu$m because this wavelength aligns with the main water-absorption spectral peak. Thus, the moisture is responsible for the attenuation of the beam propagating in free space. In order to accurately determine the free-space loss due to atmospheric absorption the power was recorded for several different positions along the optical bench, then the attenuation due to air ($\alpha_{air}$) was measured. These results are shown in Fig. 3.9.

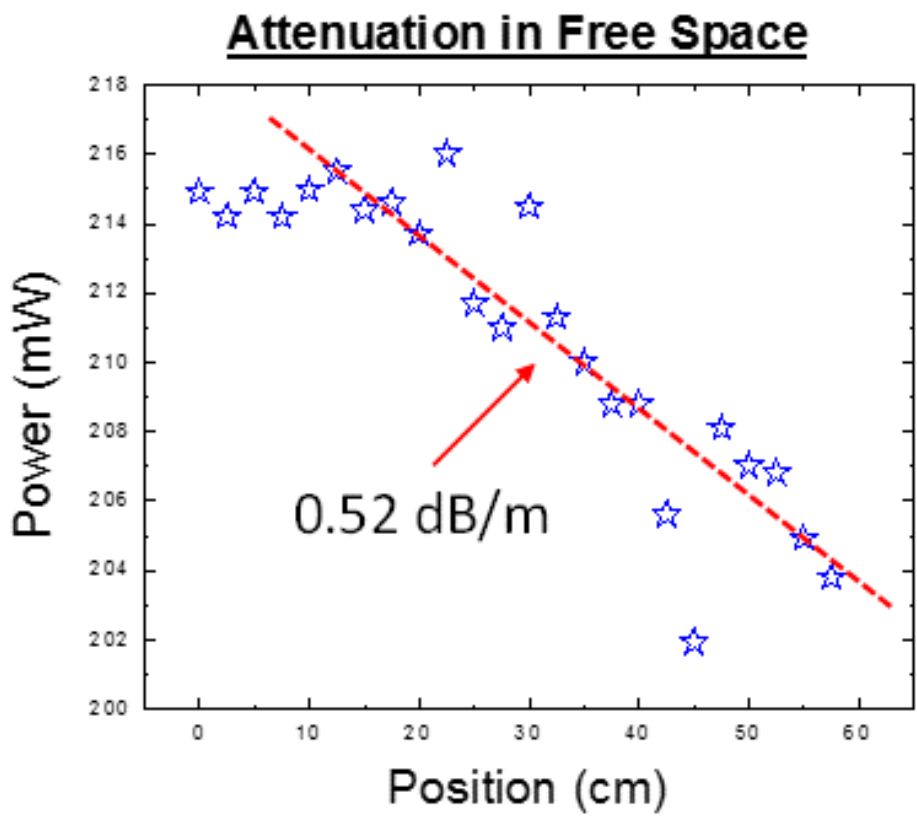

Figure 3.9: Power measurements as a function of position along the optical bench. The attenuation of the beam produced from the Er:YAG laser in free space, is calculated by the slope of these points.

For the straight HC-MOFs the total loss is $\alpha_{\text{total}}=\alpha_{\text{p}}+\alpha_{\text{air}}$ which allows for the estimation of propagation losses, which are presented in Fig. 3.10. We measured the total attenuation of the laser radiation propagating through the HC-MOF by using the cut-back method. The inherent attenuation of the beam due to the moisture in air ($\alpha_{air}$) was measured. Therefore, we can determine the propagation loss of the HC-MOFs by subtracting the attenuation value for the laser radiation propagating in free space from the total attenuation that was measured. The general trend of the plot indicates that the attenuation levels are inversely proportional to the effective air core diameter of the HC-MOF.

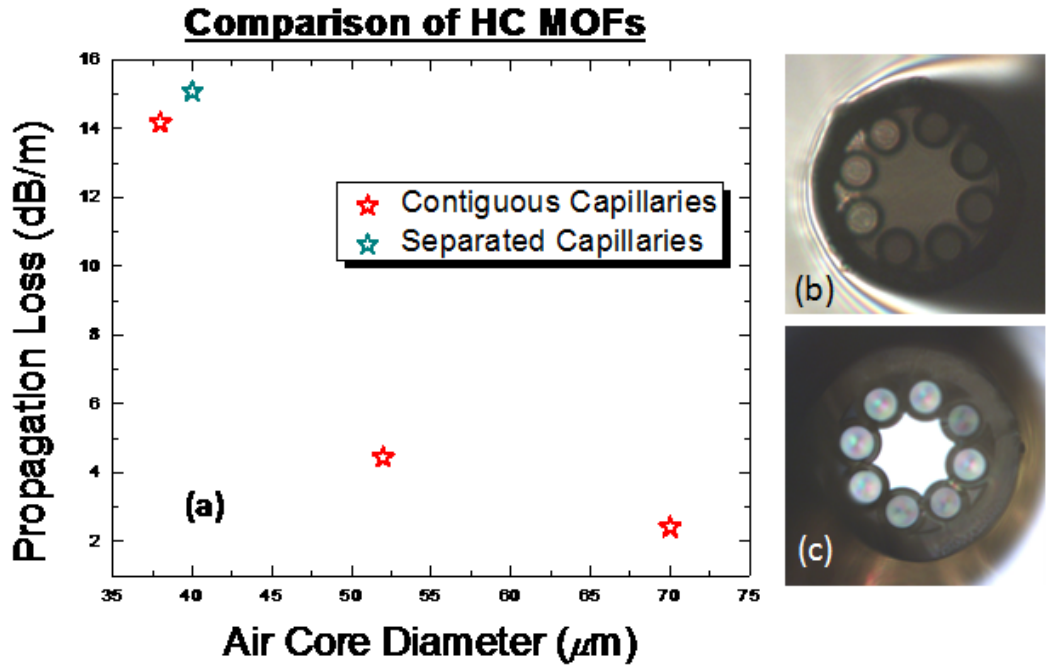

Figure 3.10: (a) Propagation loss as a function of the effective air core of the HC-MOF, the two different geometries are indicated in the legend. (b) Optical image of geometry with separated capillaries. (c) Optical image of geometry with contiguous capillaries.

Bend loss ($\alpha_b$) studies were conducted, after the propagation losses were measured. Bending loss was measured by bending a portion of the HC-MOF into a circle shown in Fig. 3.11. One end was illuminated by the Er:YAG laser and the output beam power was being measured on the other end, via power meter. The power loss (dB/m) due to the bending of the HC-MOF was studied as a function of bend radius. The total loss of the HC-MOF in the presence of a loop can be estimated as

$$10\log(P_s / P_l) = (\alpha_p + \alpha_{air})\Delta l + \pi \alpha_b D \qquad (3.2)$$

The bend loss measurements were only performed for the HC-MOF with effective air core diameters of ~50 and 70 $\mu$m. These two fibers provided sufficiently low attenuation values for the laser scalpel application, they were chosen to investigate for these bend loss studies. Both fibers proved to be impervious to bending, until the radius of curvature reached values on the order of ~5 cm, as shown in Fig. 3.11.

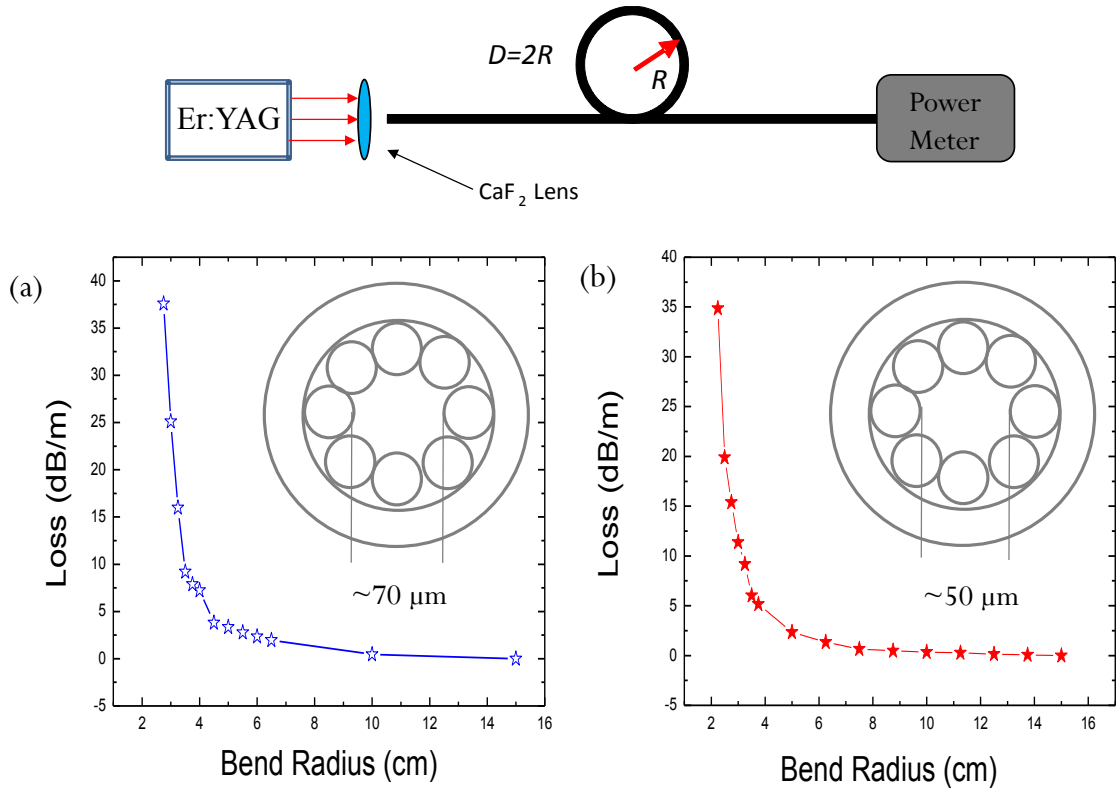

Figure 3.11: Illustration of the experimental setup for the bend loss studies. Loss as a function of bend radius for HC-MOF with effective air core diameters of (a) ~70 $\mu$m and (b) ~50 $\mu$m.

We determined that the bend losses dramatically increase with the reduction of the loop diameter from $\alpha_b = 2$ dB/m for $D$=10 cm to $\alpha_b = 35$ dB/m for $D$=4 cm. Bend losses of these HC-MOF are superior to conventional to hollow waveguides; however, the losses are higher than tradition solid core technology. We hypothesize that in practice the dominant factor of losses is expected to be the propagation loss. Total transmission of 30% should be practical in surgical applications, considering all of the losses.

### 3.4.2: Laser Scalpel Characterization

To study the focusing properties of integrated surgical systems we fixed a high-index ($n$=1.8 at $\lambda$=2.94 $\mu$m) barium titanate glass microsphere directly into the air core of HC-MOF, as illustrated in Fig. 3.12. The sphere was slightly protruding from the edge of the fiber due to the tight match between the diameter of the sphere and size of the air core, both being equal to 53 $\mu$m. The imaging of the focused beam was performed by a $CaF_2$ lens showing a peak with FWHM~4$\lambda$, as illustrated in Fig. 3.12(b). This technique is similar to standard methods of measuring the beam waist [200], however more detailed analyses of the results showed that the quality of the imaging of the focal beam suffered from an incomplete collection of light by the lens with finite NA=0.5. For this reason, the optical images obtained by such a lens can only be used for semi-quantitative characterization of the intensity profiles. In fact, such measurement can be considered as an upper limit of the beam diameter produced by the optical scalpel.

Previously in this chapter of the dissertation FDTD simulations were performed showing that a range of indices, $n$=1.6-1.9, could be suitable for the sphere used as the focusing element. In order to get more quantitative simulations we performed numerical modeling by two-dimensional (2-D) finite element method (FEM). This showed that more compact beams with diffraction-limited dimensions on the order of $\lambda/2$ can be obtained in such systems, as illustrated in Figs. 3.12(c,d) for 50 $\mu$m dielectric cylinder with index $n$=1.87. Simulations were performed using COMSOL, launch plane waves so that the electric field is polarized in the *x-y* plane, propagating along the *x*-axis. The wavelength of the illuminating light was chosen to match the Er:YAG laser radiation ($\lambda$=2.94 $\mu$m) used in our experiments. It should be noted that the spatial properties of a single-mode guided by HC-MOFs are different from the plane waves used in our numerical modeling. Thus, calculations presented in Fig. 3.12(c) show only a principal possibility to realize photonic jets with diffraction-limited dimensions. More rigorous calculations are required to model HC-MOF integrated with microspheres for more accurate solutions.

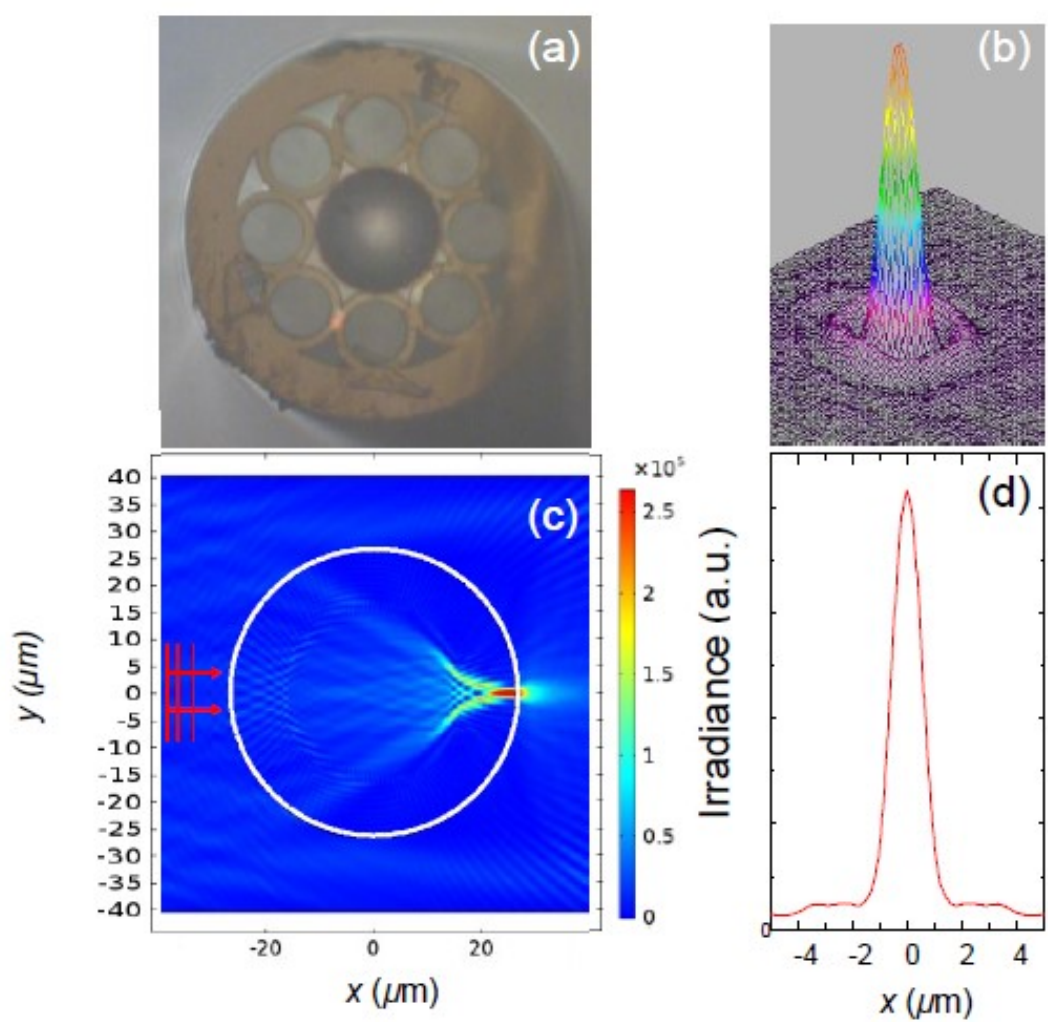

Figure 3.12: (a) HC-MOF integrated with 53 $\mu$m barium titanate-glass microsphere; (b) Imaging of the focal spot with FWHM~4 $\lambda$; (c) Calculated irradiance map of a 50 $\mu$m dielectric cylinder, $n$=1.87, illuminated by a plane wave, electric field propagating along the x-axis and oscillating in the x-y plane; (d) The transverse beam waist produced by the dielectric cylinder demonstrating diffraction-limited FWHM~ $\lambda/2$.

The transmission losses introduced by the BaTiO$_3$ microsphere consist of Fresnel reflection losses, material absorption at $\lambda$=2.94 $\mu$m, and scattering losses at the edge of HC-MOF. Experimentally, we estimated the total transmission losses introduced by the BaTiO$_3$ microsphere to be on the order of ~2 dB. The characterization of the intensity profile in Fig. 3.12(b) shows that about 20% of the total power of single-mode laser source can be coupled into the focal spot with FWHM~4$\lambda$, thus allowing the probe to produce high peak irradiance values ~ $3\text{x}10^5$ W/cm$^2$ at the center of the photonic jet. In a pulsed mode with ~100 $\mu$s pulse duration required for compact surgery [191] this system could produce pulse energy densities ~ 30 J/cm$^2$ which exceed the thresholds for ablating ophthalmic tissues by more than an order of magnitude.

## 3.5: Conclusions

In this chapter, we designed and tested focusing microprobes based on using photonic jets produced by microspheres integrated with modern mid-IR HC-MOF delivery systems. The design is optimized for laser-tissue, intraocular surgery performed using a flexible single-mode delivery system in contact conditions with strongly absorbing tissue. However, tissue studies have not been performed. Our design demonstrates high transmission, in combination with tight focusing, providing sufficient power, along with the desired transverse precision for laser-tissue surgical procedures. The single-mode design demonstrated a clear focusing and efficiency advantage over the previous multimodal systems. The design concept is not limited to intraocular surgery but could potentially be beneficial in applications of ultra-precise laser procedures in the eye, brain, or photoporation of cells.

CHAPTER 4: SUPER-RESOLUTION IMAGING THROUGH ARRAYS OF HIGH-INDEX SPHERES EMBEDDED IN TRANSPARENT MATRICES

4.1: Introduction to Microsphere-Assisted Super-Resolution Microscopy

Since the invention of the light microscope, over four-hundred years ago, scientists and engineers have attempted to improve spatial resolution capabilities of imaging systems. In the nineteenth century, Abbe identified that the resolution is limited by far-field diffraction phenomena [162]. The diffraction limit was simultaneously derived with more rigorous mathematical treatment, as shown below, by von Helmholtz [166].

$$d = \frac{\lambda}{2n\sin\theta} \tag{4.1}$$

Where $\lambda$ is the wavelength of the illuminating source for imaging, $n$ is the refractive index of the object space and $\theta$ is the half-angle over which the objective can accept light from the object. Several imaging methods are predicated off of optimizing these parameters [146-161]; however, these techniques are not adequate for many applications. One problem arises with shorter wavelengths of light; ultraviolet light for example, is strongly attenuated and can potentially damage biological samples. Also, immersion oils can increase absorption and increase chromatic dispersion which degrades the quality of the image formation.

Achieving resolution which exceeds the far-field diffraction limit would have a profound impact on numerous fields, ranging from cellular level imaging to semiconductor inspection. One mechanism to resolve fine structural features which are below the diffraction limit involves the imaging system collecting information from the objects near-field. The attempts to provide near-field assisted imaging has driven the development of various near-field probes, field-concentrators, superlenses, and surface plasmon-polariton enhanced devices [168-177, 179-182]. Another mechanism which has enabled super-resolution imaging, is rooted in constructive interference provided by an optical mask creating subwavelength focusing, and termed superoscillatory lens. Unfortunately, these techniques have not flourished due to their inherent flaws; such as, design complexity, narrow spectral bandwidth, low collection efficiency, and large commercial price tags.

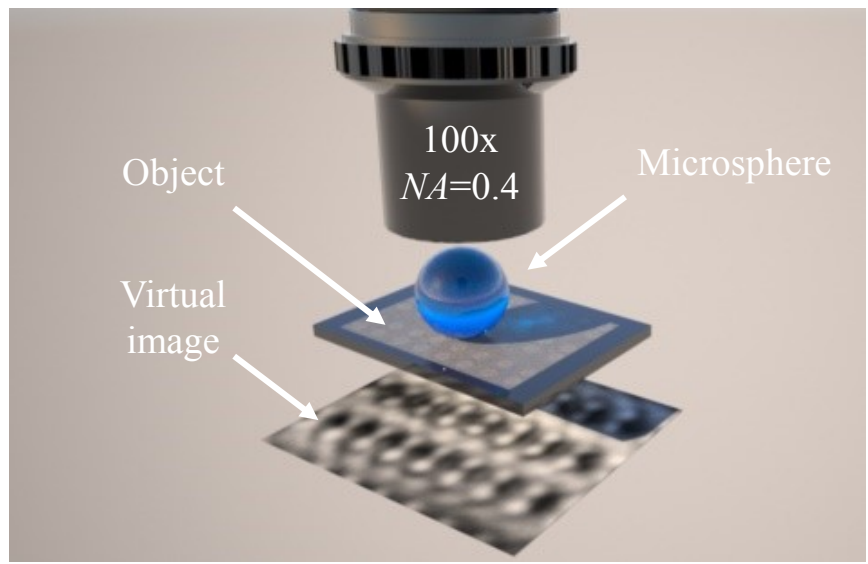

Figure 4.1: Illustration of the microsphere-assisted optical super-resolution microscopy technique.

Recently, there has been progress with the development of the microsphere-assisted optical super-resolution microscopy [51-70], illustrated in Fig. 4.1, which is promising for practical applications. Originally, in 2011, Wang *et al.* suggested the paradigm shifting idea that 2-9 $\mu$m SiO$_2$ fused glass microspheres ($n$=1.46), coupled with a standard microscope, could provide a white-light nanoscope [51]. The power of this imaging technique is that it is extremely simple and effective. After depositing the wavelength-scale spheres onto the sample, you can simply look "through the microsphere" to the virtual image plane. The resolution provided from this technique allows for sub-diffraction limited imaging. It has been determined that the super-resolution strength would be optimized at $n$~1.8 [51]. However, it has also been determined that the use of high-index spheres, $n$>1.8, would restrict this technique from producing images.

Semi-immersion of low-index microspheres was then shown to reinforce the super-resolution capabilities of this microsphere-assisted imaging technique [52]. The semi-immersion of the microsphere into a microdroplet provided more significant contrast and comparatively smaller lateral magnification factor. However, using this semi-immersion technique can be problematic due to the evaporation of the liquid. As the liquid evaporates the image contrast degrades and the magnification factor is increased. Although semi-immersion proved some promising optical enhancement of this microsphere-assisted imaging technique, it was argued [52] that this super-resolution effect would cease to exist if the liquid layer is increased to full-immersion of the microsphere ($n$=1.46)

Eventually, the Mesophotonics Laboratory demonstrated in 2012, that high-index BaTiO$_3$ fused glass microspheres ($n$~1.9-2.1) in full liquid-immersion of isopropyl alcohol (IPA, $n$=1.37) could provide super-resolution imaging [53]. Actually, high-index microsphere-assisted imaging provided superior resolution, magnification and overall image quality. Resolution for wavelength-scale spheres was estimated to be ~$\lambda$/7, while larger spheres (50<$D$<200 $\mu$m) provided resolution of ~$\lambda$/4.

Demonstrating the capabilities of this microsphere-assisted optical super-resolution technique, Li *et al*. in September 2013, successfully imaged label-free viruses, which have typical sizes ranging from 100-150 nm [63]. They performed this imaging during total liquid-immersion with 100 $\mu$m BaTiO$_3$ fused glass spheres in water and successfully resolved an adenovirus with lateral dimensions of ~75 nm. These results show, that using this simple imaging technique with a standard optical white light microscope, nanoscale biological samples can be investigated without labels such as fluorescent dyes. This opens up the opportunity to study interactions of viruses, cells, bacteria, and drugs for a more complete understanding.

Then the influence of the liquid for full-immersion was studied, by Lee *et al*. in October 2013 [58]. They demonstrated that BaTiO$_3$ fused glass spheres in water solution with 40% sugar, is a more suitable environment for super-resolution imaging, in comparison to standard immersion oil, and water without sugar. These claims were supported by Mie theory calculations in the field of the Poynting vector.

In an attempt to address practical application, Krivitsky *et al*. in November 2013, show the locomotion of low-index microspheres provided by a glass micropipette [57]. Using vacuum suction the silica sphere ($D$=6.1 $\mu$m, $n$~1.47) is attached to the tip of the micropipette. Their method allows for positioning with an accuracy of ~20 nm. However, this method does not use the advantages which are provided by high-index spheres and limits the imaging to a single microsphere.

It was shown, by the Mesophotonics Laboratory in January 2014, that microsphere-assisted imaging is advantageous, in terms of resolution, in comparison to confocal microscopy and solid-immersion lenses [65]. The letter systematically demonstrates the superior resolution of microsphere-assisted optical super-resolution, discerning minimal feature sizes of $\sim\lambda/7$.

Most recently, Yan *et al*. February 2014, presented a new technique which combined low-index 2.5, 5, and 7.5 $\mu$m fused-SiO$_2$, and 5 $\mu$m polystyrene, microspheres with a scanning laser confocal microscope (SLCM) which even further improved the lateral resolution [66]. Coupling the microsphere-assisted optical super-resolution imaging technique with SLCM, at a wavelength of 408 nm, they claim $\sim\lambda/17$ resolution. However, this resolution claim is logically fallacious in comparison to the "textbook definition".

In this dissertation, we propose a practical novel microscopy component for an optical super-resolution imaging platform. With a design concept which is based on a two-dimensional array of high-index BaTiO$_3$ fused glass ($n\sim$1.9-2.2) microspheres which are embedded in the intrinsically flexible, mechanically robust and optically transparent ($n\sim$1.4) polydimethylsiloxane (PDMS). This provides the ability to have a range of sphere diameters, $1 <D <212$ $\mu$m as illustrated in Fig. 4.2, and potentially different materials, if desired. This is achieved by inserting a metallic probe, which is connected to a hydraulic micromanipulation controller. The ability to simultaneously capture super-resolution images through the two-dimensional array of spheres during widefield microscopy allows for precise nano-positioning, limited by the mechanical aspects of the micromanipulator. This gives rise to an imaging platform with extraordinary, unprecedented, nano-positioning and super-resolution imaging capabilities.

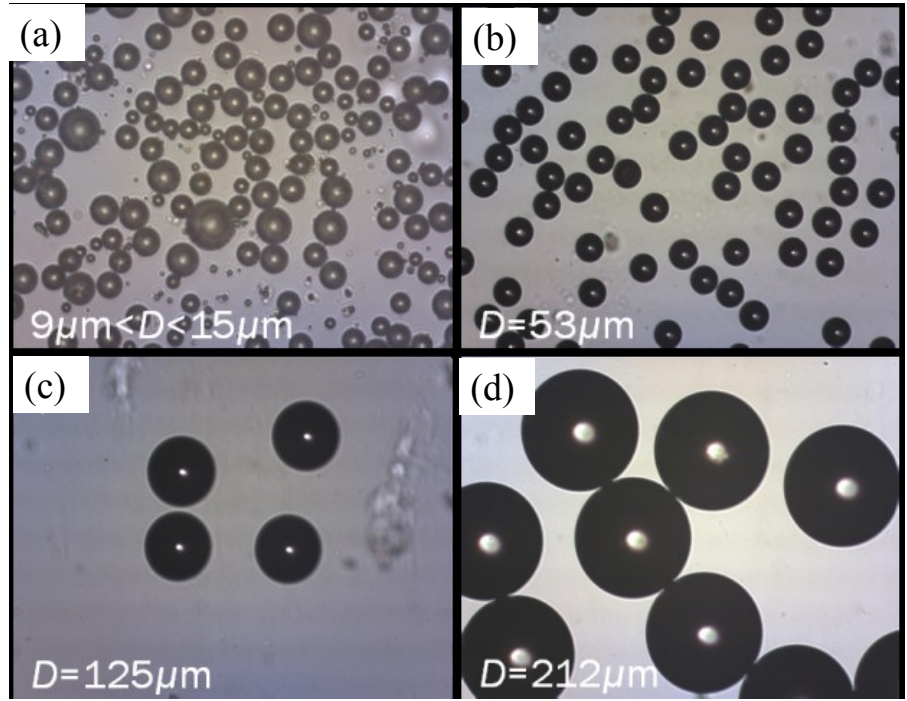

Figure 4.2: High-index barium titanate glass spheres embedded in transparent matrices. (a) Spheres ranging from 9 to 15 $\mu$m, (b) 53 $\mu$m, (c) 125 $\mu$m, and (d) 212 $\mu$m.

In this dissertation, we also address the proper treatment of resolution for objects of arbitrary geometries. We will demonstrate that some of the recent resolution claims [66] such as $\lambda/17$ have been exaggerated. These exaggerated claims are a result of incorrect treatment of images. According to the textbook definition [162-166], the resolution of the optical system is defined for point objects whereas in practice all objects have a finite size. We demonstrate a 2-dimensional convolution simulation based on the shape of the experimental object and Gaussian point spread functions. A fitting procedure is used to determine the point spread function of the imaging system, which can be applied to objects of various shapes. The point spread function can then be traced back to the "textbook definition" of resolution, using Houston's criteria [165]. This ensures that claims of resolution are justified and fit the conventions of the standard definitions.

## 4.2: Fabrication of Thin-Films

Previously it was shown, that high-index spheres ($n$~1.9-2.1) which were fully submerged in a liquid environment ($n$~1.37) proved to be advantageous over low-index spheres in air, with the same refractive index contrast [53]. However, there are some practical issues which makes this approach of full liquid-immersion undesirable. First, the specimen of interest should be chemically compatible with the liquid. Second, the specimen of interest might be influenced by the liquid environment so that it becomes dynamic which would make it difficult to capture an image. Third, the spheres themselves might become mobile in the liquid environment. Fourth, and lastly, after the desired microsphere-assisted optical super-resolution imaging is completed there is the chance that the spheres would contaminate the sample, bonding at the sample interface so that it cannot be removed. In a pragmatic approach, the high-index spheres are embedded in transparent matrices and provide the same refractive index contrast as the full-immersion case ($n$~1.4-1.5) which has been proven successful for microsphere-assisted optical super-resolution imaging.

An obvious candidate for the transparent matrices is polydimethylsiloxane (PDMS) which has several appealing properties [201, 202], namely:

*(i)*     Optically transparent ($n$~1.4).

*(ii)*    Visco-elascitiy: The PDMS is highly flexible; however, it is durable enough to be manipulated across macroscopic scales.

*(iii)*   High chemical inertness.

*(iv)*    Bio-compatibility.

*(v)*     Reversible adhesion properties, when applied to a rigid inert substrate.

PDMS is an organic based polymer that has found wide-spread applications, ranging but not limited to: MEMS, microfluidic device fabrication, soft lithography, low-index contact lenses, and the encapsulation of devices [201].

Prior to the curing process by thermal treatment, PDMS is a highly-viscous liquid which enables it to be casted and molded to a master sample. In order to provide structural rigidity, curing agents are mixed with the PDMS. Such curing agents act as branch points, providing the formation of internal cross-linking, inevitably transforming the PDMS into a flexible solid [202]. The flexibility of the PDMS in its solid form is a tunable parameter seeing as the modulus of elasticity is a function of the concentration of the curing agent, and other parameters such as temperature and length of time that the PDMS is thermally treated.

Our objective is to embed the high-index spheres ($n$~1.9-2.2) into a film of PDMS (Sylgard 184, Dow Corning), with a shallow surface relief provided by the back surface of the spheres. Achieving this insures that there will be an extremely small separation distance between the spheres and the objects, enabling the projection of the objects near-field into the far-field by the sphere. This is achieved through a 3-step process, as illustrated in Fig. 4.3. First, in Fig. 4.3(a), the high-index spheres are deposited onto a bare glass substrate. Second, in Fig. 4.3(b), the PDMS in its highly-viscous liquid form is cast over the spheres. And finally, in Fig. 4.3(c), the spheres undergo thermal treatment in an oven for ~15 minutes at a temperature of ~125 °C causing the highly viscous liquid to cure into a flexible solid, which has reversible adhesion properties. A Mesophotonics Laboratory member, Navid Farahi, was heavily involved in the development of the fabrication technology of these films with embedded microspheres.

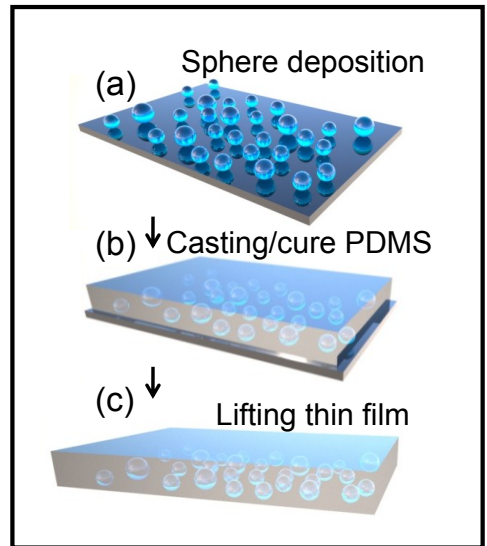

Figure 4.3: Thin-film fabrication process: (a) high-index sphere deposition onto a bare substrate, (b) casting of PDMS to be cured by thermal treatment, and (c) removing the thin-film to be used for microsphere-assisted optical super-resolution.

After these high-index spheres are embedded in the optically transparent PDMS film we characterized the back surface. The characterization was performed by an atomic force microscope (AFM) in tapping mode. By scanning the back surface of the films it was apparent that the spheres provided a shallow surface relief, as shown in Fig. 4.4, on the order of ~100 nm. The significance of the slight protrusion of the sphere is that we can be reasonably confident that the microsphere is in contact with the object of interest. We hypothesize that this contact between the sphere and the sample can play a critical role in the imaging process. The photonic nanojet produced by the sphere could potentially excite surface states with larger lateral k-vectors which then are projected into the far-field by the sphere, containing information from the higher spatial frequencies. This could be the underlying physical mechanism which leads to microsphere-assisted optical super-resolution imaging.

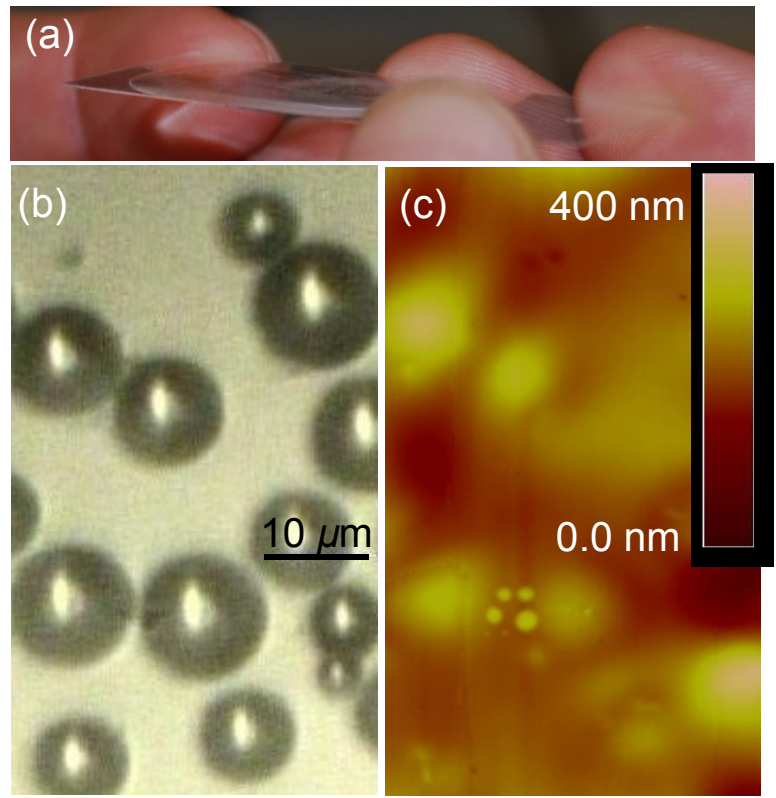

Figure 4.4: (a) Thin-film prior to being removed from the substrate. (b) Optical micrograph of the back surface of film showing the spatial distribution of spheres in the plane. (c) AFM image of the back surface showing the height variation, demonstrating a shallow surface relief due to the spheres.

## 4.3: Optical Studies

In this section, 4.3, of the dissertation, the optical studies will be elaborated on in detail. First, in section 4.3.1, I will describe the imaging systems which were implemented for the optical studies. Second, in section 4.3.2, the samples used for the experiments are described. Third, in section 4.3.3, the translation of these thin-films in a practical setting will be described. Fourth, in section 4.3.4, lateral magnification studies will be described. And finally, in section 4.3.5 the resolution capabilities of this microsphere-assisted optical super-resolution imaging technique will be shown.

### 4.3.1: Imaging Systems

Two imaging systems were implemented for the experimental optical studies. First, we used a FS70 Mitutoyo microscope which is coupled with a halogen lamp. The halogen lamp exhibits broadband spectral emission, with the system's spectral peak centered at $\lambda$=550 nm. The microscope was equipped with several Mitutoyo objectives: 100x(NA=0.90), 100x(NA=0.70), 100x(NA=0.50), 20x(NA=0.40), and 10x(NA=0.15). Images were projected onto a CCD camera (Edmund Optics, Barrington, NJ) in reflection mode. The images registered 1600x1200 pixels.

The second imaging system used was the LEXT OLS4000® (Olympus Imaging America, Inc.). This imaging system is a laser scanning confocal microscope, equipped with a $\lambda$=405 nm semiconductor laser and a white light LED. A photomultiplier is used for the LEXT OLS4000® detector in laser mode, and 1/1.8-inch 2-megapixel single-panel CCD camera is used during white-light mode. The system is capable of 108x~17,280x total magnification, 1-8x optical zoom and 1-8x electronic zoom. The *x-y* sample stage is 100x100 mm and motorized. LEXT OLS4000® is equipped with 5 LEXT objectives: 100x(NA=0.95, FOV=128-16 $\mu$m, WD=0.35 mm), 50x(NA=0.95, FOV=256-32 $\mu$m, WD=0.35 mm), 20x(NA=0.60, FOV=640-80 $\mu$m, WD=1.00 mm), 10x(NA=0.60, FOV=1280-160 $\mu$m, WD=11.0 mm), 5x(NA=0.15, FOV=2,560-320 $\mu$m, WD=20.0 mm). Seeing as the confocal microscopy technique suppresses the ambient background of the images in the optical micrograph, the LEXT OLS4000® in laser scanning confocal mode is appealing for optical studies of resolution.

### 4.3.2: Samples

In order to demonstrate the abilities of this novel optical microscopy component, we investigated samples of Au nanoplasmonic arrays of dimers and bowties, and Blu-ray disks. All of these samples have nanometric scale structural features. In order to discern the finer structural features of these samples it is necessary for the imaging system to have resolution capabilities superior to conventional optical microscopes ($\sim\lambda/2$). Also, in order to locate the nanostructures on the sample's surface the component should have extremely precise, micrometer level, maneuverability. These are precisely the reasons that such challenging samples were strategically selected to demonstrate the capabilities of this novel optical microscopy component.

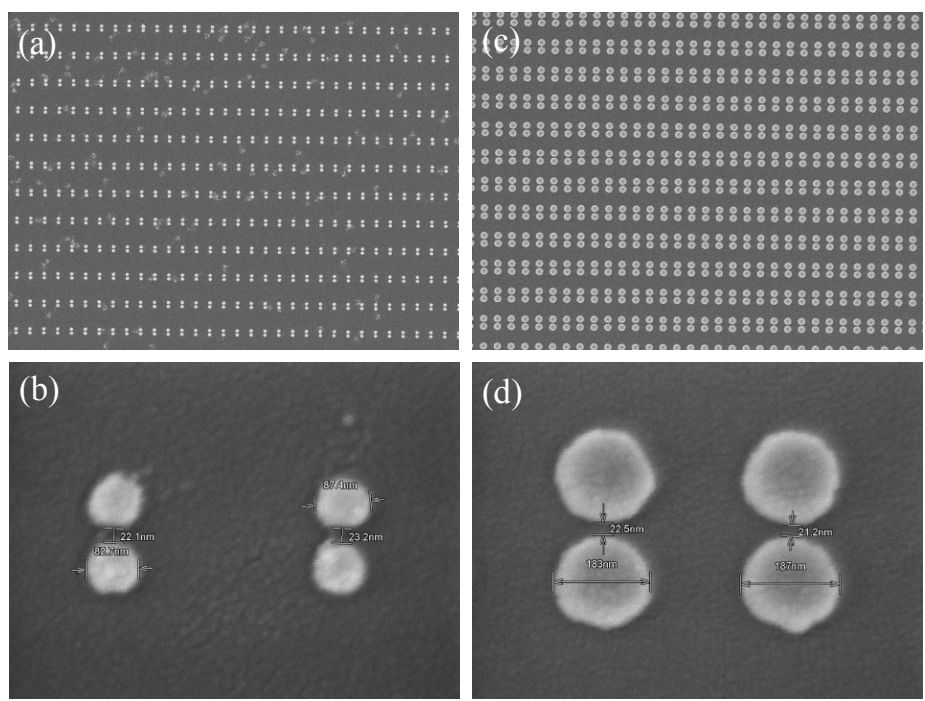

Figure 4.5: Scanning electron microscope images of (a)-(b) Au nanoplasmonic dimers with diameters of ~70 nm with a ~20 nm gap. (c)-(d) Au nanoplasmonic dimers with diameters of ~180 nm with a ~20 nm gap.

The array of Au nanoplasmonic dimers, shown in Fig. 3.5, were fabricated by our collaborators at the Air Force Research Laboratory Sensors Directorate, by Dr. Dennis Walker Jr., by an electron beam lithography process. This procedure can produce linewidths ~20 nm. The substrate for this process is sapphire, there is a Ti buffer layer (~10 nm in height), and the features are Au (~40 nm in height). Our design for the array had two variables: the diameter of the dimers and their center-to-center separation. The diameters ranged from ~50-180 nm and their center-to-center separations varied from ~100-300 nm. Vertically the rows of dimers are separated by 0.70 $\mu$m and the horizontal columns of dimers are separated by 0.35 $\mu$m. Two samples were fabricated with these parameters. One of the samples was left in this form, while the other sample was sputter coated by Au. The sputter coat of gold resulted in a ~20-30 nm conformal layer of Au on top of the sample. For the results presented in this chapter, we used the sample which was sputter coated with Au, seeing as we were able to characterize it with the SEM. The conformal conducting layer of Au allowed images to be acquired without charge build up, which degrades the SEM image.

The Au bowties were fabricated by our collaborator at MIT's Lincoln Laboratory, Dr. Vladimir Liberman. These Au bowties were fabricated on a semiconductor substrate. The length of the bowtie was ~250 nm and the width was ~100 nm, as shown in Fig. 3.6. Two triangles with a gap of ~15 nm formed the nanoplasmonic bowtie. These bowties were in arrays separated by 3 $\mu$m along the horizontal and also 3 $\mu$m along the vertical over an area of 6x9 $\mu$m$^2$.

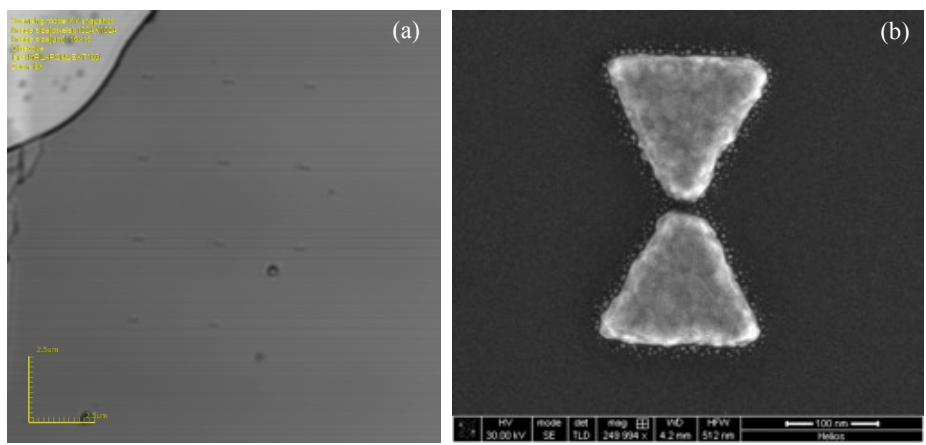

Figure 4.6: (a) Optical micrograph from the scanning laser confocal microscope showing the Au nanoplasmonic bowtie array. (b) Scanning electron microscope image of an individual Au nanoplasmonic bowtie.

Blu-ray® disks (Verbatim Americas, LLC, Charlotte, NC) are a commercially available product. This sample provides us with uniform linear objects, as illustrated in Fig. 4.7. The standard nominal track pitch is 300 nm; where the stripe width is ~ 170 nm and the stripes are separated by ~130 nm. Prior to performing the experiments a scalpel was used to remove the 100 $\mu$m thick clear polymer protective layer.

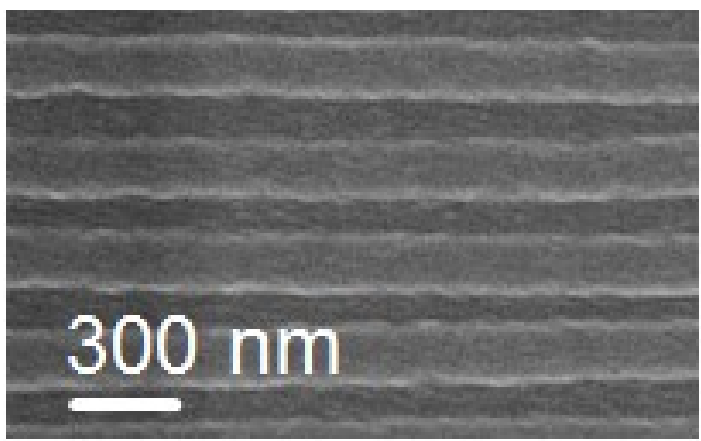

Figure 4.7: Scanning electron microscope image of a Blu-ray® disks, showing stripe widths of ~170 nm and separations of ~130 nm.

### 4.3.3: Translation of Thin-Films

The ability to translate the high-index spheres, embedded in the transparent matrices, is vital for this technologies' success as a novel optical microscopy component to be used in a practical setting. We propose that these transparent matrices containing microspheres enables the user to controllably maneuver the thin-film and locate desired nanostructures on their sample, as illustrated in Fig. 4.8.

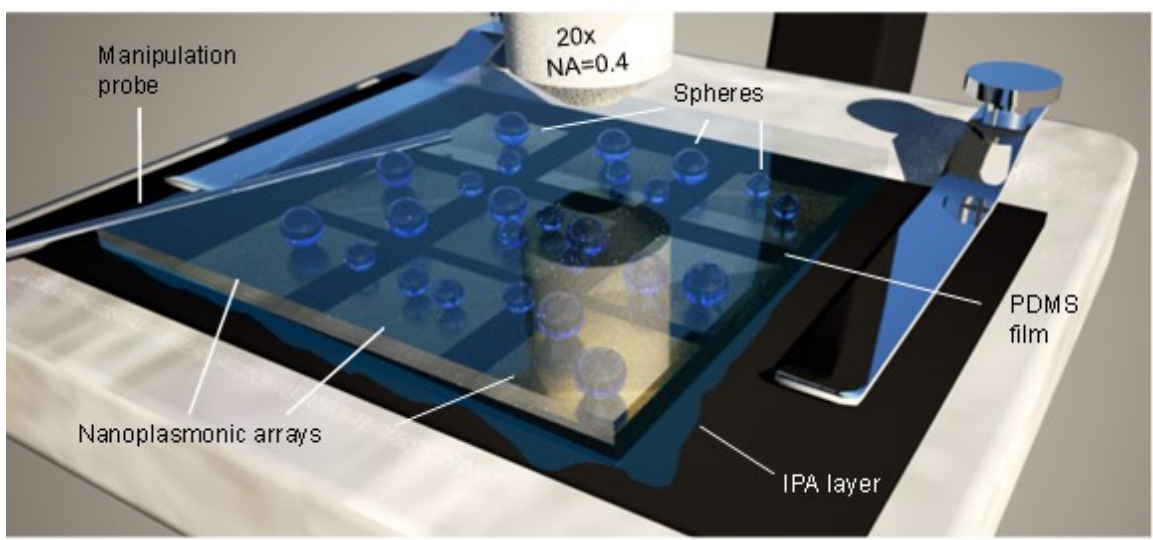

Figure 4.8: Illustration of the thin-film containing microspheres, on top of the sample, lubricated by an IPA layer, attached with an *x-y-z* manipulation probe, for practical microscopy applications requiring super-resolution.

The advantage of our thin-film, in comparison to previous studies of fully liquid-immersed spheres, is the ability to translate across the sample with phenomenal precision, as shown in Fig. 4.9. Since the array of spheres is embedded in a thin-film of an elastomeric substance, instead of a liquid, they have a fixed spatial relationship in the plane along the bottom surface of the thin-film. An index matching liquid layer of IPA can be introduced to provide lubrication, as shown in Fig. 4.8. With the proper lubrication provided between the thin-film and sample interface, translation of the thin-film is accomplished with ease, Fig. 4.9. Considering that the two-dimensional array of spheres have a fixed spatial relationship and can simultaneously provide super-resolution images, interesting features of sub-diffraction limited dimensions can be identified by using the hydraulic micromanipulation controller to translate the thin-film across the sample surface. With identified features the thin-film can then be guided by super-resolution imaging to provide micron level translation with nano-metric precision.

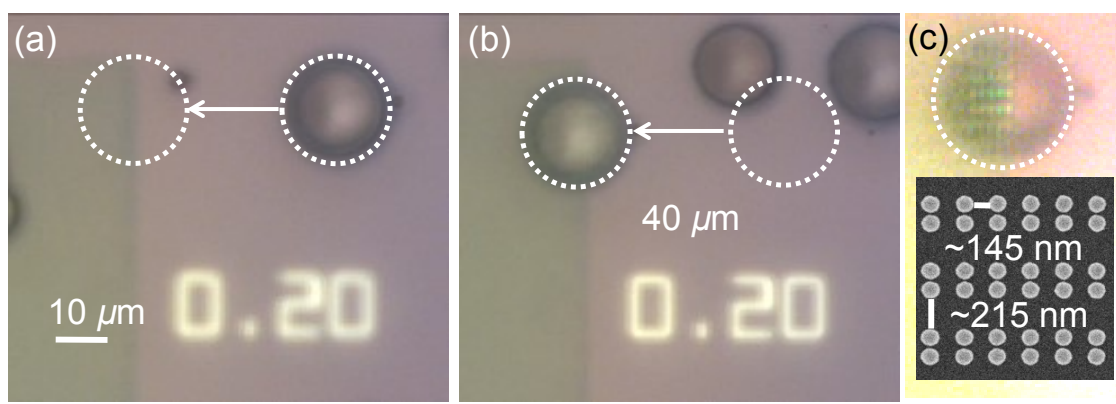

Figure 4.9: Optical micrographs demonstrating the translation of the thin-film across the samples surface. (a) Shows the original position of the sphere, which is 40 $\mu$m from the edge of a nanoplasmonic Au dimer array. (b) Shows that the sphere can be maneuvered with micron level precision to locate nanostructures of interest on the samples surface. (c) Optical micrograph of the nanostructures using the microsphere-assisted super-resolution imaging technique.

This translation technique was used for the thin-film to inspect the Au nanoplasmonic dimer array, Fig. 4.9, using the FS70 Mitutoyo optical microscope in reflection mode through the 20x(NA=0.40) objective lens. After the interface between the thin-film and sample was properly lubricated with IPA the thin-film was applied to the sample's surface. However, the original position of the thin-film did not correspond to any spheres being located on top of the Au dimer array of interest, as illustrated in the optical micrograph in Fig. 4.9(a). The thin-film was then translated, under the control of the hydraulic micromanipulator, to the edge of the Au nanoplasmonic dimer array, as illustrated in Fig. 4.9(b). After the thin-film was maneuvered to the desired location on the sample's surface the microsphere-assisted optical super-resolution imaging technique was used to visualize the nanostructures, shown in Fig. 4.9(c). It can be seen that structural features on the order of ~100 nm can be resolved, although the diffraction limit of the FS70 Mitutoyo imaging system with the 20x(NA=0.40) objective lens would be ~687.5 nm according to the Fraunhofer, far-field, diffraction formula.

In this section, it was clearly demonstrated that these thin-films of high-index $BaTiO_3$ microspheres embedded in transparent matrices can be translated across the sample under the proper conditions. Furthermore, it was demonstrated that fine structural details (~100 nm) of the nanostructures located on the sample's surface can be resolved, going beyond the resolution predicted by the classical far-field diffraction formula.

### 4.3.4: Magnification Studies

In this section of the dissertation the observations on lateral magnification will be discussed. As discussed in the previous section, 4.3.3, an IPA layer was deposited on the sample to provide lubrication at the sample-film interface. This IPA layer introduces a gap between the back surface of the microsphere which is protruding from the film and the object of interest, schematically shown in Fig. 4.10.

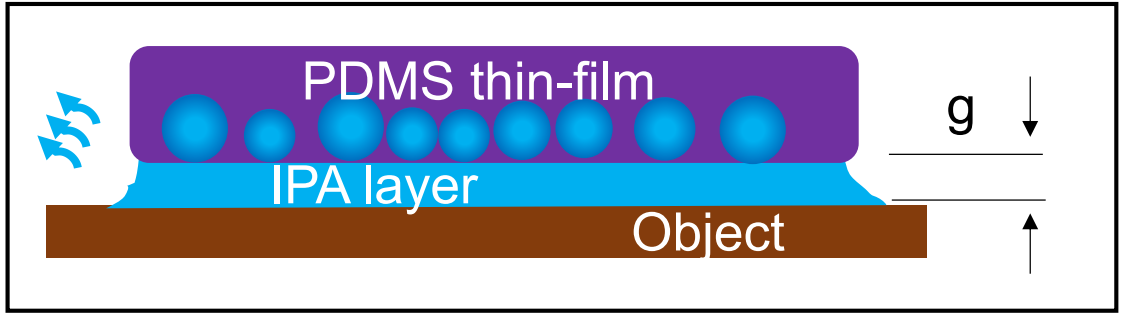

Figure 4.10: Schematic illustration of the lubricating IPA layer introducing a gap, *g*, between the back surface of the spheres and the object.

As a geometrical consequence this gap (*g*), between the back surface of the microsphere and the object, increases the lateral magnification (*M*) of the virtual image. Using the framework of geometrical optics an analytical expression can be deduced as a model for the magnification's dependence on the gap. This analytical solution was independently developed by Mesophotonics Laboratory member Navid Farahi and myself. This magnification is expressed mathematically in terms of the radius of the sphere, the gap separating the back surface from the object, and the refractive index contrast in the expression below:

$$M(n', r, g) = \frac{n'}{2(n'-1)\left(\frac{g}{r}+1\right)-n'} \tag{4.2}$$

Where *n'* is the refractive index contrast, *r* is the radius of the sphere, and *g* is the gap between the back surface of the sphere and the object. It can be seen that as the liquid layer evaporates the thickness is diminishing and the gap is reduced, and for a sphere with a fixed radius and refractive index contrast, the resulting virtual images magnification decreases, as shown in Fig. 4.11.

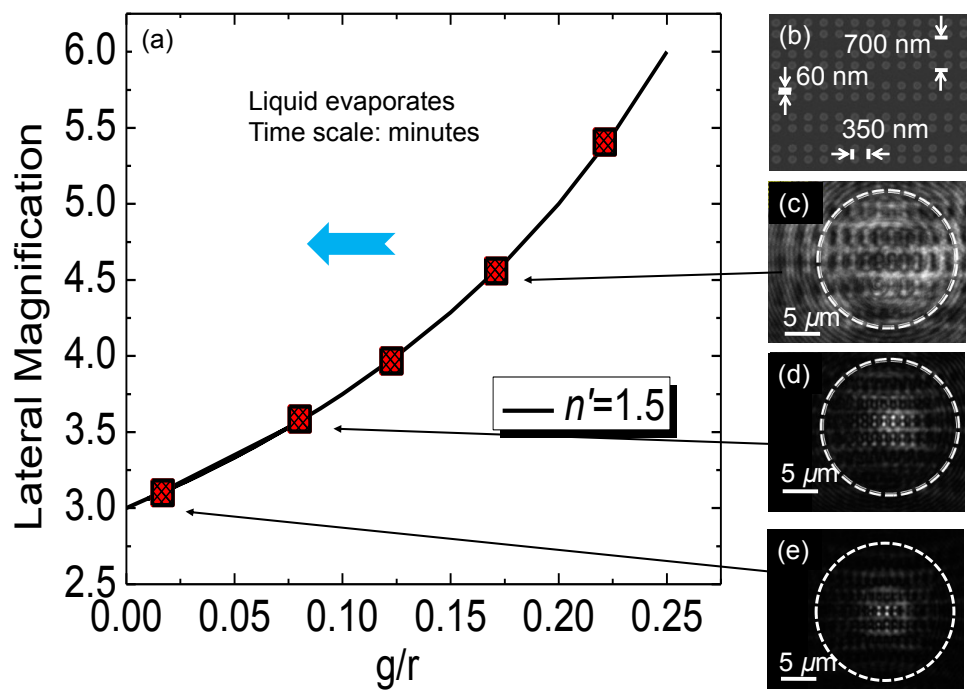

Figure 4.11: (a) Measured magnification values on the analytical curve predicted by geometrical optics, as a function of the gap to radius ratio. (b) SEM image of the structure studied for the magnification studies. (c)-(e) Dynamic evolution of the optical images, showing a reduction in magnification.

Empirically we can use this analytical model to deduce the gap between the sphere and the object. To this end, we measured a series of images of the same structure in the course of evaporation of the IPA layer, see Fig. 4.11(c-e). The main point is that we observed a slow reduction of magnification on the scale of minutes and eventually, hours. We did not present the time in this figure. However, we were able to estimate the gaps by presenting these dynamical magnification measurements along with the magnification calculated as a function of parameter g/r, as shown in Fig. 4.11(a). Originally, for the first image presented in Fig. 4.11(c) the gap is ~1.8 $\mu$m. After 10 minutes, Fig. 4.11(d), during this dynamic evaporation process, the gap was reduced to ~1 $\mu$m. Then after several days it appears that the IPA layer has almost completely evaporated and the gap has approached close to zero (~100 nm). It should be noted, however, that in this method only large gaps (typically >1 μm) can be determined with a reasonable accuracy. For smaller gaps on the order of 100 nm the accuracy is greatly diminished.

### 4.3.5: Resolution Capabilities of Thin-Films

Following the magnification studies presented in the previous section, we observed that in this region of such nano-metric scale gaps, there was an enhancement of the spatial resolution capabilities of the microsphere, as shown in Fig. 4.12. In order to properly discuss this observed effect, in a quantitative manner, a brief discussion is required in order to determine the lateral resolution of the imaging system. First, consider the nanoplasmonic array which was being imaged. It has 3 spatial dimensions of interest, the period between rows of dimers (700 nm), the spacing between individual dimers (350 nm), and the edge-to-edge separation for the dimer (60 nm), which are shown in Fig. 4.11(b). Naturally, it is reasonable to determine a criteria so that the resolution can be approached in a quantitative fashion. Seeing as there is a dip in the irradiance line profile, through the cross-section of the dimer, it is intuitive to claim that you have resolved the edge-to-edge separation. These types of observations have been claimed by many groups [51, 53, 65, 66] and the corresponding resolution has often been referred to as a "minimal discernable feature sizes". It should be noted, however, that the classical text book definition of resolution is based on delta-functions for objects. These definition do not allow for simplistic references to the visibility due to some features of the objects with finite size to determine the optical resolution of the system. We believe that this difficulty led to the claims of exaggerated resolution as high as $\lambda/17$ published most recently [66].

In order to address this problem we developed a rigorous procedure for defining the resolution of such objects. It is based on a well-known fact that the image is a result of convolution of the object with a point-spread function [167]. This procedure has already been used previously for reconstructing intensity profiles of the super-resolved images [65]. In our work, we developed this procedure for reconstructing the whole 2-D image by performing the convolution of a 2-D object with a 2-D point-spread function. This step was critical for increasing the accuracy of determination of resolution because this enabled us to compare the calculated images with the experimental images in their entirety, as 2-D images. Fine structural features of the images such as slopes of the intensity profiles along different cross-sections and some secondary intensity lobes become more distinguishable in the case of 2-D representation compared to just single line intensity profiles. In order to consider the finite size of our object we approach the problem from Fourier optics theory. The image formed ($i$) is a direct result of the convolution of the object ($o$) with the point spread function ($PSF$) of the imaging system. The size of the image produced is scaled by the magnification of the system. However, this linear magnification factor is not considered in our formulation. This can be mathematically expressed in the integral form as,

$$i(x,y) = \iint_{-\infty}^{\infty} o(x,y) PSF(X-x, Y-y) dx dy. \tag{4.3}$$

The objects which we are experimentally imaging are well characterized by scanning electron microscopy (SEM). In principle, we could use SEM images directly as the objects. We found, however, that these images contain many features which are not reproducible from dimer to dimer such as roughness of the edges of metallic circles or other shapes, some variations of the brightness determined by some charging effects in the cause of SEM characterization, et al. For this reason, as an object we simply selected an idealized (drawn) image with the main geometrical characteristics determined by SEM. As an example, in this approach the nanoplasmonic dimer is represented by two perfect circles with the diameters and separation determined by SEM. Then by using computational methods we can convolve the known object with a series of Gaussian PSFs with different widths termed FWHM. The result is a calculated convolved image. These series of images can be used to match the experimentally observed image. Since the experimental and calculated convoluted images were obtained from the same object, the PSF of the optical system can be determined as a result of such comparison providing the best match between the experimental and computed images.

After the PSF is obtained, if Houston's criteria is considered, which states that the separation between two points is greater than the FWHM of the PSF the object is resolved, we can claim the FWHM of the PSF to be the resolution of the image obtained through the microsphere. Using this methodology it can be seen that when the microsphere has a gap of ~1 $\mu$m, the resolution is ~$\lambda$/4. However, when the gap was reduced down to a nano-metric scale, in the near-field vicinity, there was an observed boost in resolution to ~$\lambda$/5, Fig. 4.12. This suggests that when the microsphere is in the near-field vicinity of the object, it converts some of the higher spatial frequencies of the evanescent waves into propagating waves which can be captured by the far-field imaging system.

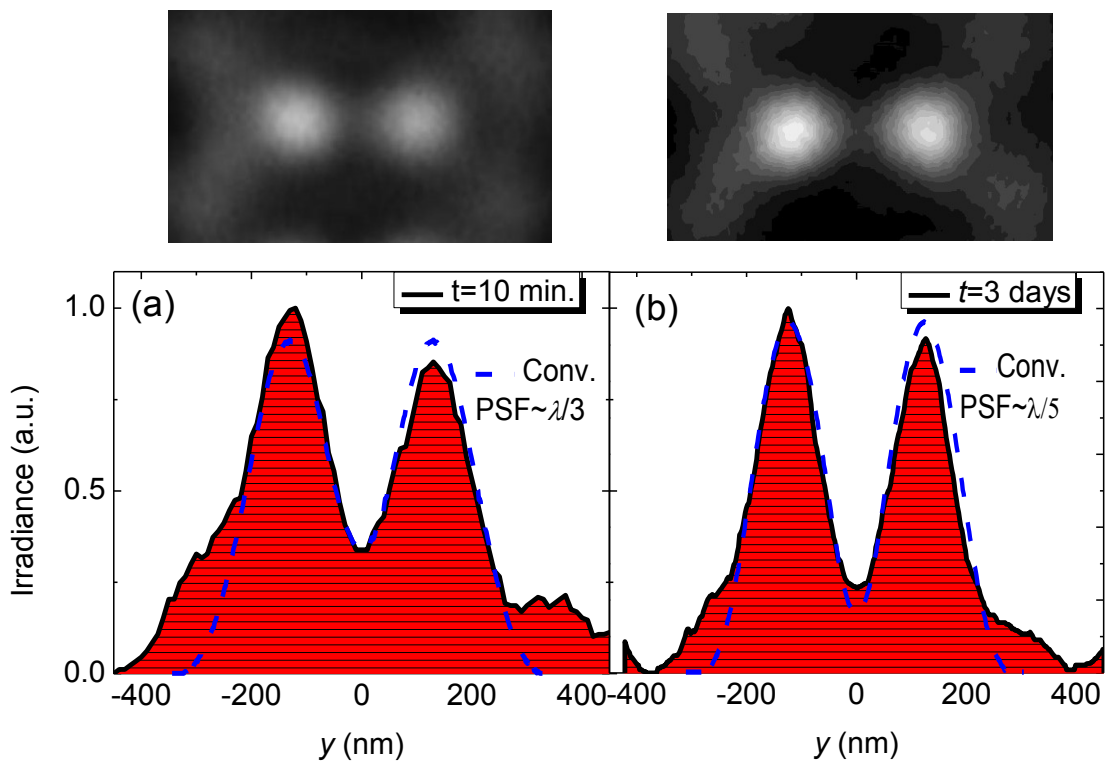

Figure 4.12: Line profiles through the cross-section of Au nanoplasmonic dimers (a) ten minutes and (b) three days, after lubricating the sample with IPA.

To investigate the ultimate resolution of such thin-films we placed them on top of the nanoplasmonic arrays of dimers and bow-ties without IPA, as shown in Fig. 4.13. Since there is not a liquid layer the PDMS adheres strongly to the bare and rigid sample, insuring a minimal gap. Using the convolution calculation the resolution was determined to be $\sim\lambda/7$ for the Au dimer, imaged through a $\sim 5$ $\mu$m $BaTiO_3$ sphere, Fig. 4.13(g). Also the resolution of the Au bow-tie, imaged through a 53 $\mu$m $BaTiO_3$ sphere, was calculated to be $\sim\lambda/5$, Fig. 4.13(h), which is consistent with previous results obtained by our group using larger microspheres.

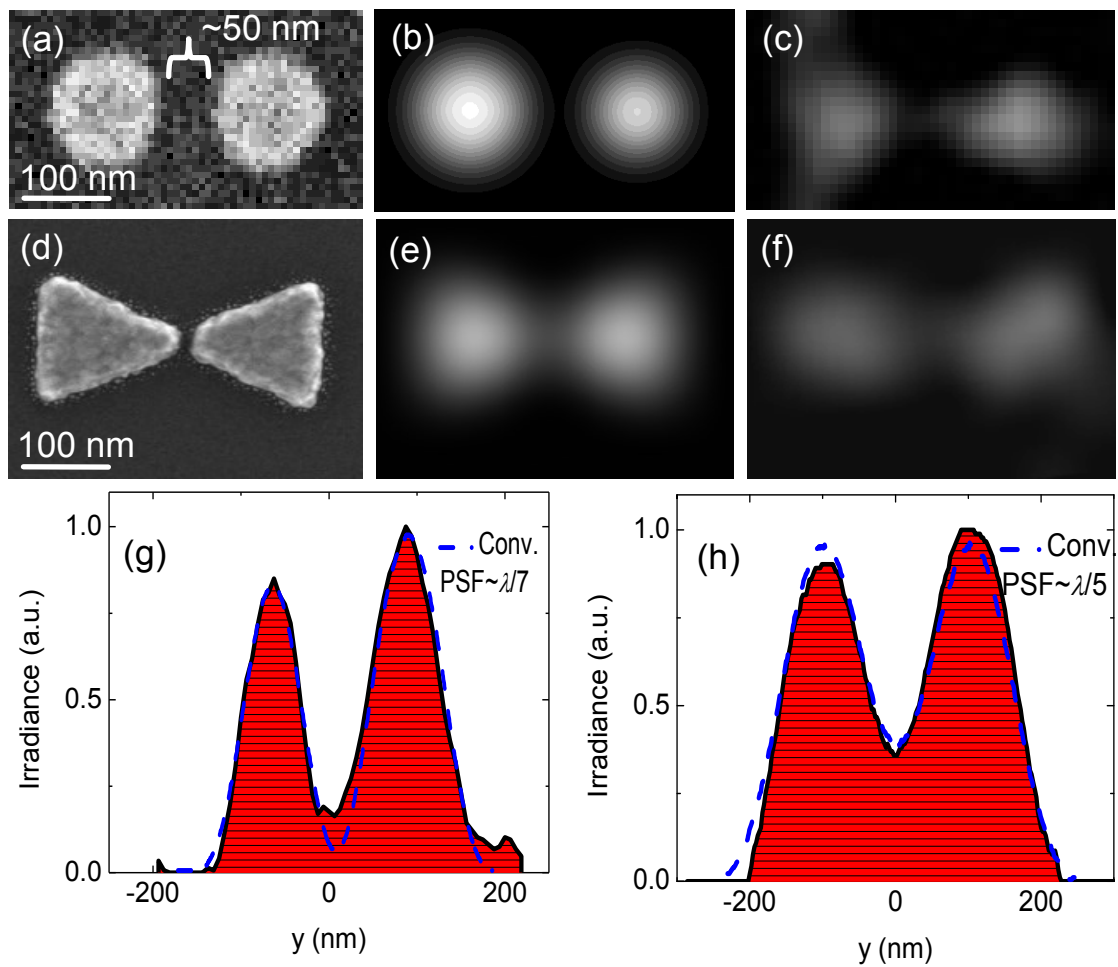

Figure 4.13: SEM, calculated, and experimental images of (a)-(c) Au dimers and (d)-(f) Au bow-ties. Line profiles through the cross-section of the experimental images compared with theoretical profiles for (g) Au dimers and (h) Au bow-ties.

## 4.4: Conclusions

In this chapter, we designed and tested arrays of high-index barium titanate glass microspheres embedded in transparent PDMS matrices. We show that such thin films containing spheres are a unique tool for microsphere-assisted optical super-resolution microscopy since different spheres can be easily aligned with the object of study by shifting the thin-film along the surface. Many questions about the mechanism of super-resolution imaging by microspheres still require further investigation. These include the role of the nanometric gap between the object and spheres and the role of surface polariton-plasmons in the metallic nanostructures. Our results, however, show that spheres embedded in elastomeric PDMS matrices can behave in these applications similar to liquid-immersed spheres studied previously. Due to resolution on the order of ~$\lambda$/7, excellent manufacturability of such thin films and their easy-of-use, such optical components can find broad applications in biology, medicine, microelectronics, and nanoplasmonics. For developing biomedical imaging applications, such matrices can be immersed in the liquid samples or cultures and micromanipulated inside such solutions.

CHAPTER 5: PHOTONIC JETS FOR STRAINED LAYER SUPERLATTICE
INFRARED PHOTODETECTOR ENHANCEMENT

## 5.1: Introduction to Infrared Photodetectors Enhancement

In recent years, enhancement techniques have been explored to increase the collection efficiency of photons to be converted into charge carriers by photodetectors. Plasmonic structures have been proposed as a solution for this problem, seeing as they have the capability to increase the optical near-field intensity by magnitude of ~$10^2$-$10^3$ with respect to the incident field [188]. This has stimulated studies of two-dimensional subwavelength hole arrays [189], photonic crystal cavities [207], nanoattennas such as dipoles and bowties [188, 208-210]. However, seeing as the field enhancement is a resonant phenomenon these techniques only operate over a narrow spectral bandwidth [188, 189, 207-210]. Photonic jets, produced by dielectric microspheres, have been suggested as an alternative to these resonant structures, in order to achieve broadband enhancements without polarization sensitivity [185]. However, experimentally this approach to the best of our knowledge has not been realized in previous studies [72].

## 5.2: Device Structure

The infrared photodetector structure used for the work presented in this chapter of the dissertation is a MWIR *n*B*n* structure grown on a GaSb substrate with InAs/InAsSb type-II strained-layer superlattice (SLS) as the active materials. SLS technology is a rapidly emerging interband technology and this demonstration of coupling photonic jets produced by dielectric microspheres with SLS detectors for enhancing photosignal further expands its range of capabilities. However, it should be mentioned that this same technique can be implemented with any other detector materials and structures such as InSb, HgCdTe, etc. As shown in Fig. 5.1, the $n$B$n$ detector structure consists of an $n$-type bottom contact layer, $n$-type SLS absorber, a wide bandgap barrier layer, and an $n$-type top contact layer. The total thickness of the device structure is ~3 $\mu$m.

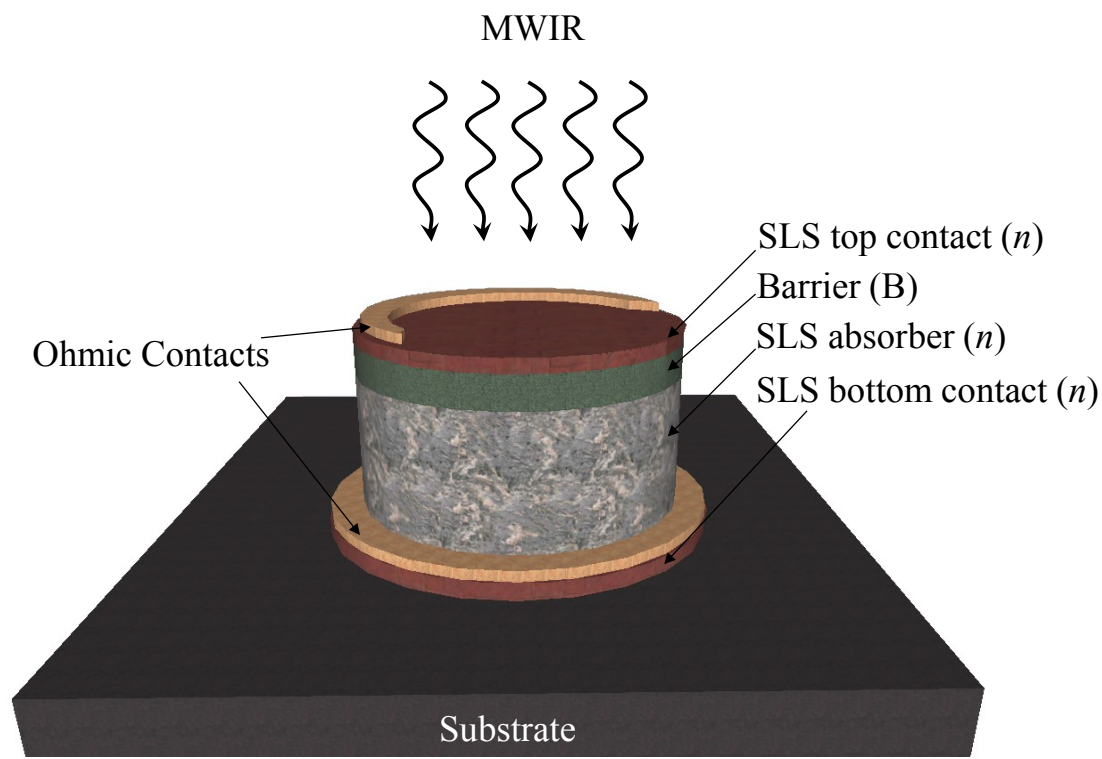

Figure 5.1: Illustration of the SLS device structure consisting of a SLS top contact layer (n), a wide bandgap barrier layer (B), SLS MWIR absorbing layer (n), SLS bottom contact layer (n), on top of a GaSb substrate. Ohmic contacts are on the top and bottom of the device structure.

The MWIR SLS absorbing (n) layer is 1.5 $\mu$m thick, the real part of the refractive index is 3.934, and the absorption coefficient can be found in Fig. 5.2 as a function of the illuminating wavelength. The barrier (B) layer is 0.3 $\mu$m thick, with a refractive index of 3.934, and transparent to the MWIR radiation. The bottom contact is 1.5 $\mu$m thick, with a real refractive index component of 3.913, and is transparent to the MWIR radiation.

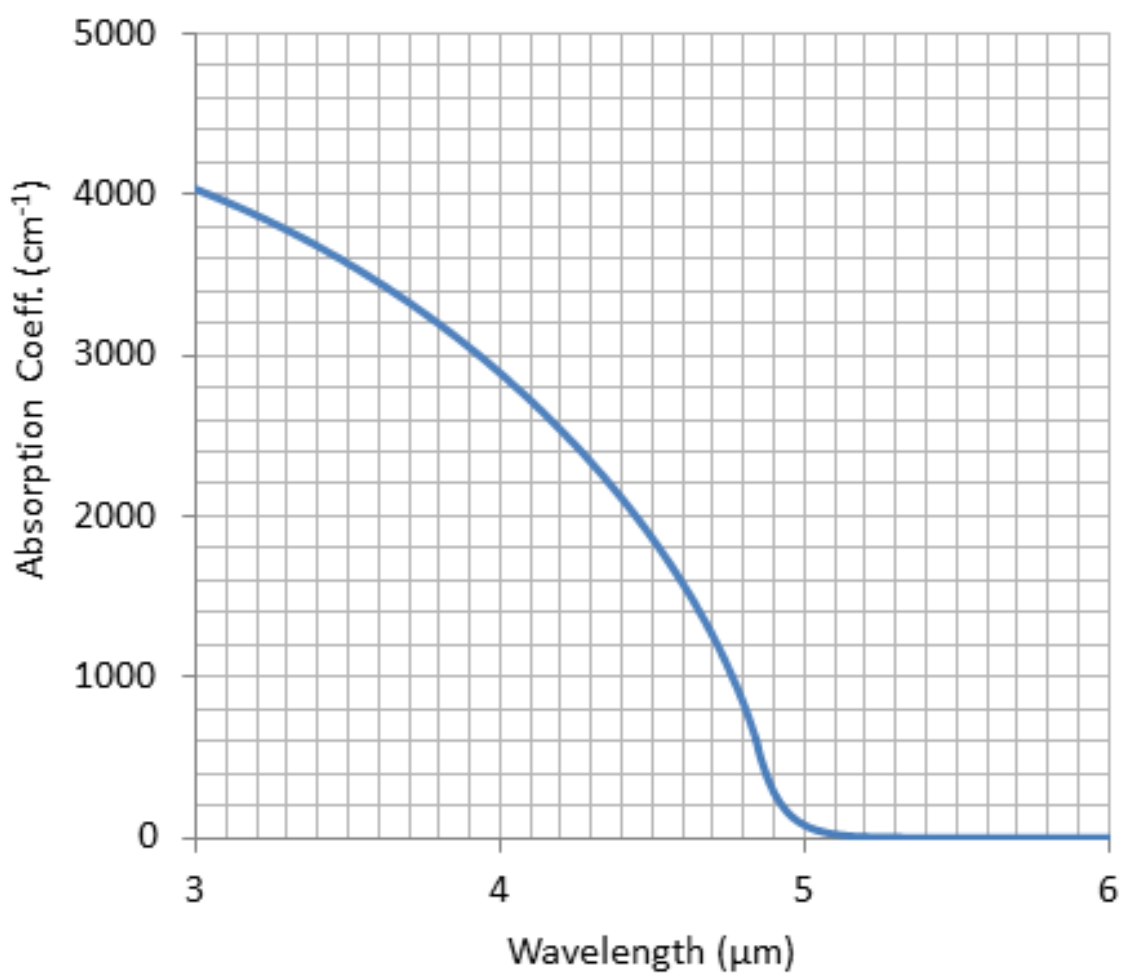

Figure 5.2: Absorption coefficient curve of the SLS MWIR absorbing (*n*) layer as a function of wavelength, courtesy of Joshua M. Duran.

5.3: Dielectric Sphere Integration

Experimental visualization was performed with the FS70 Mitutoyo microscope coupled with a halogen lamp. The halogen lamp exhibits broadband spectral emission, with the spectral peak of the combined light source/camera system centered on $\lambda$=0.55 $\mu$m. Optical images were produced through a 10x(NA=0.15) objective lens and captured with a CCD camera (Edmund Optics, Barrington, NJ). Positioning was provided by a hydraulic micromanipulation stage connected to a tapered metallic probe, with a ~5 $\mu$m tip. Visualization and manipulation was performed at $\lambda$=0.55 $\mu$m which is ~4-9x shorter wavelength than the detection wavelength, 2-5 $\mu$m, providing additional precision for the alignment process. This integration process is demonstrated below in Fig. 5.3.

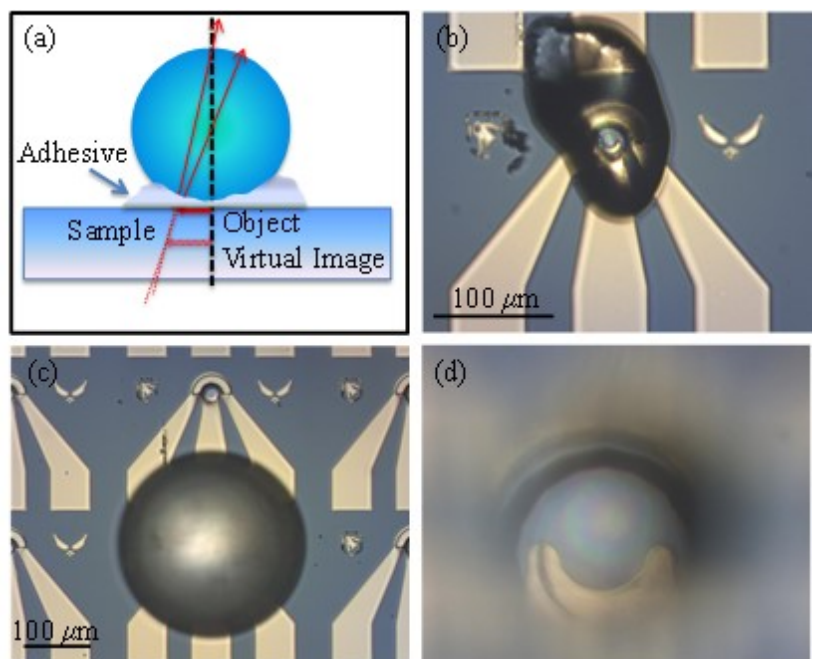

Figure 5.3: (a) Illustration of the microsphere-assisted optical super-resolution microscopy technique during semi-immersion. Optical micrographs of (b) the adhesive layer used to integrate the microsphere to the mesa, (c) microsphere after being properly positioned and (d) the image obtained through the sphere which is used to guide the positioning.

Alignment of the microsphere with the geometric center of the mesa, $d$=35 and 40 $\mu$m, is critical to provide efficient coupling of the photonic jet into the photosensitive region. This feat is exceedingly more challenging as the microsphere $D$ is increased, from 90-300 $\mu$m, because of mechanical stability. To achieve micron level alignment we implement a microsphere-assisted optical super-resolution microscopy technique [51-70], shown in Fig. 5.3. Looking through the microsphere, semi-immersed in the silicone adhesive, a virtual image of the photodetector can be visualized with sufficient resolution, while simultaneously boosting the magnification, as illustrated in Fig. 5.3(d). Micromanipulation with the tapered metallic tip while performing this imaging technique allows for spheres of 90≤$D$≤300 $\mu$m, and 1.46≤$n$≤1.71, to be positioned in alignment with the geometric center of the detectors mesa, $d$=35 and 40 $\mu$m, with micron level precision.

## 5.4: Infrared Characterization

After the spheres were aligned and integrated on top of the detector mesa the photoresponse at $\lambda$=4.5 $\mu$m was measured. Prior to the integration of the microsphere, all of the individual SLS detectors were characterized. Photoresponse under different bias voltage at 80 K was measured using a 500$^0$C calibrated blackbody source with a field-of-view (FOV) ~0.05 rad. The measured photoresponse without spheres is shown in the semitransparent grey box in Figs. 5.4(a) and 5.4(b) for $d$=35 and 40 $\mu$m, respectively. The curves outside of the boxes in Figs. 5.4(a) and 5.4(b) indicate the bias response after spheres were integrated with the SLS detectors. Significant responsivity enhancement is observed throughout the range of bias voltages from -0.5 to 0.1 V.

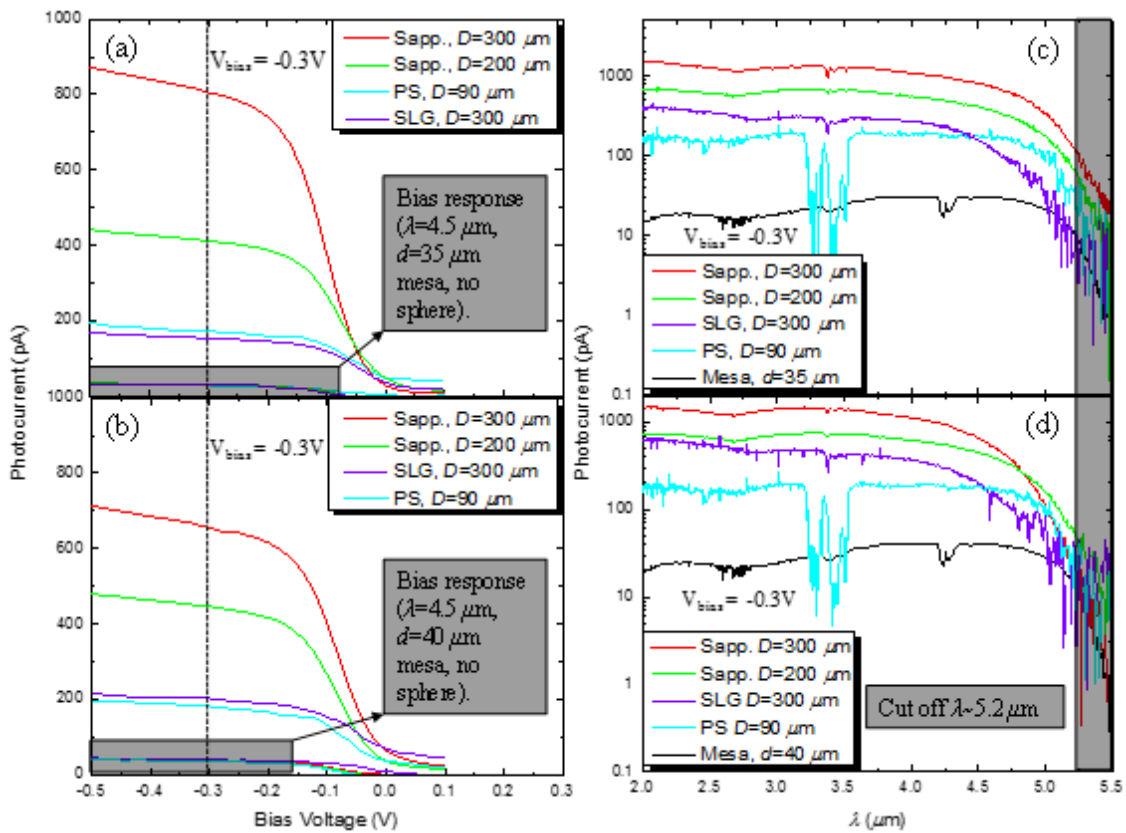

Figure 5.4: (a), (b) Photocurrent response to bias voltages from -0.5 to 0.1V for $d$=35 and 40 $\mu$m, respectively. (c), (d)  FTIR photocurrent spectra at a negative bias of 0.3V with the sphere and the reference for detector $d$=35, 40 $\mu$m, calibrated by the blackbody results shown in (a), (b).

The spectral response of the detectors before and after integrating microspheres was measured at -0.3 V bias using a Fourier transform infrared (FTIR) spectrometer. Optical settings remained constant for blackbody photoresponse and photocurrent spectra measurements in Fig. 5.4. Blackbody photoresponse measurements in Figs. 5.4(a) and 5.4(b) were used to calibrate FTIR photocurrent spectra measurements, in Figs 5.4(c) and 5.4(d). Enhancement is observed in Fig 5.4(c) and 5.4(d) when photonic jets couple to the photosensitive regions of the SLS detectors.

This enhancement factor was quantified by dividing the FTIR photocurrent spectra for the detector after integrating the sphere by the spectra of the same detector prior to sphere integration, represented as black curves in Figs. 5.4(a) and 5.4(b). Enhancement factors of ~100x are directly observed for the mesa $d$=35 $\mu$m integrated with a $D$=300 $\mu$m sapphire sphere ($n$=1.71) at $\lambda$=2.0 $\mu$m, illustrated in Fig. 5.5(a). Enhancement factors of the other $d$=35 $\mu$m detectors were observed at $\lambda$=2 $\mu$m to be 45, 26, 12x for sapphire ($D$=200 $\mu$m, $n$=1.71), soda-lime glass ($D$=300 $\mu$m, $n$=1.47), and polystyrene ($D$=90 $\mu$m, $n$=1.56), respectively. Similarly, for the $d$=40 $\mu$m detectors, enhancement factors at $\lambda$=2 $\mu$m of 73, 36, 31, and 10x are observed for sapphire ($D$=300 $\mu$m, $n$=1.71), sapphire ($D$=200 $\mu$m, $n$=1.71), soda-lime glass ($D$=300 $\mu$m, $n$=1.47), and polystyrene ($D$=90 $\mu$m, $n$=1.56), respectively.

Considering the spectral range $\lambda$=2-5 $\mu$m, there is a gradual decrease for the photocurrent enhancements. Except for polystyrene $D$=90 $\mu$m spheres, where there is no significant reduction in the enhancement factor. This gradual reduction of the enhancement factor is not as pronounced for the sapphire spheres, as it is for the soda-lime glass spheres. One factor contributing to this reduction is the decreased transmission properties of sapphire (~0.1 dB/mm) and soda-lime glass (~0.5 dB/mm) for $\lambda$ increasing from 2 to 5 $\mu$m. Material absorption alone is not significant enough to explain the reduction of the enhancement factors shown in Figs. 5.5(a) and 5.5(b). However, there is also another factor regarding the coupling efficiency of photonic jets through the mesa.

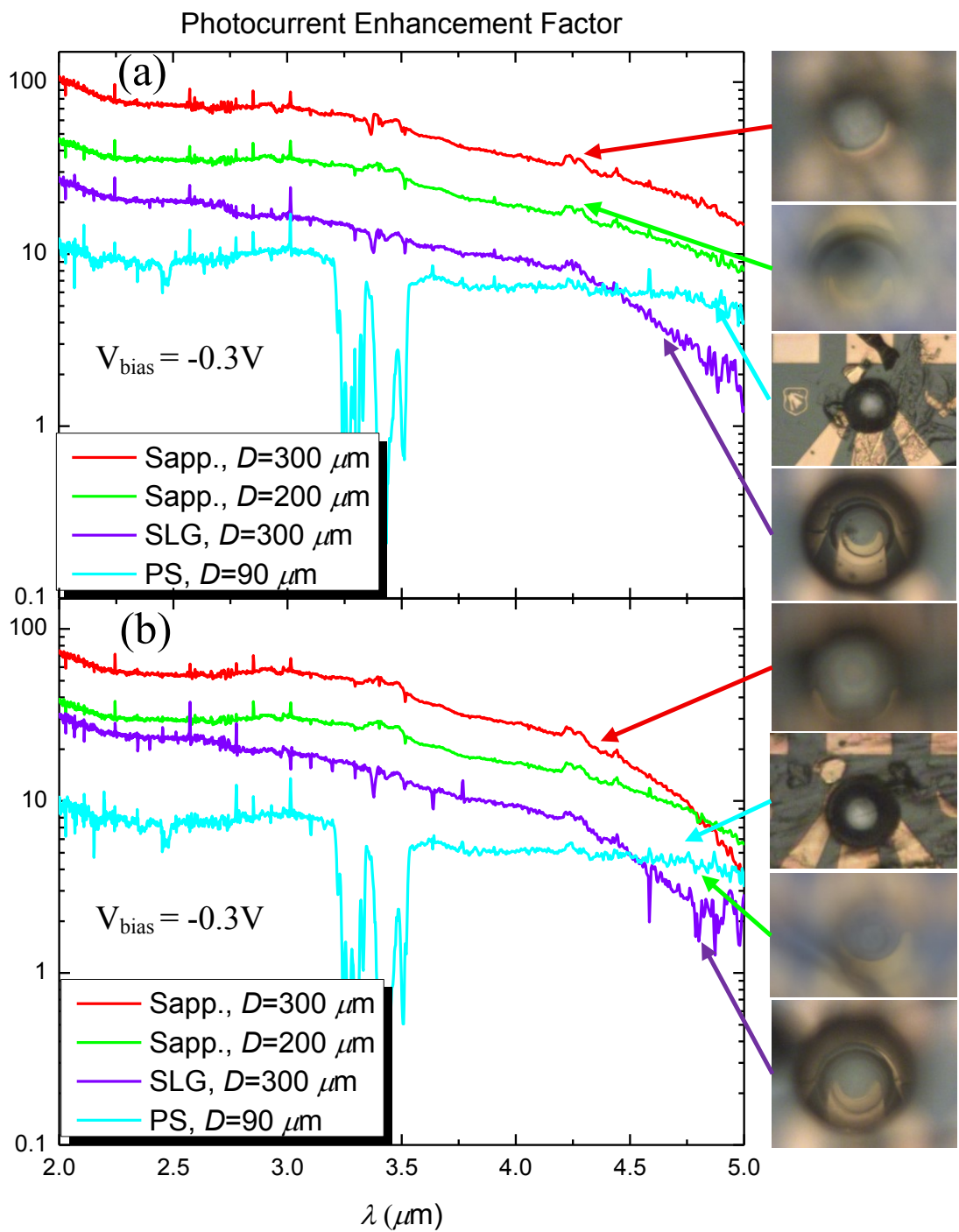

Figure 5.5: Photocurrent enhancement factors, obtained by the division of FTIR photocurrent spectra at a negative bias of 0.3V with the sphere and the reference (a), (b) Photocurrent enhancement spectra of a $d$=35, 40 $\mu$m after integrating the microspheres.

5.5: FDTD Simulations

In order to gain insight into this gradual reduction of the enhancement factor as the wavelength is increased, we performed FDTD simulations, using Lumerical FDTD solutions. The FDTD computational grid, 475x475 $\mu$m$^2$, was chosen to be 2-dimensional through the central cross-section of the model, in the *x-y* plane. To obtain an accurate solution a conformal mesh of maximum element size $\lambda/24$ was used in the FDTD computational grid. The source is a Gaussian modulating a sinusoid with a spectral width from $\lambda$=2-5 $\mu$m, launched along the *y*-axis and polarized along *x*-axis. A simulation time of 8,000 femtoseconds was used to insure all of the electromagnetic energy had reached the perfectly matched layers of the FDTD computational grid upon completion. The device geometry was modeled so that the proper refractive indices and thickness were chosen for the layers of the SLS detector, which resulted in a total thickness of 3 $\mu$m and $n$~3.93, the detector *d* was 40 $\mu$m. On the top side, gold electrodes were included as a semi-circle of 180$^0$, with an outer radius of 20 $\mu$m, and inner radius of 12 $\mu$m, with a thickness of 0.3 $\mu$m, these parameters were obtained from optical characterization of the detectors which were used for the experimental studies. The detector was modeled on top of a thick substrate with $n$=3.962, that extended beyond the perfectly matched layers of the FDTD computational grid in depth (250 $\mu$m, *y*-axis) and width (500 $\mu$m, *x*-axis).

Three different dielectric microspheres were studied directly on top of the geometric center of the detector in this model architecture: (*i*) $n$=1.71, *D*=300 $\mu$m, as sapphire, (*ii*) $n$=1.47, *D*=300 $\mu$m, as soda-lime glass, and (*iii*) $n$=1.56, *D*=90 $\mu$m, as polystyrene.

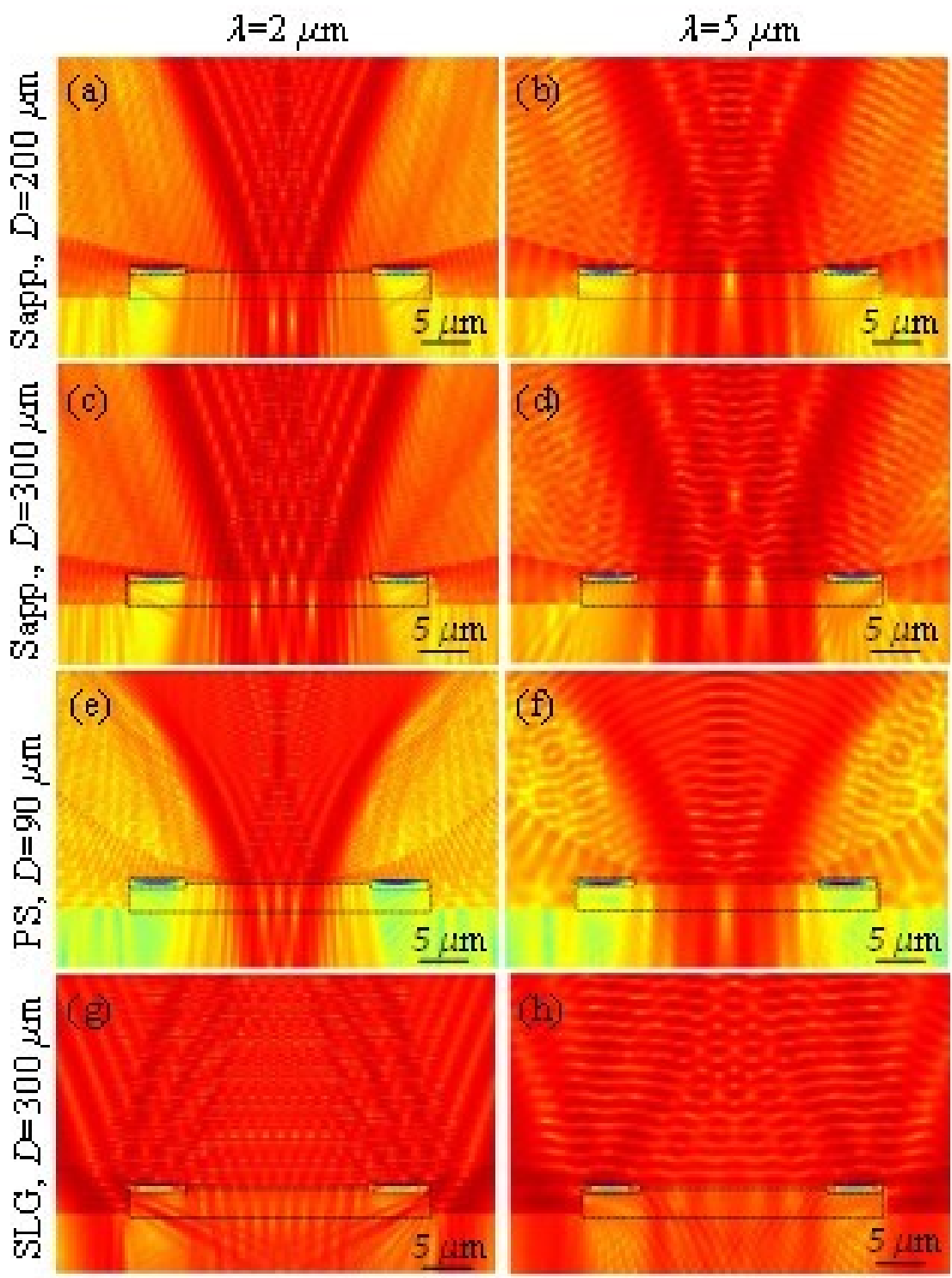

Figure 5.6: Electric field maps on the same logarithmic scale calculated from FDTD simulations at $\lambda=2$ and 5 $\mu$m for (a), (b) $n=1.71$, $D=200$ $\mu$m, as sapphire, (c), (d) $n=1.71$, $D=300$ $\mu$m, as sapphire, (e),(f) $n=1.56$, $D=90$ $\mu$m, as polystyrene, and (g),(h) $n=1.47$, $D=300$ $\mu$m, as soda-lime glass, respectively.

The first case illustrates that the concentrated light beam, photonic jet, is directed through the detector for $\lambda$=2 $\mu$m, as shown in Fig. 5.6(a), without significant loss although it is defocused at the microsphere-detector interface. However, as can be seen in Fig. 5.6(b) for $\lambda$=5 $\mu$m the photonic jet is broadened at the interface between the back surface of the microsphere and the detector such that some portion of the flux of light is reflected and scattered by the gold electrodes. For the second case illustrating polystyrene $D$=90 $\mu$m, it can be seen in Figs. 5.6(c) and 5.6(d), the photonic jet passes through the detector area over the entire spectral range from $\lambda$=2-5 $\mu$m. The last case, representing soda-lime glass, it can be seen that for $\lambda$=2 and 5 $\mu$m the photonic jet is broadened significantly at the microsphere-detector interface, Figs. 5.6(e) and 5.6(f). Some portion of the light beam is still coupled through the top-side of the detector mesa; however, the main coupling region is through the side-wall. For $\lambda$=2 $\mu$m the photonic jets expanded lateral dimensions at the microsphere-detector interface allows for significant coupling through the side-wall, as shown in Fig. 5.6(g). In the case $\lambda$=5 $\mu$m, Fig. 5.6(h), the photonic jet has expanded past the side-wall of the detector, so that the only coupled light is the small portion from the top-side which has also been subtly diminished.

These FDTD simulations give valuable insight to the experimentally obtained FTIR photocurrent spectra. For sapphire spheres ($n$~1.71, $D$=300 and 200 $\mu$m) the subtle decrease in the photocurrent spectra is explained by the defocused photonic jet being slightly blocked by the gold electrode at longer wavelengths approaching 5 $\mu$m. Polystyrene spheres ($n$~1.56, $D$=90 $\mu$m) have relatively lower refractive indices which extends the distance of the focal point from the back surface of the sphere; however, the smaller diameter compensates for that extension seeing as the focal point also scales as a function of diameter. This interplay of parameters resulted in a photonic jet which is directed through the microsphere-detector interface without being blocked by the gold electrodes, explaining the flat spectral enhancement. As for the soda-lime glass spheres ($n$~1.47, $D$=300 $\mu$m) the coupling of the photonic jet appears to have two components. Coupling is provided through the side-wall of the detector mesa and through the top-side. Side-wall coupling was completely extinguished when $\lambda$ was extended from 2 to 5 $\mu$m, which clearly manifests itself in the FTIR photocurrent spectra. Other losses can be attributed to the misalignments from the geometric center of the SLS infrared photodetector mesa.

## 5.6: Discussion

The studies presented in this chapter of the dissertation have been devoted to a proof of concept for photonic jet enhancement of photoresponse for infrared photodetectors at the single-channel level. This concept at the single-channel level can be extended to new methods and systems for providing enhancement on infrared photodetector focal plane arrays (FPAs). The technology would allow for changes in the detectors architecture leading to reduced dark current, and increased speed of the infrared photodetectors. The proposed methods and systems are based on assembly of an array of dielectric microspheres at the top of the photodetector FPAs in such a way that individual microspheres are positioned above the photosensitive regions of FPAs. These regions can be represented by *pin* junctions containing quantum wells, quantum dots or other materials with the light absorption properties in the desired spectral range.

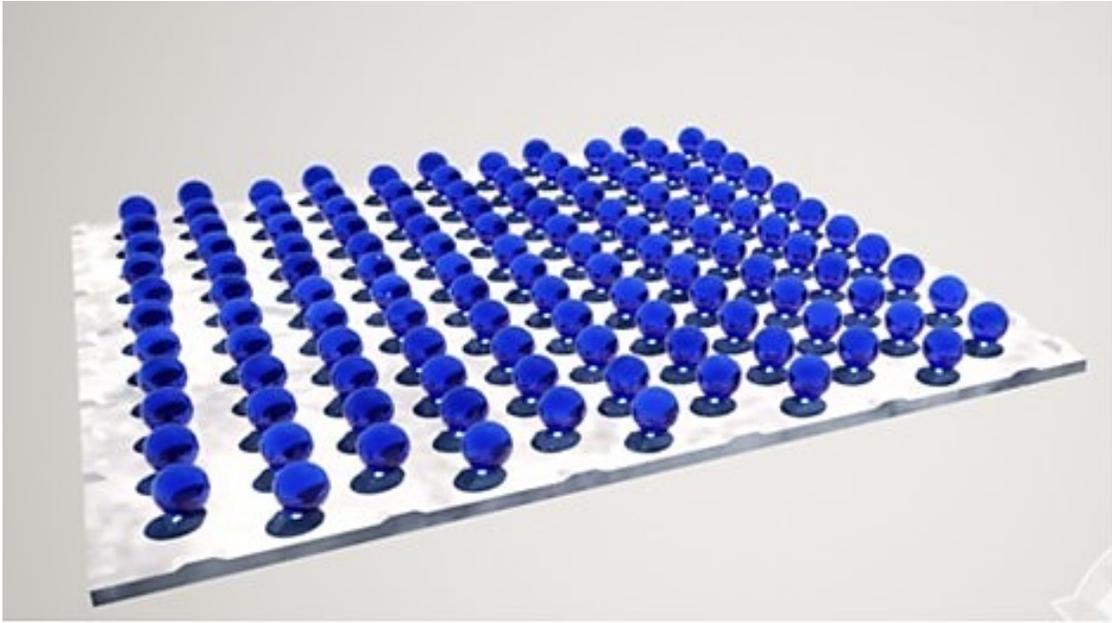

Figure 5.7: Illustration of microspheres integrated at the top of photodetector focal plane arrays.

Just as shown earlier in this chapter dielectric microspheres provide strong concentration of electromagnetic power, photonic jet effect, directly into photosensitive regions of FPA. This is provided through the mesas fabricated at the surface of FPA. This leads to improved efficiency of collection of light in such structures. Tightly focused beam widths of photonic jets allows using mesas with wavelength-scale dimensions which should result in reduced dark current and increased frequency response of FPAs. The parameters of microspheres should be optimized for a given photodetector FPA to achieve the best focusing properties at the optimal depth inside the structure. The typical values of the index of refraction ($n$) and diameter ($D$) of microspheres are within $1.4 < n < 2.0$ and $2\lambda < D < 100\lambda$ ranges. The positioning of large number of microspheres can be performed by various self-assembly, as shown in Fig. 5.8 below.

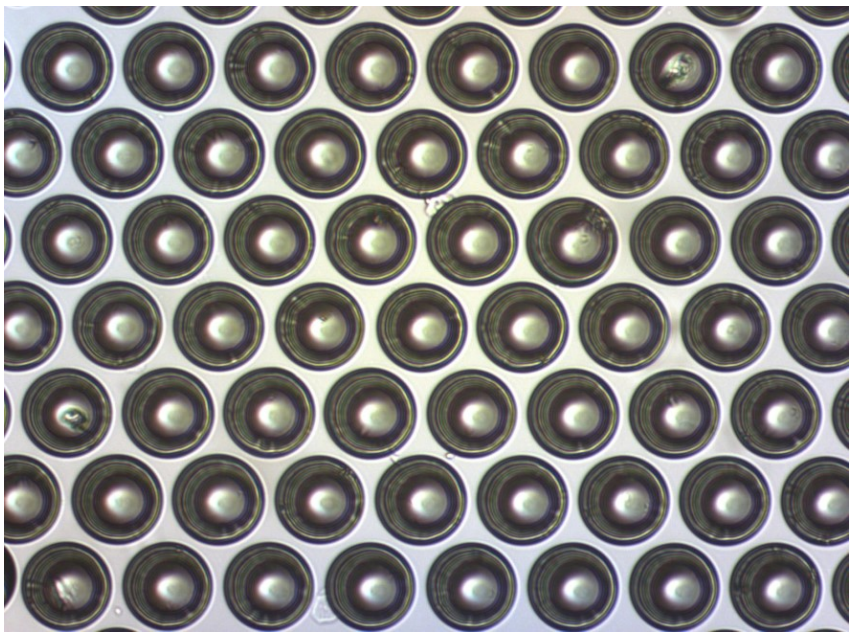

Figure 5.8: Optical micrograph of a polystyrene microsphere array, $D$=90 $\mu$m, self-assembled in prefabricated dents which could potentially be fabricated on top of the photodetector FPAs.

## 5.7: Conclusions

In conclusion, dielectric microspheres were integrated as hierarchical components to couple photonic jets into SLS infrared photodetectors. This integration was guided by the microsphere-assisted optical super-resolution imaging, discussed in Chapter 4. Concentration of light by microspheres led to broadband photocurrent enhancement over a bandwidth $\lambda$=2-5 $\mu$m. An extraordinary photocurrent enhancement factor of ~100x was observed for a sapphire sphere ($D$=300 $\mu$m) on a $d$=35 $\mu$m SLS detector at $\lambda$=2 $\mu$m. FDTD simulations give rise to a qualitative explanation to the experimental FTIR photocurrent spectra. This work can be applied for the development of high-speed MWIR single-channel detectors with increased speed and sensitivity. Integration of photodetector arrays with microspherical arrays can be used for enhancing the sensitivity of mid-IR imaging systems.

# CHAPTER 6: CONCLUSIONS AND FUTURE DIRECTIONS

Chapter 2 of the dissertation concentrated heavily on resonant and non-resonant mechanisms of propagation of light in microsphere-chain waveguides. We observed a gradual transition from the regime of light propagation, described by geometrical optics in chains with $D>20\lambda$, to the wave optics propagation effects dominating transmission properties of chains with $D<20\lambda$. Although the effects of "beam tapering" and small losses have been observed previously in chains formed by small spheres [13, 14], we showed for the first time, the evolution of these properties as a function of $D$. We showed that the chains formed by the smallest spheres ($D$=2 $\mu$m) demonstrate extraordinarily small attenuation levels of ~0.076 dB/sphere, in distant portions of the chain $35 \leq N \leq 40$. During the process of light coupling to the distant portions of the chain, the beam develops progressively smaller waists at the interface between adjacent spheres in the MCW. These effects are present only for wavelength-scale spheres at $n$=1.59, in the visible regime. As the sphere diameters are scaled up to $D>20\lambda$ the optical properties converge back to the theoretical predictions of geometrical optics.

It was also shown by FDTD simulations that spheres with identical positions of WGM resonances, "photonic atoms," under subwavelength separation, can be coupled by their evanescent fields and resonantly transport light through clusters forming "photonic molecules". The spectra of the photonic molecule is dependent on the coupling strength between the photonic atoms, the spectral characteristics of the photonic atoms, and their spatial configuration. Our results are useful for understanding the resonant optical transport properties of coupled-cavity networks. They can also be used in developing sensors and counterfeit technologies.

Future directions that can be investigated:

(1) Assembling MCWs from spheres made from nonlinear and/or active materials, where the spatially reproduced tightly focused beams could effectively excite parametric phenomena.

(2) By adjusting illumination parameters such as polarization, spatial beam profile, etc., explore the possibility of engineering NIMs.

(3) Experimentally investigate photonic molecules.

(4) Perform FDTD simulations to investigate complex networks of clusters of spheres with degenerate WGM resonant position, to observe percolation thresholds, or phase transitions.

In Chapter 3, we designed and tested focusing microprobes based on using photonic jets produced by dielectric microspheres integrated with modern mid-IR HC-MOF delivery system. The design is optimized for laser-tissue intraocular surgery performed using a flexible single-mode delivery system in contact conditions with strongly absorbing tissue. Our design demonstrates high transmission, in combination with tight focusing providing sufficient power along with desired transverse precision for laser-tissue surgical procedures. We expect that these single-mode laser scalpels will be able to cut the tissue with 20-30 $\mu$m depth and comparable lateral width. The speed of cutting can be sufficiently high if a high frequency (>1kHz) source is used. The main advantages of this technique, in comparison with conventional steel instruments, are better protection for the healthy retinal tissue, reduced bleeding, and post-operational complications. The underlying objective proposed in the single-mode design is complementary to the previously studied multimodal designs. The availability of powerful single-mode laser sources is critical for developing such single-mode designs. They can provide significant advantages over multimodal designs due to their high efficiencies, reaching 20-30% in our experiments, and more robust, and simple focusing component represented by a single high-index microsphere. All applications of such devices are difficult to envision at this point; however, we envision the use of the devices will not be limited to intraocular surgery but can be extended to brain surgery, and photoporation of cells, or generally any application which requires ultra-precise delivery of mid-IR power.

Future directions that can be investigated:

(1) Explore different materials of spheres such as sapphire, ruby, silica, and soda-lime glass.

(2) Perform FDTD simulations in order to engineer HC-MOFs with the desired effective air core and transmission bands corresponding to Er:YAG laser radiation.

(3) Ex- and in-vivo laser-tissue interaction studies.

In Chapter 4, we developed a novel practical approach to optical super-resolution imaging. This is a very attractive approach to super-resolution based on near-field interactions of dielectric microspheres with surface excitations in various nanostructures. Due to its inherent simplicity, and broadband spectral nature, the potential impact of this technology is tremendous. However, one of the major bottleneck problems in this area is developing sample scanning capability. In our work, we followed ideas expressed in the recent patent of the Mesophotonics Laboratory [219], and for the first time experimentally realized, and studied the thin films with embedded spheres which can be translated along the surface to investigate structures, enabled by the scanning functionality.

We demonstrate that two-dimensional arrays of high-index ($n$~1.9-2.2) spheres, embedded in mechanically robust thin-films of PDMS ($n$~1.4), possess the lateral spatial resolution of ~$\lambda$/7. We developed a rigorous procedure of finding the resolution of the optical system, which is important for our work, but it is also important for many groups working in this area. It is a standard process, allowing to establish a relationship between the imaging of nanostructures with various characteristic sizes and shapes to the textbook definition of super-resolution, given for point-objects. There was a growing number of super-resolution claims in this area, which are not justified from the point of view of this rigorous procedure. For this reason, we believe that this methodology will eventually be used by different groups as a unifying approach, allowing for the comparison of results obtained by different microscopy techniques, on different structures. We show that the translation of such thin-films with embedded microspheres can be facilitated by their surface lubrication. It is observed that as the liquid layer evaporates, the gap between the object and thin-film is reduced to a nanometric scale into the near-field vicinity, there is an enhancement in spatial resolution. The combination of superb nano-positioning and super-resolution imaging enables us to discern fine structural features of nanoplasmonic arrays of Au dimers and bow-ties. Such thin-films can be used in applications ranging from the investigation of microelectronics, dynamical process in biological samples, and the inspection of metamaterial surfaces.

Future directions that can be investigated:

(1) Embedding spheres into transparent matrices with fixed spatial relationships and implementing image registration techniques to capture large field-of-view super-resolution images.

(2) Investigate the role of the nanoplasmonic structures to the image formation. Using confocal microscopy to image using two wavelengths so that one is on resonance and one is off resonance, to study the resolution dependence on plasmons.

(3) Study the fundamental role of the gap separating the object and microspherical lens. The variation of the gap can be achieved by various techniques including the precise nanopositioning or specially designed experimental geometries with the gap variation.

In Chapter 5, dielectric microspheres were integrated with the SLS infrared photodetectors, in such a way that the photonic jets were coupled precisely into the device mesa. This integration was guided by microsphere-assisted optical super-resolution imaging, as described in Chapter 4. Photonic jets produced by the microspheres, coupled to SLS detectors, led to broadband photocurrent enhancement over a bandwidth $\lambda$=2-5 $\mu$m. An extraordinary photocurrent enhancement factor of ~100x was observed for a sapphire sphere ($D$=300 $\mu$m) on a $d$=35 $\mu$m SLS detector at $\lambda$=2 $\mu$m. FDTD simulations provide a qualitative explanation to the experimental FTIR photocurrent spectra. This work can be applied for the development of high-speed MWIR single-channel detectors, or the technology could be extended to focal plane arrays.

Future directions that can be investigated:

(1) Develop focal plane arrays using the strained-layer superlattice detectors and massive scale self-assembly of microspheres.

(2) Using a Schottky barrier type detector, to develop a high-speed single-channel detector. With an RC time constant limited detector, the surface area is proportional to the capacitance. Microspheres producing photonic jets enable the surface area to be significantly reduced, in turn reducing the capacitance, and increasing the overall speed of the detector.

(3) Investigate photonic jet coupling, produced by microspheres, to plasmonic meshes on top of the detector mesa, illustrated in Fig. 6.1. Increasing the effective aerial fill factor, and exciting resonances in the mesh could lead to significant enhancement of the photocurrent while increasing the quantum efficient, which is currently a problem with quantum dot infrared photodetectors (QDIPs).

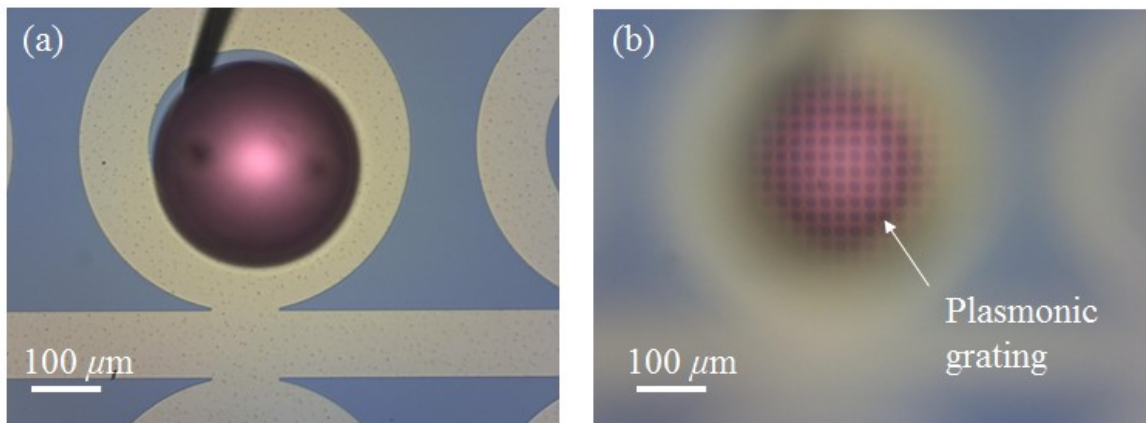

Figure 6.1: $D$=300 $\mu$m ruby microsphere integrated with a quantum dot infrared photodetector with a 2-D plasmonic grating covering the top of the detector mesa.

REFERENCES

[1]     K. W. Allen, A. Darafsheh, F. Abolmaali, N. Mojaverian, N. I. Limberopoulos, A. Lupu, and V. N. Astratov., "Microsphere-chain waveguides: Transport and focusing properties," Applied Physics Letters, 105, 021112 (2014).

[2]     K. W. Allen, "Light Focusing and Transport Properties of Mesophotonic Structures," University of North Carolina at Charlotte, (2012).

[3]     K. W. Allen, A. Darafsheh, and V. N. Astratov, "Photonic nanojet-induced modes: from physics to applications," IEEE proc. of Transparent Optical Networks (ICTON), 2011 13th International Conference, 1-4 (2011).

[4]     K. W. Allen, A. Darafsheh, and V. N. Astratov, "Beam tapering effect in microsphere chains: from geometrical to physical optics," International Society for Optics and Photonics, SPIE LASE, 823622-823622-7 (2012).

[5]     A. Darafsheh, N. Mojaverian, N. I. Limberopoulos, K. W. Allen, A. Lupu, and V. N. Astratov, "Formation of polarized beams in chains of dielectric spheres and cylinders," Optics Letters, 38(20), 4208-4211 (2013).

[6]     V. N. Astratov, A. Darafsheh, M. D. Kerr, K. W. Allen, N. M. Fried, A. N. Antoszyk, and H. S. Ying, "Photonic nanojets for laser surgery," SPIE Newsroom, 12, 32-34 (2010).

[7]     A. Darafsheh, M. D. Kerr, K. W. Allen, N. M. Fried, A. N. Antoszyk, H. S. Ying, and V. N. Astratov, "Integrated microsphere arrays: light focusing and propagation effects," International Society for Optics and Photonics, SPIE OPTO, 76050R-76050R-9 (2010).

[8]     V. N. Astratov, A. Darafsheh, M. D. Kerr, K. W. Allen, and N. M. Fried, "Focusing microprobes based on integrated chains of microspheres," Proc. of Progress in Electromagnetics Research Symposium, 5-8 (2010).

[9]     A. Darafsheh, K. W. Allen, A. Fardad, N. M. Fried, A. N. Antoszyk, H. S. Ying, and V. N. Astratov, "Focusing capabilities of integrated chains of microspheres in the limit of geometrical optics," International Society for Optics and Photonics, SPIE LASE, 79131A-79131A-7 (2011).

[10]   A. Darafsheh, M. D. Kerr, K. W. Allen, and V. N. Astratov, "Integrated microsphere arrays as a compact focusing tool for biomedical and photonics applications," IEEE proc. of Laser and Electro-Optics (CLEO) and Quantum Electronics and Laser Science Conference (QELS), 1-2 (2010).

[11]   T. C. Hutchens, A. Darafsheh, A. Fardad, A. N. Antoszyk, H. S. Ying, V. N. Astratov, and N. M. Fried, "Characterization of novel microsphere chain fiber optic

tips for potential use in ophthalmic laser surgery," Journal of Biomedical Optics, 17(6), 068004 (2012).

[12]  Z. Chen, A. Taflove, and V. Backman, "Highly efficient optical coupling and transport phenomena in chains of dielectric microspheres," Optics Letters, 31(3), 389-391 (2006).

[13]  A. M. Kapitonov and V. N. Astratov, "Observation of nanojet-induced modes with small propagation losses in chains of coupled spherical cavities," Optics Letters, 32(4), 409-411 (2007).

[14]  S. Yang and V. N. Astratov, "Photonic nanojet-induced modes in chains of    size disordered microspheres with an attenuation of only 0.08 dB per sphere," Applied Physics Letters, 92(26), 261111(3pp) (2008).

[15]  C. Y. Liu, "Ultra-high transmission of photonic nanojet induced modes in chains of core-shell microcylinders," Physics Letters A, 376(45), 3261-3266 (2012).

[16]  T. Mitsui, Y. Wakayama, T. Onodera, Y. Takaya, and H. Oikawa, "Light Propagation within Colloidal Crystal Wire Fabricated by a Dewetting Process," Nano Letters, 8(3), 853-858 (2008).

[17]  M. Gerlach, Y. P. Rakovich, and J. F. Donegan, "Nanojets and directional emission in symmetric photonic molecules," Optics Express, 15(25), 17343-17350 (2007).

[18]  T. Mitsui, Y. Wakayama, T. Onodera, T. Hayashi, N. Ikeda, Y. Sugimoto, T. Takamasu, and H. Oikawa, "Micro-demultiplexer of Coupled Resonator Optical Waveguide Fabricated by Microspheres," Advanced Materials, 22(28), 3022-3026 (2010).

[19]  O. Lecarme, T. P. Rivera, L. Arbez, T. Honegger, K. Berton, and D. Peyrade, "Collodial optical waveguides with integrated local light sources built by capillary force assembly," Journal of Vacuum Science and Technology B, 28, C6011 (2010).

[20]  T. Mitsui, T. Onodera, Y. Wakayama, T. Hayashi, N. Ikeda, Y. Sugimoto, T. Takamasu, and H. Oikawa, "Influence of micro-joints formed between spheres in coupled-resonator optical waveguide," Optics Express, 19(22), 22258-22267 (2011).

[21]  C. Perez-Rodriguez, S. Rios, I. R. Martin, P. Haro-Gonzalez, and D. Jaque, "Upconversion emission obtained in $Yb^{3+}$-$Er^{3+}$doped fluoroindate glasses using silica microspheres as focusing lens," Optics Express, 21(9), 10667-10675 (2013).

[22]    A. K. Tiwari, B. Chandra, R. Uppu, and S. Mujumdar, "Collective lasing from a linear array of dielectric micropsheres with gain," Optics Express, 20(6), 6598-6603 (2012).

[23]    Y. Ku, C. Kuang, X. Hao, Y. Xue, H. Li, and X. Liu, "Superenhanced three-dimensional confinement of light by compound metal-dielectric microspheres," Optics Express, 20(15), 16981-16991 (2012).

[24]    A. K. Tiwari, R. Uppu, and S. Mujumdar, "Experimental demonstration of small-angle bending in an active direct-coupled chain of spherical microcavities," Applied Physics Letters, 103, 171108 (2013).

[25]    Y. Li, K. W. Allen, F. Abolmaali, A. V. Maslov, and V. N. Astratov, "Spectral finger-prints of photonic molecules," IEEE proc. of Transparent Optical Networks (ICTON), 1-4 (2014).

[26]    S. Schiller and R. L. Byer, "High-resolution spectroscopy of whispering gallery modes in large dielectric spheres," Optics Letters, 16(15), 1138-1140 (1991).

[27]    C. C. Lam, P. T. Leung, and K. Young, "Explicit asymptotic formulas for the positions, widths, and strengths of resonances in Mie scattering," Journal of the Optical Society of America B, 9(9), 1585-1592 (1992).

[28]    S. L. McCall, A. F. J. Levi, R. E. Slusher, S. J. Pearton, and R. A. Logan, "Whispering-gallery mode microdisk lasers," Applied Physics Letters, 60, 289 (1992).

[29]    J. C. Knight, G. Cheung, F. Jacques, and T. A. Birks, "Phase-matched excitation of whispering-gallery-mode resonances by a fiber taper," Optics Letters, 22(15), 1129-1131 (1997).

[30]    V. S. Ilchenko, P. S. Volikov, V. L. Velichansky, F. Treussart, V. Lefèvre-Seguin, J.-M. Raimond, and S. Haroche, "Strain-tunable high-$Q$ optical microsphere resonator," Optics Communications, 145(1-6), 86-90 (1998).

[31]    M. L. Gorodetsky and V. S. Ilchenko, "Optical microsphere resonators: optical coupling to high-$Q$ whispering-gallery modes," Journal of the Optics Society of America B, 16(1), 147-154 (1999).

[32]    A. N. Oraevsky, "Whispering-gallery waves," Quantum Electronics, 32(5), 377400 (2002).

[33]    V. N. Astratov, "Fundamentals and applications of microsphere resonator circuits," in [Photonic microresonator research and applications], I. Chremmos, O. Schwelb, and N. Uzunoglu Eds., Springer Series in Optical Sciences 156, 423-457, New York, NY (2010).

[34]  A. Chiasera, Y. Dumeige, P. Féron, M. Ferrari, Y. Jestin, G. Nunzi Conti, S. Pelli, S. Soria, and G. C. Righini, "Spherical whispering-gallery-mode microresonators," Laser & Photonics Reviews, 4(3), 457-482 (2010).

[35]  G. C. Righini, Y. Dumeige, P. Féron, M. Ferrari, G. Nunzi Conti, D. Ristic, and S. Soria, "Whispering gallery mode microresonators: Fundamentals and applications," Rivista Del Nuovo Cimento, 34(7), 435-488 (2011).

[36]  S. Arnold, M. Khoshsima, I. Teraoka, S. Holler, and F. Vollmer, "Shift of whispering-gallery modes in microspheres by protein adsorption," Optics Letters, 28(4), 272-274 (2003).

[37]  K. J. Vahala, "Optical microcavities," Nature, 424(6950), 839-846 (2003).

[38]  K. R. Hiremath and V. N. Astratov, "Perturbation of whispering gallery modes by nanoparticles embedded in microcavities," Optics Express, 16(8), 5421-5426 (2008).

[39]  F. Vollmer and S. Arnold, "Whispering-gallery-mode biosensing: label-free detection down to single molecules," Nature Methods, 5(7), 591-596 (2008).

[40]  J. Ward and O. Benson, "WGM microresonators: sensing, lasing and fundamental optics with microspheres," Laser & Photonics Reviews, 5(4), 553-570 (2011).

[41]  S. Arnold, S. I. Shopova, and S. Holler, "Whispering gallery mode bio-sensor for label-free detection of single molecules: thermo-optic vs. reactive mechanism," Optics Express, 18(1), 281-287 (2010).

[42]  A. B. Matsko, A. A. Savchenkov, D. Strekalov, V. S. Ilchenko, and L. Maleki, "Review of applications of whispering-gallery mode resonators in photonics and nonlinear optics," IPN Progress Report, 42-162 (2005).

[43]  A. B. Matsko and V. S. Ilchenko, "Optical resonators with whispering-gallery modes-Part I: Basics," IEEE journal of selected topics in Quantum Electronics, 12(1), 3-14 (2006).

[44]  V. S. Ilchenko and A. B. Matsko, "Optical resonators with whispering-gallery modes-Part II: Applications," IEEE journal of selected topics in Quantum Electronics, 12(1), 15-32 (2006).

[45]  O. Svitelskiy, D. Sun, A. Darafsheh, M. Sumetsky, A. Lupu, M. Tchernycheva, and V. N. Astratov, "Characterization of high index microsphere resonators in fiber integrated microfluidic platforms," Proceedings of SPIE, 7913, 791314(7pp) (2011).

[46]    O. Svitelskiy, Y. Li, A. Darafsheh, M. Sumetsky, D. Carnegie, E. Rafailov, and V. N. Astratov, "Fiber coupling to BaTiO3 glass microspheres in aqueous environment," Optics Letters, 36(15), 2862-2864 (2011).

[47]    T. Lu, H. Lee, T. Chen, S. Herchak, J.-H. Kim, S. E. Fraser, R. C. Flagan, and K. Vahala, "High sensitivity nanoparticle detection using optical microcavities," Proceeding of the National Academy of Sciences of the United States of America, 108(5), 5976-5979 (2011).

[48]    M. Gerlach, Y. P. Rakovich, and J. F. Donegan, "Radiation-pressure-induced mode splitting in a spherical microcavity with an elastic shell," Optics Express, 15(6), 3597-3606 (2007).

[49]    Y. Li, O. V. Svitelskiy, A. V. Maslov, D. Carnegie, E. Rafailov, and V. N. Astratov, "Giant resonant light forces in microsperical photonics," Light: Science and Applications, 2, e64 (2013).

[50]    A. V. Maslov and V. N. Astratov, "Microspherical photonics: Sorting resonant photonic atoms by using light," Applied Physics Letters, 105, 121113 (2014).

[51]    Z. Wang, W. Guo, L. Li, B. Luk'yanchuk, A. Khan, Z. Liu, Z. Chen, and M. Hong, "Optical virtual imaging at 50 nm lateral resolution with a white-light nanoscope," Nature Communications, 2, 218 (2011).

[52]    X. Hao, C. Kuang, X. Liu, H. Zhang, and Y. Li, "Microsphere based microscope with optical super-resolution capability," Applied Physics Letters, 99, 203102 (2011).

[53]    A. Darafsheh, G. F. Walsh, L. D. Negro, and V. N. Astratov, "Optical super-resolution by high-index liquid-immersed microspheres," Applied Physics Letters, 101, 141128 (2012).

[54]    Y. Duan, G. Barbastathis, and B. Zhang, "Classical imaging theory of a microlens with super-resolution," Optics Letters, 38(16), 2988-2990 (2013).

[55]    Y. Ben-Aryeh, "Superresolution observed from evanescent waves transmitted through nano-corruagted metallic films," Applied Physics B, 109, 165-170 (2012).

[56]    Y. H. Lin and D. P. Tsai, "Near-field scanning optical microscopy using a super-resolution cover glass slip," Optics Express, 20(15), 16205-16211 (2012).

[57]    L. A. Krivitsky, J. J. Wang, Z. Wang, and B. Luk'yanchuk, "Locomotion of microspheres for super-resolution imaging," Scientific Reports, 3, 3501 (2013).

[58] S. Lee, L. Li, Y. Ben-Aryeh, Z. Wang, and W. Guo, "Overcoming the diffraction limit induced by microsphere optical nanoscopy," Journal of Optics, 15(12), 125710 (2013).

[59] R. Ye, Y. H. Ye, H. F. Ma, J. Ma, B. Wang, J. Yao, S. Liu, L. Cao, H. Xu, and J. Y. Zhang, "Experimental far-field imaging properties of a ~5-μm diameter spherical lens," Optics Letters, 38(11), 1829-1831 (2013).

[60] H. Yang, N. Moullan, J. Auwerx, and M. A. M. Gijs, "Super-Resolution Biological Microscopy Using Virtual Imaging by a Microsphere Nanoscope," Small, 10(9), 1712-1718 (2014).

[61] S. Lee, L. Li, Z. Wang, W. Guo, Y. Yan, and T. Wang, "Immersed transparent microsphere magnifying sub-diffraction-limited objects," Applied Optics, 52(30), 7265-7270 (2013).

[62] X. Huang, X. N. He, W. Xiong, Y. Gao, L. J. Jiang, L. Liu, Y. S. Zhou, L. Jiang, J. F. Silvain, and Y. F. Lu, "Contrast enhancement using silica microspheres in coherent anit-Stokes Raman spectroscopic imaging," Optics Express, 22(3), 2889-2896 (2014).

[63] L. Li, W. Guo, Y. Yan, S. Lee, and T. Wang, "Label-free super-resolution imaging of adenoviruses by submerged microsphere optical nanoscopy," Light: Science and Applications, 2, e104 (2013).

[64] B. Sandnes, T. A. Kelf, H. Liu, and A. V. Zvyagin, "Retroemission by a glass bead monolayer for high-sensitivity, long-range imaging of upconverting phosphors," Optics Letters, 36(15), 3009-3011 (2011).

[65] A. Darafsheh, N. I. Limberopoulos, J. S. Derov, D. E. Walker Jr., and V. N. Astratov, "Advantages of microsphere-assisted super-resolution imaging technique over solid immersion lens and confocal microscopies," Applied Physics Letters, 104, 061117 (2014).

[66] Y. Yan, L. Li, C. Feng, W. Guo, S. Lee, and M. Hong, "Microsphere-Coupled Scanning Laser Confocal Nanoscope for Sub-Diffraction-Limited Imaging at 25 nm Lateral Resolution in the Visible Spectrum," ACS Nano, 8(2), 1809-1816 (2014).

[67] R. Ye, Y. H. Ye, H. F. Ma, L. Cao, J. Ma, F. Wyrowski, R. Shi, and J. Y. Zhang, "Experimental imaging properties of immersion microscale spherical lenses," Scientific Reports, 4, 3769 (2014).

[68] S. Lee, L. Li, and Z. Wang, "Optical resonances in microsphere photonic nanojets," Journal of Optics, 16(1), 015704 (2014).

[69]    L. Yao, Y. H. Ye, H. F. Ma, L. Cao, and J. Hou, "Role of the immersion medium in the microscale spherical lens imaging," Optics Communications, 335, 23-27 (2014).

[70]    K. W. Allen, N. Farahi, Y. Li, N. I. Limberopoulos, D. E. Walker Jr., A. M. Urbas, and V. N. Astratov, "Super-resolution Imaging by Arrays of High-Index Spheres Embedded in Transparent Matrices," IEEE proc. of NAECON, in press (2014).

[71]    K. W. Allen, A. F. Kosolapov, A. N. Kolyadin, A. D. Pryamikov, N. Mojaverian, N. I. Limberopoulos, and V. N. Astratov, "Photonic jets produced by microspheres integrated with hollow-core fibers for ultraprecise laser surgery," IEEE proc. of Transparent Optical Networks (ICTON), 1-4 (2013).

[72]    K. W. Allen, J. M. Duran, G. Ariyawansa, J. H. Vella, N. I. Limberopoulos, A. M. Urbas, and V. N. Astratov, "Photonics Jets for Strained-Layer Superlattice Infrared Photodetector Enhancement," IEEE proc. of NAECON, in press (2014).

[73]    Z. Chen, A. Taflove, and V. Backman, "Photonic nanojet enhancement of backscattering of light by nanoparticles: a potential novel visible-light ultramicroscopy technique," Optics Express, 12(7), 1214-1220 (2004).

[74]    X. Li, Z. Chen, A. Taflove, and V. Backman, "Optical analysis of nanoparticles via enhanced backscattering facilitated by 3-D photonic nanojets," Optics Express, 13(2), 526-533 (2005).

[75]    A. Heifetz, S.-C. Kong, A. V. Sahakian, A. Taflove, and V. Backman, "Photonic nanojets," Journal of Computational and Theoretical Nanoscience, 6(9), 1979-1992 (2009).

[76]    S. Lecler, Y. Takakura, and P. Meyrueis, "Properties of a three-dimensional photonic jet," Optics Letters, 30(19), 2641-2643 (2005).

[77]    A. V. Itagi and W. A. Challener, "Optics of photonic nanojets," Journal of the Optical Society of America A, 22(12), 2847-2858 (2005).

[78]    Z. Chen, A. Taflove, X. Li, and V. Backman, "Superenhanced backscattering of light by nanoparticles," Optics Letters, 31(2), 196-198 (2006).

[79]    P. Ferrand, J. Wenger, A. Devilez, M. Pianta, B. Stout, N. Bonod, E. Popov, and H. Rigneault, "Direct imaging of photonic nanojets," Optics Express, 16(10), 6930-6940 (2008).

[80]    A. Devilez, B. Stout, N. Bonod, and E. Popov, "Spectral analysis of three dimensional photonic jets," Optics Express, 16(18), 14200-14212 (2008).

[81]    A. Devilez, N. Bonod, J. Wenger, D. Gérard, B. Stout, H. Rigneault, and E. Popov, "Three-dimensional subwavelength confinement of light with dielectric microspheres," Optics Express, 17(4), 2089-2094 (2009).

[82]    S.-C. Kong, A. Taflove, and V. Backman, "Quasi one-dimensional light beam generated by a graded-index microsphere," Optics Express, 17(5), 3722-3731 (2009).

[83]    S.-C. Kong, A. Taflove, and V. Backman, "Quasi one-dimensional light beam generated by a graded-index microsphere: errata," Optics Express, 18(4), 3973 (2010).

[84]    Y. E. Geints, A. A. Zemlyanov, and E. K. Panina, "Controlling the parameters of photon nanojets of composite microspheres," Optics and Spectroscopy, 109(4), 590-595 (2010).

[85]    Y. E. Geints, E. K. Panina, and A. A. Zemlyanov, "Control over parameters of photonic nanojets of dielectric microspheres," Optics Communications, 283(23), 4775-4781 (2010).

[86]    H. Ding, L. Dai, and C. Yan, "Properties of the 3D photonic nanojet based on the refractive index of surroundings," Chinese Optics Letters, 8(7), 706-708 (2010).

[87]    M.-S. Kim, T. Scharf, S. Mühlig, C. Rockstuhl, and H. P. Herzig, "Engineering photonic nanojets," Optics Express, 19(11), 10206-10220 (2011).

[88]    M.-S. Kim, T. Scharf, S. Mühlig, C. Rockstuhl, and H. P. Herzig, "Gouy phase anomaly in photonic nanojets," Applied Physics Letters, 98(19), 191114(3pp) (2011).

[89]    Y. Liu, B. Wang, and Z. Ding, "Influence of incident light polarization on photonic nanojet," Chinese Optics Letters, 9(7), 072901(3pp) (2011).

[90]    D. McCloskey, J. J. Wang, and J. F. Donegan, "Low divergence photonic nanojets from $Si_3N_4$ microdisks," Optics Express, 20(1), 128-140 (2012).

[91]    Y. E. Geints, A. A. Zemlyanov, and E. K. Panina, "Photonic nanojet calculations in layered radially inhomogeneous micrometer-sized spherical particles," Journal of the Optical Society of America B, 28(8), 1825-1830 (2011).

[92]    Y. E. Geints, A. A. Zemlyanov, and E. K. Panina, "Photonic nanojet effect in multilayer micrometer-sized spherical particles," Quantum Electronics, 41(6), 520-525 (2011).

[93] P. Bon, B. Rolly, N. Bonod, J. Wenger, B. Stout, S. Monneret, and H. Rigneault, "Imaging the Gouy phase shift in photonic jets with a wavefront sensor," Optics Letters, 37(17), 3531-3533 (2012).

[94] Y. E. Geints, A. A. Zemlyanov, and E. K. Panina, "Photonic jets from resonantly excited transparent dielectric microspheres," Journal of the Optical Society of America B, 29(4), 758-762 (2012).

[95] H. Guo, Y. Han, X. Weng, Y. Zhao, G. Sui, Y. Wang, and S. Zhuang, "Near-field focusing of the dielectric microsphere with wavelength scale radius," Optics Express, 21(2), 2434-2443 (2013).

[96] A. D. Kiselev and D. O. Plutenko, "Mie scattering of Leguerre-Gaussian beams: Photonic nanojets and near-field optical vortices," Physics Review A, 89, 043803 (2014).

[97] M. Hasan and J. J. Simpson, "Three-dimension subwavelength confinement of a photonic nanojet using a plasmonic nanoantenna GAP," Microwave and Optical Technology Letters, 56(11), 2700-2706 (2014).

[98] T. Jalali and D. Erni, "Highly confined photonic nanojet from elliptical particles," Journal of Modern Optics, 61(13), 1069-1076 (2014).

[99] T. Matsui and A. Okajima, "Finite-difference time-domain analysis of photonic nanojets from liquid-crystal-containing microcylinders," Japanese Journal of Applied Physics, 53(1S), 01AE04 (2014).

[100] C-Y. Liu Y-H. Wang, "Real-space observation of photonic nanojet in dielectric microspheres," Physica E: Low-dimensional Systems and Nanostructures, 61, 141-147 (2014).

[101] M. Bayer, T. Gutbrod, J. P. Reithmaier, A. Forchel, T. L. Reinecke, P. A. Knipp, A. A. Dremin, and V. D. Kulakovskii, "Optical Modes in Photonic Molecules," Physics Review Letters, 81, 2582 (1998).

[102] S. V. Boriskina, "Theoretical prediction of a dramatic $Q$-factor enhancement and degeneracy removal of whispering gallery modes in symmetrical photonic molecules," Optics Letters, 31(3), 338-340 (2006).

[103] Y. P. Rakovich, J. F. Donegan, M. Gerlach, A. L. Bradley, T. M. Connolly, J. J. Bolland, N. Gaponik, and A. Rogach, "Fine structure of coupled optical modes in photonic molecules," Physics Review A, 70, 051801(R) (2004).

[104] S. V. Boriskina, "Spectrally engineered photonic molecules as optical sensors with enhanced sensitivity: a proposal and numerical analysis," Journal of the Optical Society of America B, 23(8), 1565-1573 (2006).

[105]  S. Ishii and T. Baba, "Bistable lasing in twin microdisk photonic molecules," Applied Physics Letters, 87, 181102 (2005).

[106]  S. V. Boriskina, "Coupling of whispering-gallery modes in size-mismatched microdisk photonic molecules," Optics Letters, 32(11), 1557-1559 (2007).

[107]  A. Nakagawa, S. Ishii, and T. Baba, "Photonic molecule laser composed of GaInSbP microdisks," Applied Physics Letters, 86, 041112 (2005).

[108]  S. V. Boriskina, "Photonic molecules and spectral engineering," Chapter 16, Photonic Microresonator Research and Applications Springer Series in Optical Sciences, 156, 393-421 (2010).

[109]  Y. P Rakovich, J. J. Boland, and J. F. Donegan, "Tunable photon lifetimes in photonic molecules: a concept for delaying an optical signal," Optics Letters, 30(20), 2775-2777 (2005).

[110]  S. V. Boriskina and L. D. Negro, "Self-referenced photonic molecule bio(chemical)sensor," Optics Letters, 35(14), 2496-2498 (2010).

[111]  Y. F. Lu, L. Zhang, W. D. Song, Y. W. Zheng, and B. S. Luk'yanchuk, "LASER WRITING OF SUB-WAVELENGTH STRUCTURE ON SILICON (100) SURFACES WITH PARTICLE ENHANCED OPTICAL IRRADIATION," Pis'ma v ZhETF, 72(9), 658-661 (2000).

[112]  A. Darafsheh and V. N. Astratov, "Periodically focused modes in chains of dielectric microspheres," Applied Physics Letters, 101(6), 061123 (2012)

[113]  V. N. Astratov, Y. Li, O. V. Stivelskiy, A. V. Maslov, M. I. Bakunov, D. Carnegie, and E. Rafailov, "Microspherical Photonics: Ultra-High Resonant Propulsion Forces," Optics and Photonics News, 24(12), 40-40 (2013).

[114]  A. Yariv., Y. Xu, R. K. Lee, and A. Scherer, "Coupled-resonator optical waveguide: A proposal and analysis," Optics Letters, 24, 711–713 (1999).

[115]  M. Bayindir, B. Temelkuran, and E. Ozbay, "Tight-binding description of the coupled defect modes in three-dimensional photonic crystals," Physics Review Letters, 84, 2140–2143 (2000).

[116]  B. M. Möller, U. Woggon, and M. V. Artemyev, "Bloch modes and disorder phenomena in coupled resonator chains," Physics Review B, 75, 245327 (2007).

[117]  A. V. Kanaev, V. N. Astratov, and W. Cai, "Optical coupling at a distance between detuned spherical cavities," Applied Physics Letters, 88, 111111 (2006).

[118]  R. K. Chang and A. J. Campillo, "Optical processes in microcavities," World Scientific, Singapore (1996).

[119]  N. Le Thomas, U. Woggon, W. Langbein, and M. V. Artemyev, "Effect of a dielectric substrate on whispering-gallery mode sensors," Journal of the Optical Society of America B, 23, 2361-2365 (2006).

[120]  S. Yang and V. N. Astratov, "Spectroscopy of coherently coupled whispering-gallery modes in size-matched bispheres assembled on a substrate," Optics Letters, 34, 2057-2059 (2009).

[121]  S. Yang and V. N. Astratov, "Spectroscopy of photonic molecular states in supermonodispersive bispheres," Published in Laser Resonators and Beam Control XI, In" A. V. Ku-dryashov, A. H. Paxton, V. S. Ilchenko, and L.. Aschke Proc. of SPIE, 7194, 719411-1 (2009).

[122]  H. C. Van de Hulst, "Light scattering by small particles," Wiley, New York (1981).

[123]  C. Arnaud, M. Boustimi, and P. Feron, "Wavelength shifts in erbium doped glass microspherical whispering gallery mode lasers," Proc. of the International Workshop on Photonics and Applications, Hanoi, Vietnam, 209-220 (2004).

[124]  J. A. Harrington, "A Review of IR Transmitting, Hollow Waveguides," Fiber and Integrated Optics, 19, 211-217 (2000).

[125]  F. Luan, A. K. George, T. D. Hedley, G. J. Pearce, D. M. Bird, J. C. Knight, and P. St. J. Russell, "All-solid photonic bandgap fiber," Optics Letters, 29(20), 2369-2371 (2004).

[126]  B. Mangan, L. Farr, A. Langford, P. J. Roberts, D. P. Williams, F. Couny, M. Lawman, M. Mason, S. Coupland, R. Flea, H. Sabert, T. A. Birks, J. C. Knight, and P. St. J. Russell, "Low loss (1.7dB/km) hollow core photonic bandgap fiber," Optical Fiber Communication Conference, 1-55752-767-9 (2004).

[127]  J. C. Knight, "Photonic crystal fibres," Nature, 424, 847-851 (2003).

[128]  P. St. J. Russell, "Photonic-crystal fibers," Journal of Lightwave Technology, 24(12), 4729-4749 (2006).

[129]  J. D. Joannopoulos, P. R. Villeneuve, and Shanhui Fan, "Photonic crystals, putting a new twist on light," Nature, 386, 143-149 (1997).

[130]  B. Temelkuran, S. D. Hart, G. Benoit, J. D. Joannopoulos, and Y. Fink, "Wavelength-scalable hollow core fibres with large photonic bandgaps for $CO_2$ laser transmission," Nature, 420, 650-653 (2002).

[131]  P. St. J. Russell, "Photonic Crystal Fibers," Science, 17(5605), 358-362 (2003).

[132]  J. M. Dudley and R. Taylor, "Ten years of nonlinear optics in photonic crystal fibre," Nature Photonics, 3, 85-90 (2009).

[133]  T. Ritari, J. Tuominen, H. Ludvigsen, J. C. Petersen, T. Sorensen, T. P. Hansen, and H. R. Simonsen, "Gas sensing using air-guiding photonic bandgap fibers," Optics Express, 12(17), 4080-4087 (2004).

[134]  F. M. Cox, A. Argyros, and M. C. J. Large, "Liquid-filled hollow core microstructured polymer optical fiber," Optics Express, 14(9), 4135-4140 (2006).

[135]  Y. Huang, Y. Xu, and A. Yariv, "Fabrication of functional microstructured optical fibers through a selective-filling technique," Applied Physics Letters, 85, 5182 (2004).

[136]  A. D. Pryamikov, A. S. Biriukov, A. F. Kosolapov, V. G. Plotnichenko, S. L. Semjonov, and E. M. Dianov, "Demonstration of a waveguide regime for a silica hollow-core microstructured optical fiber with a negative curvature of the core boundary in the spectral region > 3.5 μm," Optics Express, 19(2), 1441-1448 (2011).

[137]  R. F. Cregan, B. J. Mangan, J. C. Knight, T. A. Birks, P. St. J. Russell, P. J. Roberts, and D. C. Allan, "Single – mode photonic band gap guidance of light in air," Science, 285(5433), 1537–1539 (1999).

[138]  P. J. Roberts, F. Couny, H. Sabert, B. J. Mangan, D. P. Williams, L. Farr, M. W. Mason, A. Tomlinson, T. A. Birks, J. C. Knight, and P. St. J. Russell, "Ultimate low loss of hollow-core photonic crystal fibres," Optics Express, 13(1), 236–244 (2005).

[139]  P. J. Roberts, D. P. Williams, B. J. Mangan, H. Sabert, F. Couny, W. J. Wadsworth, T. A. Birks, J. C. Knight, and P. St. J. Russell, "Realizing low loss air core photonic crystal fibers by exploiting an antiresonant core surround," Optics Express, 13(20), 8277–8285 (2005).

[140]  J. D. Shephard, W. N. Macpherson, R. P. Maier, J. D. C. Jones, D. P. Hand, M. Mohebbi, A. K. George, P. J. Roberts, and J. C. Knight, "Single-mode mid-IR guidance in a hollow-core photonic crystal fiber," Optics Express, 13(18), 7139–7144 (2005).

[141]  F. Benabid, "Hollow–core photonic band gap fibre: new light guidance for new science and technology," Philosophy Transactions of the Royal Society London, Ser. A 364(1849), 3439–3462 (2006).

[142]  A. Argyros and J. Pla, "Hollow-core polymer fibres with a kagome lattice: potential for transmission in the infrared," Optics Express, 15(12), 7713–7719 (2007).

[143]  F. Gérôme, R. Jamier, J.-L. Auguste, G. Humbert, and J.-M. Blondy, "Simplified hollow-core photonic crystal fiber," Optics Letters, 35(8), 1157–1159 (2010).

[144]  A. F. Kosolapov, A. D. Pryamikov, A. S. Biriukov, V. S. Shiryaev, M. S. Astapovich, G. E. Snopatin, V. G. Plotnichenko, M. F. Churbanov, and E. M. Dianov, "Demonstration of $CO_2$-laser power delivery through chalcogenide-glass fiber with negative-curvature hollow core," Optics Express, 19(25), 25723–25728 (2011).

[145]  G. Renversez, P. Boyer, and A. Sagrini, "Antiresonant reflecting optical waveguide microstructured fibers revisited: a new analysis based on leaky mode coupling," Optics Express, 14(12), 5682–5687 (2006).

[146]  S. M. Mansfield and G. S. Kino, "Solid immersion microscope," Applied Physics Letters, 57, 2615 (1990).

[147]  B. D. Terris, H. J. Mamin, D. Rugar, W. R. Studenmunt, and G. S. Kino, "Near-field optical data storage using a solid immersion lens," Applied Physics Letters, 65, 388-390 (1994).

[148]  M. Baba, T. Sasaki, M. Yoshita, and H. Akiyama, "Aberrations and allowances for errors in a hemisphere solid immersion lens for submicron-resolution photoluminescence microscopy," Journal of Applied Physics, 85, 6923-6925 (1999).

[149]  S. M. Mansfield, "Solid Immersion Microscopy," Ph.D. Dissertation, Standford University (1992).

[150]  T. D. Milster, "Near-field optical data storage: avenues for improved performance," Optical Engineering, 40(10), 2255-2260 (2001).

[151]  G. S. Kino, "The solid immersion lens," Proceedings of SPIE, 3740, 2-5 (1999).

[152]  Q. Wu, L. P. Ghislain, and V. B. Elings, "Imaging with solid immersion lenses, spatial resolution, and applications," Proceedings of the IEEE, 88(9), 1491-1498 (2000).

[153]  K. Karrai, X. Lorenz, and L. Novotny, "Enhanced reflectivity contrast in confocal solid immersion lens microscopy," Applied Physics Letters, 77(21), 3459-3461 (2000).

[154]  T. D. Milster, F. Akhavan, M. Bailey, J. K. Erwin, D. M. Felix, K. Hirota, S. Koester, K. Shimura, and Y. Zhang, "Super-resolution by combination of a solid immersion lens and an aperture," Japanese Journal of Applied Physics, 40(3B), 1778-1782 (2001).

[155]  S. Moehl, H. Zhao, B. Dal Don, S. Wachter, and H. Kalt, "Solid immersion lens enhanced nano-photoluminescence: Principle and applications," Journal of Applied Physics, 93(10), 6265-6272 (2003).

[156]  Y. Zhang, "Design of high-performance supersphere solid immersion lenses," Applied Optics, 45(19), 4540-4546 (2006).

[157]  K. A. Serrels, E. Ramsay, R. J. Warburton, and D. T. Reid, "Nanoscale optical microscopy in the vectorial focusing regime," Nature Photonics, 2(5), 311-314 (2008).

[158]  M.-S. Kim, T. Scharf, M. T. Hag, W. Nakagawa, and H. P. Herzig, "Subwavelength-size solid immersion lens," Optics Letters, 36(19), 3930-3932 (2011).

[159]  A. K. Bates, M. Rothschild, T. M. Bloomstein, T. H. Tedynyshyn, R. R. Kunz, V. Liberman, and M. Switkes, "Review of technology for 157-nm lithography," IBM Journal of Research and Development, 45(5), 605-614 (2001).

[160]  I. Lang-Pauluzzi, "The behavior of plasma membrane during plasmolysis: a study by UV microscopy," Journal of Microscopy, 198(3), 188-198 (2001).

[161]  J. A. N. Scott, A. R. Procter, B. J. Fergus, and D. A. I. Goring, "The Application of Ultraviolet Microscopy to the Distribution of Lignin in Wood Description and Validity of the Technique," Wood Science and Technology, 3, 73-92 (1969).

[162]  E. Abbe, "Beiträge zur theorie des mikroskops und der mikroskopischen wahrnehmung," Archiv für Mikroskopische Anatomie, 9(1), 413-418 (1873).

[163]  Lord Rayleigh, "Investigations in optics, with special reference to the spectroscope," Philosophical Magazine, 8(49), 261-274 (1879).

[164]  C. M. Sparrow, "On spectroscopic resolving power," Astrophysical Journal, 44(2), 76-86 (1916).

[165]  W. V. Houston, "A compound interferometer for fine structure work," Physical Review, 29(3), 478-484 (1927).

[166]  H. Helmholtz, "On the Limits of the Optical Capacity of the Microscope," The Monthly microscopical journal, 16(1), 15-39 (1876).

[167] A. Lipson, S. G. Lipson, and H. Lipson, "Optical Physics, 4th Edition," Cambridge University Press, Cambridge, UK (2010).

[168] X. Hao, C. Kuang, Z. Gu, Y. Wang, S. Li, Y. Ku, Y. Li, J. Ge, and X. Liu, "From microscopy to nanoscopy *via* visible light," Light: Science and Applications, 2, e108 (2013)

[169] E. A. Ash, and G. Nicholls, "Super-resolution aperture scanning microscope," Nature, 237, 510–512 (1972).

[170] G. Binning, H. Rohrer, C. Gerber, and E. Weibel, "Surface studies by scanning tunneling microscopy," Physics Review Letters, 49, 57–61 (1982).

[171] G. Binnig, C. F. Quate, and C. Gerber, "Atomic force microscope," Physics Review Letters, 56, 930–933 (1986).

[172] J. B. Pendry, "Negative refraction makes a perfect lens," Physics Review Letters, 85, 3966–3969 (2000).

[173] N. Fang, H. Lee, C. Sun, and X. Zhang, "Sub-diffraction-limited optical imaging with a silver superlens," Science, 308, 534–537 (2005).

[174] X. Zhang, and Z. W. Liu, "Superlenses to overcome the diffraction limit," Nature Mater, 7, 435–441 (2008).

[175] L. Novotny, and B. Hecht, "Principles of Nano-optics," Cambridge, Cambridge University Press, (2006).

[176] Z. W. Liu, S. Durant, H. Lee, Y. Pikus, N. Fang, Y. Xiong, C. Sun, and X. Zhang, " Far-field optical superlens," Nano Letters, 7, 403–408 (2007).

[177] Z. W. Liu, H. Lee, Y. Xiong, C. Sun, and X. Zhang, "Far-field optical hyperlens magnifying subdiffraction-limited objects," Science, 315, 1686–1686 (2007).

[178] R. S. Hegde, M. A. Fiddy, and W. J. R. Hoefer, "The reconstruction of finite extent objects with the superlens," Applied Physics A, 116, 969-975 (2014).

[179] J. R. Sambles, G. W. Bradbery, and F. Z. Yang, "Optical-excitation of surface-plasmons—an introduction," Contemporary Physics, 32, 173–183 (1991).

[180] I. I. Smolyaninov, C. C. Davis, and A. V. Zayats, "Image formation in surface plasmon polariton mirrors: applications in high-resolution optical microscopy," New Journal of Physics, 7, 175 (2005).

[181]  I. I. Smolyaninov, C. C. Davis, J. Elliott, and A. V. Zayats, "Resolution enhancement of a surface immersion microscope near the plasmon resonance," Optics Letters; 30, 382–384 (2005).

[182]  X. Hao, X. Liu, C. F. Kuang, Y. H. Li, Y. L. Ku, H. Zhang, H. Li, and L. Tong, "Far-field super-resolution imaging using near-field illumination by micro-fiber," Applied Physics Letters, 102, 013104–013104 (2013).

[183]  J. Y. Lee, B. H. Hong, W. Y. Kim, S. K. Min, Y. Kim, M. V. Jouravlev, R. Bose, K. S. Kim, I. Hwang, L. J. Kaufman, C. W. Wong, P. Kim, and K. S. Kim, "Near-field focusing and magnification through self-assembled nanoscale spherical lenses," Nature, 460, 498–501 (2009).

[184]  Z. L. Liau, "Annular solid-immersion lenslet array super-resolution optical microscopy," Journal of Applied Physics, 112, 083110–083117 (2012).

[185]  M. Hasan, and J. J. Simpson, "Photonics nanojet-enhanced nanometer-scale germanium photodiode," Applied Optics, 52(22), 5420-5425 (2013).

[186]  L. Esaki, and R. Tsu, "Superlattice and negative differential conductivity in semiconductors," IBM Journal of Research and Development, 14(61), 5 (1970).

[187]  D. Z. Ting, A. Soibel, S. A. Keo, S. B. Rafol, J. M. Mumolo, J. K. Liu, C. J. Hill, A. Khoshakhlagh, L. Hoglund, E. M. Luong, and S. D. Gunapala, "Development of quantum well, quantum dot, and type II superlattice infrared photodetectors," Journal of Applied Remote Sensors, 8(1), 084998 (2014).

[188]  L. Tang, S. E. Kocabas, S. Latif, A. K. Okyay, D. S. L. Gagnon, K. C. Saraswat, and D. A. B. Miller, "Nanometre-scale germanium photodetector enhanced by a near-field dipole antenna," Nature Photonics, 2, (2008).

[189]  P. Vasinajindakaw, J. Vaillancourt, G. Gu, R. Liu, Y. Ling, and X. Lu, "A Fano-type interference enhanced quantum dot infrared photodetector," Applied Physics Letters, 98, 211111 (2011).

[190]  S. Arnold, J. Comunale, W. B. Whitten, J. M. Ramsey, and K. A. Fuller, "Room-temperature microparticle-based persistent hole-burning spectroscopy," Journal of the Optical Society of America B, 9(5), 819-824 (1992).

[191]  D. X. Hammer, G. D. Noojin, R. J. Thomas, C. E. Clary, B. A. Rockwell, C. A. Toth, and W. P. Roach, "Intraocular laser surgical probe for membrane disruption by laser-induced breakdown," Applied Optics, 36(7) 1684-1693 (1997).

[192]  G. M. Hale and M. R. Querry, "Optical constants of water in the 200 nm to 200 μm wavelength region," Applied Optics, 12(3), 555-563 (1973).

[193]  Y. Xiong, Z. Liu, C. Sun, X. Zhang, "Two-dimensional Imaging by far-field superlens at visible wavelengths," Nano Letters, 7, 3360–3365 (2007).

[194]  J. A. Harrington, "Infrared hollow waveguides: An update," Proceedings of SPIE, Optical Fibers and Sensors for Medical Diagnostics and Treatment Applications VIII, 6852, 68520M (2008).

[195]  A. Darafsheh, A. Fardad, N. M. Fried, A. N. Antoszyk, H. S. Ying, and V. N. Astratov, "Contact focusing multimodal microprobes for ultraprecise laser tissue surgery," Optics Express, 19(4), 3440-3448 (2011).

[196]  J. G. Sousa, D. Welford, and J. Foster, 'Efficient 1.5 W CW and 9 mJ quasi-CW TEM00 mode operation of a compact diode-laser-pumped 2.94-μm Er:YAG laser," Proceedings of SPIE, Solid State Lasers XIX: Technology and Devices, 7578, 75781E (2010).

[197]  G. L. Spaeth, H. Danesh-Meyer, I. Goldberg, and A. Kampik, "Ophthalmic Surgery: Principles and Practice, 4th Edition" Saunders, Philadelphia, PA (2011).

[198]  G. A. Peyman, "Method for modifying corneal curvature," US Patent US4840175A (1989).

[199]  Scheaumann Laser Inc., "User Manual for Er:YAG laser diode," Sheaumann Laser Inc., acquired (2012).

[200]  R. D. Niederriter, J. T. Gopinath, and M. E. Siemens, "Measurement of the $M^2$ beam propagation factor using a focus-tunable liquid lens," Applied Optics, 52, 1591-1598 (2013).

[201]  I. D. Johnston, D. K. McCluskey, C. K. L. Tan, and M. C. Tracey, "Mechanical characterization of bulk Sylgard 184 for microfluidics and microengineering," Journal of Micromechanics and Microengineering, 24, 035017 (2014).

[202]  Information about Dow Corning ® Brand Silicone Encapsulants. Product Information. Dow Corning and Sylgard are registered trademarks of Dow Corning Corporation. ©2000-2008 Dow Corning Corporation. All rights reserved. Form No. 10-898I-01.

[203]  B. S. Luk'yanchuk, N. Arnold, S. M. Huang, Z. B. Wang, and M. H. Hong, "Three-dimensional effects in dry laser cleaning," Applied Physics A, 77(2), 209-215 (2003).

[204]  B. S. Luk'yanchuk, Y. W. Zheng, and Y. Lu, "Laser cleaning of solid surface: optical resonance and near-field effects," Proceedings of SPIE, High-Power Laser Ablation, 576-587 (2000).

[205]  Y. W. Zheng, B. S. Luk'yanchuk, Y. Lu, W. D. Song, and Z. H. Mai, "Dry laser cleaning of particles from solid substrates: experiments and theory," Journal of Applied Physics, 90(5), 2135-2142 (2001).

[206]  J. C. Hulteen and R. P. V. Duyne, "Nanosphere lithography: A materials general fabrication process for periodic particle array surfaces," Journal of Vacuum Science and Technology A, 13, 1553 (1995).

[207]  K. T. Posani, V. Tripathi, S. Annamalai, N. R. Weisse-Bernstein, S. Krishna, R. Perahia, O. Crisafulli, and O. J. Painter, "Nanoscale quantum dot infrared sensors with photonic crystal cavity," Applied Physics Letters, 88, 151104 (2006).

[208]  T. Ishi, J. Fujikata, K. Makita, T. Baba, and K. Ohashi, "Si Nano-Photodiode with a Surface Plasmon Antenna," Japanese. Journal of Applied Physucs, 44, L364-L366 (2005).

[209]  G. Konstantatos and E. H. Sargent, "Nanostructured materials for photon detections," Nature Nanotechnology, 5, 391-400 (2010).

[210]  H. Fischer and O. J. F. Martin, "Engineering the optical response of plasmonic nanoantennas," Optics Express, 16, 9144-9154 (2008).

[211]  J. Fahrenfort, "Attenuated total reflection: A new principle for the production of useful infra-red reflection spectra of organic compounds,"Spectrochemica Acta, 17(7), 698-709 (1961).

[212]  E. Goormaghtigh, V. Cabiaux, and J-M. Ruysschaert, " Secondary structure and dosage of soluble and membrane proteins by attenuated total reflection Fourier-transform infrared spectroscopy on hydrated films," European Journal of Biochemistry, 193(2), 409-420 (1990).

[213]  S. Frey and L. K. Tamm, "Orientation of melittin in phospholipid bilayers: A polarized attenuated total reflection infrared study," Biophysics Journal, 60, 922-930 (1991).

[214]  W. Hubner and H. H. Mantsch, " Orientation of specifically [13]C=O labeled phosphatidylcholine multilayers from polarized attenuated total reflection FT-IR spectroscopy," Biophyics Journal, 59, 1261-1272 (1991).

[215]  D. Bommannan, R. O. Potts, and R. H. Guy, "Examination of Stratum Corneum Barrier Function In Vivo by Infrared Spectroscopy," Journal of Investigative Dermatology, 95, 403-408 (1990).

[216]  J. Sanghera and I. D. Aggarwal, "Infrared Fiber Optics," CRC Press LLC, 349 (1998).

[217]  H. Tokiwa and Y. Mimuran, "Ultralow-loss fluoride-glass single-mode fiber design," Journal of Lighwave Technology, LT-4 (8), 1260-1266 (1986).

[218]  Y. Ohishi, S. Mitachi, and S. Takahashi, "Fabrication of Fluoride Glass Single-Mode Fibers," Journal of Lightwave Technology, LT-2 (5), 593-596 (1984).

[219]  V. N. Astratov and A. Darafsheh, "Methods and systems for super-resolution optical imaging using high-index of refraction microspheres and microcylinders," US patent application 14042834 filed on 10/01/2013 with priority from original application on 06/07/2012

## APPENDIX A: PEER-REVIEWED PUBLICATIONS

[1]     K. W. Allen, A. Darafsheh, F. Abolmaali, N. Mojaverian, N. I. Limberopoulos, A. Lupu, and V. N. Astratov., "Microsphere-chain waveguides: Transport and focusing properties," Applied Physics Letters, 105, 021112 (2014).

[2]     K.W. Allen, J. Duran, N.I. Limberopoulos, A. M. Urbas, and V.N. Astratov, "Photonic Jets for Strained-Layer Superlattice Infrared Photodector Enhancement," IEEE proc. of NAECON, (accepted 2014).

[3]     K.W. Allen, N. Farahi, Y. Li, D. E. Walker Jr., N.I. Limberopoulos, A. M. Urbas, and V.N. Astratov, "Super-Resolution Imaging by High-Index Spheres Embedded in Transparent Matrices," IEEE proc. of Transparent Optical Networks (ICTON), (accepted 2014).

[4]     K.W. Allen, N. Farahi, Y.Li, N.I. Limberopoulos, D. E. Walker Jr., A. M. Urbas, V. Liberman, and V.N. Astratov, "Attachable thin films with embedded high-index microspheres for super-resolution microscopy," to be submitted to Laser and Photonics Review, (submitted for clearance Fall 2014).

[5]     K.W. Allen, J. Duran, N.I. Limberopoulos, and V.N. Astratov, "Direct Observation of ~100x Photocurrent Enhancement by Photonic Jets for Infrared Photodetectors," to be submitted to Applied Physics Letters, (Draft in preparation, to be submitted Spring 2015).

[6]     K. W. Allen, "Light Focusing and Transport Properties of Mesophotonic Structures," University of North Carolina at Charlotte, (2012).

[7]     K. W. Allen, A. F. Kosolapov, A. N. Kolyadin, A. D. Pryamikov, N. Mojaverian, N I. Limberopoulos, and V. N. Astratov, "Photonic jets produced by microspheres integrated with hollow-core fibers for ultraprecise laser surgery," IEEE proc. of Transparent Optical Networks (ICTON), 15[th] International Conference, 1-4 (2013).

[8]     K. W. Allen, A. Darafsheh, and V. N. Astratov, "Photonic nanojet-induced modes: from physics to applications," IEEE proc. of Transparent Optical Networks (ICTON), 2011 13[th] International Conference, 1-4 (2011).

[9]     K. W. Allen, A. Darafsheh, and V. N. Astratov, "Beam tapering effect in microsphere chains: from geometrical to physical optics," International Society for Optics and Photonics, SPIE LASE, 823622-823622-7 (2012).

[10]    A. Darafsheh, N. Mojaverian, N. I. Limberopoulos, K. W. Allen, A. Lupu, and V. N. Astratov, "Formation of polarized beams in chains of dielectric spheres and cylinders," Optics Letters, 38(20), 4208-4211 (2013).

[11]    Y. Li, K.W. Allen, F. Abolmaali, N.I. Limberopoulos, and V.N. Astratov, "Spectral Fingerprints of Photonic Molecules," IEEE proc. of Transparent Optical Networks (ICTON), (accepted 2014).

[12]    V. N. Astratov, A. Darafsheh, M. D. Kerr, K. W. Allen, N. M. Fried, A. N. Antoszyk, and H. S. Ying, "Photonic nanojets for laser surgery," SPIE Newsroom, 12, 32-34 (2010).

[13]    A. Darafsheh, M. D. Kerr, K. W. Allen, N. M. Fried, A. N. Antoszyk, H. S. Ying, and V. N. Astratov, "Integrated microsphere arrays: light focusing and propagation effects," International Society for Optics and Photonics, SPIE OPTO, 76050R-76050R-9 (2010).

[14]    V. N. Astratov, A. Darafsheh, M. D. Kerr, K. W. Allen, and N. M. Fried, "Focusing microprobes based on integrated chains of microspheres," Proc. of Progress in Electromagnetics Research Symposium, 5-8 (2010).

[15]    A. Darafsheh, K. W. Allen, A. Fardad, N. M. Fried, A. N. Antoszyk, H. S. Ying, and V. N. Astratov, "Focusing capabilities of integrated chains of microspheres in the limit of geometrical optics," International Society for Optics and Photonics, SPIE LASE, 79131A-79131A-7 (2011).

[16]    A. Darafsheh, M. D. Kerr, K. W. Allen, and V. N. Astratov, "Integrated microsphere arrays as a compact focusing tool for biomedical and photonics applications," IEEE proc. of Laser and Electro-Optics (CLEO) and Quantum Electronics and Laser Science Conference (QELS), 1-2 (2010).

APPENDIX B: PATENTS

[1]     Vasily N. Astratov, Kenneth W. Allen Jr., Joshua M. Duran, Nicholaos I. Limberopoulos, and Augustine M. Urbas, "Methods and Systems for Enhancing the Performance of Photodetector Focal Plane Arrays," Provisional U.S. patent filed, (2014).

APPENDIX C: ORIGIN OF THE PHOTONIC NANOJET

As discussed in Chapter 1, the concept of "photonic nanojet" has an interesting history. The terminology of photonic nanojets was introduced in 2004 in the context of achieving subdiffraction-limited beam waists as a result of ultra-sharp focusing of light by dielectric cylinders and spheres. The concept was initially introduced based on FDTD numerical modeling [73, 74] and later it was studied analytically as a result of interference of zero-order transmitted and scattered waves [80]. The problem is rather complicated mathematically because of a series of factors which can play some role in such focusing depending on the specific parameters (index and size of the sphere). These factors include involvement of near-field optics in cases where the nanojet is formed sufficiently close to the back surface of the spheres [95] and a possibility of coupling of light to whispering gallery modes in spheres [12]. These studies showed that besides unusually narrow beam waists, photonic nanojets can be elongated substantially along the propagation direction. The tightly focused beam can propagate several wavelengths without significant divergence. For the definition of a photonic nanojet, however, the key property is the fact that the beam waist can be significantly smaller than the diffraction limit. In ref. [73] beam waists as small as $\lambda/3$ have been numerically demonstrated for small dielectric microspheres illuminated with plane waves propagating in air.

It should be noted that the subject of ultrasharp focusing of light, by microspheres, has been of interest for researchers well before introduction of photonic nanojets. This interest was driven in particular by applications of microspheres in dry laser cleaning [203-205] and periodical patterning [206]. It has been generally known that tiny beams as small as 80 nm at $\lambda$=240 nm can be obtained for dielectric microspheres at the top of high-index substrates [101].

In order to study this in a more detailed way, the position of the narrowest waist produced by a sphere on the substrate was determined, using FDTD calculations, for the conditions considered in Ref. [101]. A 248 nm KrF excimer laser illuminated a silica sphere $D$=0.5 $\mu$m which was in contact with a silicon (100) surface. Experimentally it has been shown that the focused beam produced hillocks of ~100 nm [101]. Our FDTD simulation predicted a 100 nm beamwidth at the interface between the silica sphere and the silicon slab, shown in the Fig. C.1 below.

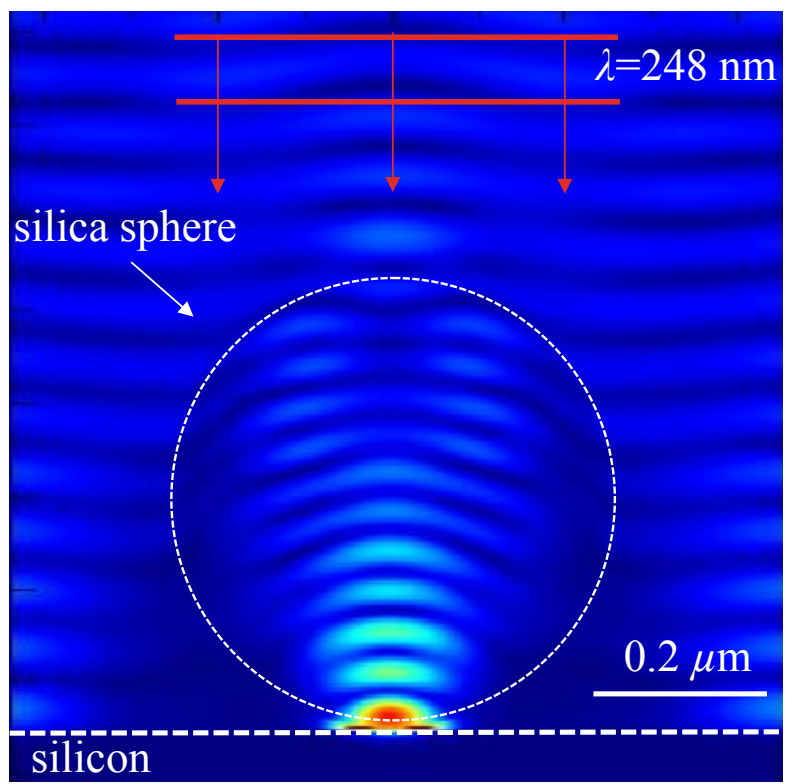

Figure C.1: FDTD calculations of a silica sphere, on top of a silicon slab, illuminated by a plane wave.

Ref. [101] claimed that Mie theory calculations predicted an 80 nm beamwidth. We found that the narrowest waist of the focused beam takes place 15 nm within the back surface of the silica sphere, and the beam FWHM=80 nm was in agreement with Ref. [101]. However, by using the index of the object-space $n{\sim}1.5$, we can estimate that this FWHM is roughly consistent with the classical diffraction limit. Indeed, the diffraction-limited beam FWHM can be estimated as $d{=}\lambda/2\mathrm{NA}$, where $\mathrm{NA}{=}n\sin\theta{\leq}n$ and $n$ is the index of the object-space. Using $n{\sim}1.5$ it can be determined that $d{\sim}\lambda/3$, which is qualitatively consistent with the numerical results of ref. [101].